\documentstyle[aps,preprint,tighten,eqsecnum]{revtex}
\draft

\newcommand{\bc}{\begin{center}}
\newcommand{\ec}{\end{center}}
\newcommand{\nin}{\noindent}
\newcommand{\be}{\begin{equation}}
\newcommand{\ee}{\end{equation}}
\newcommand{\ba}{\begin{array}}
\newcommand{\ea}{\end{array}}

\newcommand{\dosc}{{d^{\rm osc}}}
\newcommand{\Dosc}{{D^{\rm osc}}}
\newcommand{\nosc}{{n^{\rm osc}}}
\newcommand{\Nosc}{{N^{\rm osc}}}
\newcommand{\oosc}{{\omega^{\rm osc}}}
\newcommand{\Oosc}{{\Omega^{\rm osc}}}

\newcommand{\Gsc}{G_E^{\text sc}}
\newcommand{\cl}{\chi_{\scriptscriptstyle L}}
\newcommand{\cgc}{\chi^{\rm \scriptscriptstyle GC}}
\newcommand{\kf}{k_{\scriptscriptstyle F}}
\newcommand{\kb}{k_{\scriptscriptstyle B}}
\newcommand{\vf}{v_{\scriptscriptstyle F}}

\newcommand{\br}{{\bf r}}
\newcommand{\bM}{{\bf M}}
\newcommand{\gs}{{\sf g_s}}
\newcommand{\Hch}{\hat{\cal H}}

\newcommand{\hM}{\hat M}

\newcommand{\dif}{{\rm d}}
\newcommand{\bsM}{{\bf \scriptscriptstyle M}}

\newcommand{\A}{{\cal A}}
\newcommand{\C}{{\cal C}}
\newcommand{\G}{{\cal G}}

\newcommand{\bp}{{\bf p}}
\newcommand{\bq}{{\bf q}}
\newcommand{\bA}{{\bf A}}

\begin{document}

\vspace{-3.0cm}

\title{\bf ORBITAL MAGNETISM IN THE BALLISTIC REGIME:
GEOMETRICAL EFFECTS}

\vspace{1cm}

\author{Klaus Richter$^{(1,2)}$, Denis Ullmo$^{(1,3)}$, and 
Rodolfo A. Jalabert$^{(1,4)}$} 

\address{$^{(1)}$Division de Physique Th\'eorique,
Institut de Physique Nucl\'eaire, 91406 Orsay Cedex, France }

\address{$^{(2)}$Institut f\"{u}r Physik, Memminger Str. 6,
86135 Augsburg, Germany}

\address{$^{(3)}$AT\&T Bell Laboratories, 1D-265, 600 Mountain Avenue,
Murray Hill, New Jersey 07974-0636}

\address{$^{(4)}$Universit\'e Louis Pasteur, IPCMS-GEMME, 
23 rue du Loess, 67037 Strasbourg Cedex, France}

\vspace{1cm}

\date{\today}

\maketitle

\vspace{16cm}
\noindent
Submitted to Physics Reports
\hspace{7cm}   IPNO/TH 95--49

\newpage

{\tighten
\begin{abstract}

We present a general semiclassical theory of the orbital
magnetic response of noninteracting electrons confined in
two-dimensional potentials. We calculate the magnetic 
susceptibility of singly-connected and the 
persistent currents of multiply--connected geometries. We
concentrate on the geometric effects by studying confinement by perfect 
(disorder free) potentials  stressing the importance
of the underlying classical dynamics. We demonstrate that in 
a constrained geometry the standard Landau diamagnetic 
response is always present, but is dominated by finite-size
corrections of a quasi-random sign which may be orders of
magnitude larger. These corrections are very sensitive to the
nature of the classical dynamics.  Systems which are 
integrable at zero magnetic field exhibit larger 
magnetic response than those which are chaotic. This 
difference arises from the large oscillations of the density 
of states in integrable systems due to the existence of families
of periodic orbits. The connection between quantum and 
classical behavior naturally arises from the use of semiclassical
expansions. This key tool becomes particularly simple and
insightful at finite temperature, where only short classical
trajectories need to be kept in the expansion. In addition to the
general theory for integrable systems, we analyze in detail a few
typical examples of experimental relevance: circles, rings and
square billiards. In the latter, extensive numerical calculations are used
as a check for the success of the semiclassical analysis. We study 
the weak--field regime where classical trajectories remain essentially
unaffected, the intermediate field regime where we identify new 
oscillations characteristic for ballistic mesoscopic structures,
and the high-field regime where the typical de Haas--van Alphen
oscillations exhibit finite-size corrections.
We address the comparison with
experimental data obtained in high-mobility semiconductor microstructures
discussing the differences between individual and ensemble measurements,
and the applicability of the present model.

\end{abstract}
}

\narrowtext

\newpage
\tableofcontents

\newpage

\section {Introduction}
\label{sec:Intro}

\subsection {Historical perspective}

The study of orbital magnetism in an electron gas goes back to the 1930's
with the pioneering work of Landau \cite{Land,LanLip} demonstrating the
existence of a small diamagnetic response at weak fields $H$ and 
low temperatures $T$ (such that $\kb T$ exceeds the typical spacing
$\hbar \omega$, \ $\omega=eH/mc$). Three features of
this original proposal contributed to the slowness of its general 
acceptance.
First, it deals with a purely quantum result that can be expressed as a 
thermodynamic relationship without an explicit $\hbar$ dependence. In 
contrast to that the Bohr-van Leeuwen theorem \cite{vLeeu21} 
establishes the absence of magnetism for a system of 
classical particles. For finite systems the boundary currents are
shown to exactly cancel the diamagnetic contribution from cyclotron orbits
of the interior. (This result remains valid even if we consider Fermi or
Bose statistics \cite{Peierls}.) Secondly, boundary effects (so crucial in
obtaining the correct classical behavior) did not enter into Landau's
original derivation. 
Twenty years later Sondheimer and Wilson \cite{SondWil}
presented a more rigorous formulation for the magnetism of 
{\em unconstrained} electrons at weak and strong fields without 
using explicit knowledge of the energy levels, thus avoiding
complicated arguments involving 
boundary electrons. (Here we present a semiclassical derivation of Landau 
diamagnetism independent of the energy level structure and
valid for {\em constrained} geometries at arbitrary magnetic fields.)
Finally, Landau diamagnetism for standard metals yields a small effect
(one third of the Pauli spin paramagnetism) making its experimental 
observation rather difficult. Peierls \cite{Peierls,Pei33} showed shortly
after Landau's work that the diamagnetic susceptibility persists 
when electrons
are placed in a periodic potential and its value is obtained by simply 
using the effective mass instead of the free electron mass. 
But even if the effective mass is smaller than the bare mass, and the 
diamagnetic orbital response dominates over the spin paramagnetic 
susceptibility (as  typically happens in doped semiconductors), 
the detailed comparison with the experimental data on metals was 
still difficult 
\cite{SondWil}. This follows from the complicated electronic structure and
the fact that taking into account electron-electron interactions in the
same way as a periodic potential by 
renormalising the effective mass, is a too crude approximation.

While the restriction of the electron gas to a two--dimensional plane
(still in the thermodynamic limit) does not pose any new conceptual
or calculational difficulty \cite{Peierls,Shoe}, the effect of confining 
the electron system to a finite volume introduces a new energy scale in 
the problem (the typical level spacing $\Delta$) and leads to a 
modification of the Landau susceptibility. The latter point has 
therefore been the object of 
several conflicting studies\footnote{For the historical account of this 
tortuous and often contradictory chain of findings see 
Ref.~\cite{vanLthesis}.}. 
The investigation of finite-size corrections was motivated by 
experiments on small metal clusters and dealt with various 
theoretical models: thin plates
\cite{plate}, thin cylinders \cite{Dingle52}, confinement by quadratic
potentials \cite{Denton73,parabol}, circular \cite{Bog} 
and rectangular boxes 
\cite{Bivin77,LOS95}. Finite-size effects and corrections to bulk
magnetism obviously depend on the relation between
the typical size $a$ of the system and other relevant
length scales \cite{Sha}: The thermal length 
$L_T =\hbar \vf \beta / \pi$, \ ($\vf$ is
the Fermi velocity and $\beta=1/\kb T$), the elastic mean free path $l$
(with respect to impurity scattering), and the phase-coherence length
$L_{\Phi}$ (taking into account inelastic processes like electron-phonon 
scattering). Most of the above mentioned studies neglect other scattering 
mechanisms than that by the boundaries of the device,
and deal with the macroscopic (or high temperature) case of 
$L_T \! \ll \! a$. 
The first assumption severely limits the possible
comparison with the experimental data of real metal clusters, while
in the regime where the second assumption is valid the magnetic
response is dominated by its smooth component for which, as
generally shown by Robnik \cite{rob86} and Antoine \cite{antoine}, only
small corrections to the the diamagnetic bulk susceptibility are found.

Opposite to the macroscopic limit, there had been studies in the
extreme quantum limit \cite{vRuitenb}, where the temperature is low
enough to enable the resolution of individual levels 
($\kb T \! < \! \Delta$). In this regime the magnetic
susceptibility is dominated by erratic fluctuations that, as we will see
later, hinder its unequivocal determination. The purpose of the 
present work is
the study of size corrections in the {\em mesoscopic} 
regime \cite{Imrymeso}, intermediate
between the two previous limits (that is, for temperatures verifying
$L_T/a\!>\!1\!>\!\beta \Delta$) and where inelastic processes do not
inhibit quantum interference effects ($L_{\Phi}>a$). 
Nowadays this is 
an experimentally accessible regime receiving considerable attention
due to the richness of its physical properties. 
When $L_{\Phi}\!>\!a\!>\!l$ we have the mesoscopic 
{\em diffusive} regime where
the electron motion is dominated by impurity scattering, while for
$L_{\Phi}\!>\!l\!>\!a$ we enter into the {\em ballistic} regime where 
electrons are
mainly scattered off the walls of a confining potential.
A central conclusion of our work is that finite--size corrections
to the magnetic susceptibility in the {\em ballistic}
regime can be {\em orders of magnitude} larger than the bulk values.

One of the reasons for the sustained interest of the last few years in
the mesoscopic ballistic regime is the possibility of studying the relation
of the underlying classical dynamics and the quantum properties. This issue
is precisely the subject of the field known as ``quantum chaos''
\cite{gutz_book,RevBoh}. 
Since the number of electrons $N$ in a mesoscopic system is always large,
particles at the Fermi energy have a De Broglie wave length $\lambda_F$
much smaller than the typical size $a$ of the system ($a/\lambda_F \propto
\kf a \propto N^{1/d}$, $\kf$ is the Fermi wave vector, $d$ the number of
degrees of freedom), and are therefore well in the semiclassical regime.
High-mobility mesoscopic semiconductor samples provide an 
appropriate experimental testing ground in this context and have 
been recently examined with respect to the role of chaos in 
transport phenomena (for a review, see \cite{Chaos,Chaost}). 
The present work extends the connection 
between mesoscopic systems and quantum chaos to thermodynamic properties,
and analyzes recent experiments \cite{levy93,BenMailly} measuring the
magnetic response of ballistic microstructures. One main concern of our
work is to show that mesoscopic finite-size effects 
depend crucially on the classical dynamics of the ballistic billiard, 
i.e.\ whether it is integrable or chaotic, and that the magnetic
response provides an experimentally accessible 
criterion in order to distinguish between integrable and chaotic devices
(much more neatly than through the subtle differences found in the
transport problem \cite{Chaos,Chaost}).

The importance of geometrical effects for the finite-size
corrections in the above defined macroscopic limit had already been noticed
in terms of the sensitivity of the magnetic susceptibility on
the structure of the confining potential \cite{Dingle52,Denton73,parabol}.
The chosen potentials were obviously non generic but used due to their
calculational simplicity, and therefore it was not possible to
anticipate the order-of-magnitude effect that classical dynamics
might have on the susceptibility outside the macroscopic limit.
The problem of orbital magnetism from a quantum chaos point of view
was first addressed by Nakamura and Thomas \cite{NakTho88} in their
numerical study of the differences in the magnetic response of 
circular and elliptic billiards at zero temperature. The circular billiard 
is integrable
at arbitrary field, while the ellipse develops chaotic behavior at finite
fields. They found a reduction compared to the bulk susceptibility and
strong fluctuations (with varying magnetic field), and observed 
that both effects
were stronger for the elliptic billiard. As already mentioned, the
difficulty of these studies in the extreme quantum limit 
(at zero temperature)
consists in the existence of strong fluctuations arising from exact or
quasi--crossings of energy levels (depending parametrically on the magnetic
field) where the susceptibility diverges.
Similar features were obtained for other integrable systems
in the quantum limit like
the rectangular box  \cite{vRuitenb}, the Corbino disk and the
cylinder \cite{Rezak_etal91}. However,
this unphysical behavior is regularized by finite temperature that 
approximately adjusts the populations of both levels to each other
at a crossing (or anti-crossing). 

Parallel to the studies of the orbital response in finite size 
singly--connected systems, there have been important developments in the 
understanding of persistent currents (i.e. the orbital magnetism in
multiply--connected geometries)\footnote{For historical accounts on 
persistent currents see Refs.~\cite{von_thesis,LanMori}.}.
These latter studies started usually from very general considerations
without making the connection with the
Landau diamagnetism. The pioneering work of B\"{u}ttiker, Imry and Landauer
\cite{BIL} demonstrating that in the presence of magnetic flux the ground
state of a one dimensional ring has a current flow generated a large
theoretical activity, mainly directed towards generalizations of quasi-one
dimensional and diffusive rings \cite{von_thesis,RiGe}. 
The first experimental evidence
of persistent currents in an ensemble of mesoscopic copper rings 
was given by the 1990 measurement of L\'evy {\em et al} \cite{percon}.
The use of an
ensemble was motivated by experimental reasons and brought up important
issues about the differences between the canonical and grand canonical
ensembles in the mesoscopic regime \cite{BM,Imry,ensemble} that we will
review in the present work.
Later experiments achieved the measurement of persistent currents in 
single disordered \cite{Chandra91} and ballistic \cite{BenMailly} rings.
In Sec.~\ref{sec:integrable} we analyze in detail the last experiment
making the connection with the orbital magnetism of the other sections. 
The connection between classical mechanics and persistent currents has
already been explored in Refs. \cite{Illinois,RivO,BerryKeat}.

Small metallic samples at sufficiently low temperatures operate in the
diffusive mesoscopic regime, where the classical electron motion is a 
random walk through the impurity potential. This was the regime of the
original experiment on persistent currents \cite{percon} and therefore
received considerable theoretical attention. The effect of disorder on
persistent currents has been evaluated by diagrammatic perturbation
theory \cite{ensemble}. A weak disorder potential does not alter the
bulk Landau diamagnetism \cite{Che} and gives within perturbation theory 
enhancement factors proportional to $\kf l$ for finite samples 
\cite{disor}. Highly pure semiconductor heterojunctions combined with
lithographic techniques allow the realization of samples small enough to be
in the mesoscopic ballistic regime where $l>a$. This is the case of the
orbital magnetism and persistent current measurements of 
Refs.~\cite{levy93} and \cite{BenMailly}. In the ballistic regime electrons
move almost straight between collisions with the walls of the confining
potential. The small drift between collisions is due to the unavoidable
disorder potential existing in real structures. Neglecting completely the
effect of disorder, and therefore the associated drift, leads to an ideal
or {\em clean} system which describes simply an electron billiard. A
central result of our work is that the application of semiclassical
expansions at finite temperature allows one only to consider {\em short 
classical periodic trajectories}. Therefore, the clean model provides a
reasonable approach to the weak and smooth disorder of the 
ballistic regime.
We will get back to this point in this work, and in a separate paper
\cite{rod2000} we examine in detail the role of disorder in ballistic
samples. 

The previously cited developments, as well as most of the present work, 
deal with finite-size effects in the orbital response at {\em weak fields}.
At high fields the magnetic response is dominated by the occurrence of 
Landau levels (whose spacing $\hbar \omega$ is much larger than $\Delta$
or $\kb T$) yielding the well known de Haas--van Alphen effect. 
In 1938 Landau derived
(see Refs.~\cite{SondWil} and \cite{Shoe}) a complete analytical expression
for the susceptibility of a degenerate free electron gas including the
weak-field diamagnetic response and the de Haas--van Alphen oscillations.
Since the latter turned out to be a powerful technique to examine the
electron structure of metals \cite{AsMe} its study has been at the heart of
Condensed Matter Physics for various decades. Its measurement in
two-dimensional electron gases has allowed the determination of the
density of states at high field \cite{Eisen}. With the advent of Mesoscopic
Physics the question of finite-size effects on the de Haas - van Alphen
effect was naturally raised, and free electron gases on a disk 
\cite{SivanImry,jap} and confined by a parabolic potential \cite{HajShap}
were considered at high fields. 
The semiclassical theory used in the
present work provides the finite--temperature  
susceptibility at arbitrary fields and allows the identification 
of an intermediate regime characteristic for ballistic samples that 
we discuss in Sec.~\ref{sec:highB}.

Our work aims at the convergence of various seemingly disconnected fields:
Landau diamagnetism, persistent currents, de Haas - van Alphen effect,
finite-size corrections of thermodynamic functions, quantum chaos, and 
electronic properties of weakly disordered systems. 
We will show that the semiclassical analysis naturally enters in the
problem of the magnetic response of ballistic structures, provided
a model of non-interacting electrons is adequate. The expression of the
magnetic susceptibility and persistent currents in terms of classical
trajectories provides a unifying approach applicable to various
geometrical shapes, different temperatures and magnetic field strengths.

\subsection{Susceptibility of unconstrained and constrained electron
systems}

We now present the basic formulas defining the magnetic
susceptibility and then compare the unconstrained magnetic response with
the susceptibility obtained by confining the electron gas to a finite
region to illustrate the subject of our studies in this paper.
Let us consider a noninteracting electron gas confined in a volume 
(area in two dimensions) $A$ at temperature $T$ under a magnetic 
field $H$. The magnetic moment of the system in statistical equilibrium 
is given by the thermodynamic relation
	\begin{equation}
	{\cal M} = - \left(\frac{\partial \Omega}{\partial H}
	\right)_{T,\mu} 
	\label{eq:magngc}
	\end{equation}
where $\Omega(T,\mu,H)$ is the thermodynamic potential, and $\mu$ the
chemical potential of the electron gas. The
differential magnetic susceptibility is defined by
	\begin{equation}
	\cgc =  \frac{1}{A} \ 
	\left(\frac{\partial {\cal M}}{\partial H} \right)_{T,\mu} =
	- \frac{1}{A} \left(\frac{\partial^{2}\Omega}{\partial H^{2}}
	\right)_{T,\mu} \ .
	\label{eq:susgc}
	\end{equation}
The notation with the superscript GC is used in order to emphasize the
fact that we are working in the grand canonical ensemble. The choice of the
ensemble in the macroscopic limit of $N$ and $A$ $\rightarrow \infty$ is a
matter of
convenience. As it is well known by now \cite{BM,Imry,ensemble} 
the equivalence
between the ensembles may break down in the 
mesoscopic regime that interests us,
and this point will be thoroughly discussed in the remaining of the paper.
However, for the purpose of this didactical introduction 
we will work in the
grand canonical ensemble studying the magnetic response of 
electron systems with fixed chemical potentials.
The calculation advantages of the GC ensemble arise from the
simple form of the thermodynamic potential
	\begin{equation}
	\Omega(T,\mu,H) = - \frac{1}{\beta} \int \ {\rm d} E \ d(E) \
	\ln{(1+\exp{[\beta(\mu\!-\!E)]})} \ ,
	\label{eq:therpot}
	\end{equation}
in terms of the single--particle density of states 
	\be
	d(E) = \gs  \sum_{\lambda} \delta (E-E_{\lambda}) \ .
	\label{DOS}
	\ee
The factor $\gs \! = \! 2$ takes into account spin degeneracy,
$E_{\lambda}$ are the eigenenergies of the system.
The magnetic susceptibility is directly extracted from the knowledge of the
density of states. The case of a free electron gas is particularly simple
since the electron eigenstates are Landau states with energies
	\be
	E_{k} = \hbar w \ (k+1/2) \hspace{2cm} k=0,1,2,\ldots
	\label{LL}
	\ee
and degeneracies $\gs \Phi/\Phi_0$. The cyclotron frequency is $w=eH/mc$,
$\Phi = H A$ is the flux through an area $A$, and $\Phi_0 = hc/e$ is the
elemental flux quantum. 
Throughout this work we will neglect the Zeeman splitting term due 
to the electron spin.  It can however be incorporated easily when 
spin-orbit coupling is negligible \cite{Harsh}.

Landau's derivation of the magnetic susceptibility of a free 
electron system arising from
the quantization condition (\ref{LL}) can be found for the 
three--dimensional case in standard textbooks \cite{LanLip,Peierls}.
The two--dimensional case \cite{Shoe,vanLthesis} follows upon 
the same lines.
In the following we present a sketch of the latter which will be 
useful towards a semiclassical understanding of the problem. 
($H$ is now the component of the field perpendicular to the plane 
of the electrons.)

By the use of the Poisson summation formula the density of states
 related to the quantization condition (\ref{LL}) can be written as
	\be
	d(E) = \gs \frac{mA}{2 \pi \hbar^{2}} 
	        + \gs \frac{mA}{\pi \hbar^{2}}
	\sum_{n=1}^{\infty} (-1)^{n}
			\cos{\left(\frac{2\pi n E}{\hbar w}\right)} \ .
	\label{DOSLL}
	\ee
This decomposition is usually interpreted as coming from the Weyl term
(given by the volume of the energy manifold in phase space) and the
contribution of cyclotron orbits (second term, strongly energy dependent).
We stress though that in the bottom of the spectra, from which the
Landau diamagnetic component originates, this distinction is 
essentially meaningless.

In the case of a degenerate electron gas
with a weak field such that $\hbar w \ll k_{B}T \ll \mu$ 
the energy integral (\ref {eq:therpot}) is easily performed resulting in
	\begin{equation}
	\Omega(\mu) \simeq \bar \Omega(\mu) =
	 - \gs \frac{mA}{2 \pi \hbar^{2}} \ \frac{\mu^{2}}{2} +
	\gs \frac{e^2}{24 \pi mc^2} \ \frac{AH^2}{2} \ ,
	\label{therpotwf}
	\end{equation}
where $\bar \Omega$ is the smooth part (in energy) of the thermodynamic 
potential.  (Note that the second term of Eq.~(\ref{therpotwf})
comes nevertheless from the integral of the rapidly oscillating term
of the density of states.)
Thus, we obtain the two-dimensional diamagnetic Landau susceptibility
	\begin{equation}
	-\cl = - \frac{\gs e^2}{24 \pi mc^2} \ .
	\label{susLand}
	\end{equation}

For high magnetic fields, $k_{B}T \ll \hbar w$, the energy integrals are
slightly more complicated than before since the rapidly oscillating
component of $\Omega$ is not negligible any longer.  This latter
can be computed (see Appendix  \ref{app:convolution} for
the treatment of similar cases) as
	\be \label{therpotdHvA}
	\Oosc = \gs \left(\frac{m A}{\pi \hbar^2}\right)
	\sum_{n=1}^{\infty} (-1)^{n}
	\left(\frac{\hbar w}{2 \pi n}\right)^{2}
	\cos{\left(\frac{2\pi n \mu}{\hbar w}\right)}
	R_T(n) \ ,
	\end{equation}
where $R_T(n)$ is a temperature dependent damping factor
	\begin{equation}
	R_T(n) = \frac{2 \pi^2 n k_{B}T/\hbar w}
	{\sinh{(2 \pi^2 n k_{B}T/\hbar w)}} \ .
	\label{R:dHvA}
	\end{equation}
With $\Omega = \bar \Omega + \Oosc$, we have the Landau and de Haas--van 
Alphen contributions to the magnetic susceptibility
	\begin{equation}
	\frac{\cgc}{\cl} =- 1  - 24 \left(\frac{\mu}{\hbar w}\right)^{2}
	\sum_{n=1}^{\infty} (-1)^{n}
	\cos{\left(\frac{2\pi n \mu}{\hbar w}\right)}
	R_T(n) \ .
	\label{intro:susdHvA}
	\end{equation}
The second term exhibits the characteristic oscillations
with period $1/H$ and is exponentially damped with temperature (and the
summation index $n$).\footnote{For high fields we cannot in 
principle separate
the orbital and spin effects. The de Haas--van Alphen 
oscillations are given
only by the orbital component, that is the only one that interests 
us for our model of spinless electrons.}

While going from the bulk two-dimensional case (macroscopic regime) to the
constrained case (ballistic mesoscopic) two important changes
take place: i) the confining energy appears as a relevant scale and
Eq.~(\ref {LL}) no longer provides the quantization condition; ii) since we
are not in the thermodynamic limit of $N$ and $A \rightarrow \infty$, the
constraint of a constant number of electrons in [isolated] microstructures
is no longer equivalent to having a fixed chemical potential. 
These two effects will be thoroughly discussed in the paper. 
For didactical purposes we restrict
ourselves in this introductory section to only the changes
(i) due to the confinement, and we anticipate some of the results 
that will be later discussed in detail.

We imagine a mesoscopic square of size $a$ connected to an electron 
reservoir with chemical potential $\mu$. Direct numerical diagonalization 
in the presence of a magnetic field (Fig.~\ref{f1}a) allows us to 
obtain $\cgc$ (solid line in
Fig.~\ref{f1}b). In the high field region ($2r_c\!<\!a$, \  we note 
$r_c=  \vf / \omega$ the cyclotron radius) 
the characteristic de Haas~-~van Alphen oscillations are obtained, 
although not with the amplitude expected from calculations in the bulk 
(Eq.~(\ref {intro:susdHvA})).
For lower fields the discrepancy between our numerical results and the bulk
Landau diamagnetism is quite striking. Thus, confining deeply alters the
orbital response of an electron gas. Without entering into details at this
point we remark the fact that the whole curve is quite well reproduced 
by a finite-temperature semiclassical theory (dashed line) that 
takes into account
only one type of trajectory (see insets) in each of the three regimes: 
a) the interference-like regime, dominated by the shortest trajectories 
with the largest enclosed area for squares at zero magnetic field; 
b) the transition regime dominated by the bending of bouncing--ball 
trajectories between parallel
sides of the square; 
c) the de Haas~-~van Alphen regime dominated by
cyclotron orbits. It is remarkable how an exceedingly
complicated spectrum as that of Fig.~\ref{f1}a can be understood within
such a simple semiclassical picture once finite temperature acts as a 
filter  selecting only few types of trajectories.

\subsection {Overview of this work}

The purpose of this paper is to provide an [essentially comprehensive]
theory of the orbital magnetic properties of non-interacting spinless
electrons in the mesoscopic ballistic regime. We restrict ourselves
to the clean limit, where the different behavior of the magnetic
response arises as a geometrical effect (shape of the microstructure). 
We will make extensive use of semiclassical techniques since they appear 
to be perfectly suited for these problems.
For the smooth components (such as in Eq.~(\ref{therpotwf})) we will
use the general techniques developed by Wigner to obtain higher $\hbar$
 corrections to the Weyl term which are field dependent.   
For the oscillating components, we will rely on the so called
semiclassical trace formulas, which provide simple and intuitive
expressions for the density of states as a sum over Fourier-like
components associated to closed classical orbits.

In this respect it will be seen that the nature of the classical
dynamics, i.e.\ integrable versus chaotic (and more precisely
existence versus absence of continuous families of periodic orbits),
plays a major role.  Although we will present a complete
formalism for both cases, our main emphasis, and in particular all
the examples treated explicitly, will concern integrable
geometries.
The reason for this choice is twofold.  First, as we will make clear
in the sequel, one expects a much larger magnetic response
for integrable systems than for chaotic ones, yielding a
more striking effect easier to observe.  
The second point is that, contrarily
to what might seem natural a priori, integrable geometries
present a few conceptual difficulties in their treatment which are not
present for chaotic systems.  Indeed integrable systems  lack 
of structural stability, which means that under any small
perturbation (such as the one provided by the presence of a magnetic
field) they generically do not remain integrable.
  Chaotic systems on the contrary remain chaotic
under a small perturbation.  Therefore, as shown in Ref.\cite{aga94},
the Gutzwiller trace formula \cite{gutz_book,gut71},
valid for chaotic 
systems, can be used at finite fields without further complications.  
For integrable geometries however, the Berry-Tabor \cite{ber76,ber77} 
or Balian-Bloch \cite{bal69} trace formulae valid for integrable systems
usually do not apply in the presence of a perturbing magnetic field.
It will therefore be necessary, following Ozorio de Almeida
\cite{ozor86,ozor:book}, to consider the more complicated case of nearly
integrable systems, which we will do in detail here.

To perform this program, the present work is organized as follows.
In the next section  we present the thermodynamic formalism
appropriate for working in the canonical and grand canonical ensembles,
stressing its semiclassical interpretation and incorporating the changes 
 due to the constancy of the number of electrons in the experimentally 
relevant  microstructures.  
In Sec.~\ref{sec:LanGen} we consider the smooth magnetic response
and show that the Landau diamagnetism is present 
in any confined geometry at arbitrary fields. 
In Sec.~\ref{sec:integrable} we address the
magnetic response (susceptibility and persistent currents) in the
simplest possible geometries: circles and rings billiards that are 
integrable with and without magnetic field.
In Sec.~\ref{sec:square} we present the calculation of the 
magnetic susceptibility for the experimentally relevant case of the square
billiard \cite{levy93} whose integrability at zero field is broken by the 
effect of an
applied magnetic field. 
An initial study along these lines was presented in Refs.~\cite{URJ95}
and \cite{JRU94} and independently proposed by von 
Oppen \cite{vO95}. This geometry and the corresponding experiment have
also been analyzed from a completely different point of view by
Gefen, Braun and Montambaux \cite{gef94} stressing the importance of the
residual disorder (see also Ref.~\cite{altgef95}).
We consider  in Sec.~\ref{sec:general} the generic magnetic response
of both integrable and chaotic geometries, stressing the similarities and
differences in their behavior and calculating the line-shape of the
average magnetization in generic chaotic systems. 
In Sec.~\ref{sec:highB} we demonstrate how the semiclassical formalism
we have developed applies not only to the weak--field limit,
but also to higher field and in particular to the high field 
regime of the de Haas~-~van Alphen oscillations.  We treat
explicitly the example of the square geometry, including an
intermediate field regime dominated by bouncing--ball orbits as depicted
in Fig.~\ref{f1}.
We discuss our conclusions and their experimental relevance 
in Sec.~\ref{sec:concl}.    The modifications of our results due to the 
effect of a weak disordered potential are discussed in a separate 
publication \cite{rod2000}.

To keep the focus on the physical concepts developed in the text,
a few  technical derivations have been relegated to some
appendices.   Appendix~\ref{app:convolution} presents 
the generic case of the convolution of a
rapidly oscillating function with the derivative of the Fermi
function.
Appendix~\ref{app:wigner} gives the calculation of the first 
field-dependent term of the heat Kernel in an $\hbar$ expansion.
In Appendix~\ref{app:ring_g} we compute  the action integrals
associated with the dynamics of circular and ring billiards 
needed to define the energy 
manifold in action space. Appendix~\ref{app:D_M} presents the calculation
of the prefactor of the Green function for an integrable system,
while in
Appendix~\ref{app:highB} we show how to compute the semiclassical Green
function at a focal point, and apply the obtained result to the
particular case of cyclotron motion.

\newpage

%
\section {Thermodynamic formalism}
\label{sec:TherFor}

One main subject of the present work is the introduction of semiclassical
concepts into the thermodynamics of mesoscopic systems.
In this section we provide the basic formalism allowing one to obtain
the thermodynamic properties (grand potential, free energy)
from the quasiclassically calculated single--particle density of states
and hence the susceptibility. We begin with general
definitions and relations between grand canonical and canonical 
quantities.

For a system of electrons in a volume (area in two dimensions)
$A$ connected to a reservoir of particles with chemical potential
$\mu$ (grand canonical ensemble) the magnetic susceptibility
is obtained, as given by Eq.~(\ref{eq:susgc}), as
	\[
	\cgc = 	- \frac{1}{A}
	   \left(\frac{\partial^{2}\Omega}{\partial H^{2}} \right)_{T,\mu} \ .
	\]
$\Omega(T,\mu,H)$ is the thermodynamic potential, which can be
expressed for non--interacting electrons in terms of the single--particle 
density of states through Eq.~(\ref{eq:therpot}).

For actual microstructures, the number $\bf N$ of particles
inside the device might be large but is fixed in contrast to the 
chemical potential $\mu$.
As discussed in the introduction, it will be necessary in some cases,
namely when considering the average susceptibility of an ensemble of
microstructures, to take explicitly into account the conservation
of $\bf N$, and to work within the canonical ensemble.
For such systems with a fixed number $\bf N$ of particles, the relevant
thermodynamic function is not the grand potential $\Omega$, but its
Legendre transform, the free energy\footnote{In standard 
thermodynamics, Eq.~(\ref{eq:free}) just represents the definition of the 
grand potential.  It should be borne in mind however that from a
statistical physics point of view this is not an exact relation,
but the result of a stationary--phase evaluation of the average over
the occupation number, valid only when $k_BT$ is larger than the 
typical level spacing. Therefore, we are entitled
to use this relation in the mesoscopic regime that interests us, but
not in the microscopic regime, where features on the scale of a 
mean spacing become relevant.}
	\begin{equation} \label{eq:free}
	F(T,H,{\bf N}) = \mu {\bf N} + \Omega(T,H,\mu) \ .
	\end{equation}
In particular, the magnetic susceptibility of a system of $\bf N$ electrons
is
	\begin{equation} \label{eq:sus}
	\chi = - \frac{1}{A}
	\left(\frac{\partial^{2}F}{\partial H^{2}} \right)_{T,{\bf N}}
	\ .
	\end{equation}

Except for the calculation of the Landau contribution performed in
the following section all the computations of the magnetic response of the 
microstructures to be considered 
will involve two clearly separated parts.  In the first one
the (oscillating part of the) density of states will be
calculated semiclassically. Depending on the underlying
classical dynamics (integrable versus chaotic, with or without breaking
of the invariant tori, with or without focal points, etc.), the 
results as well as their derivation will vary noticeably.
In the second stage the integrals over energy yielding the desired
thermodynamic properties have to be performed in a leading order in $\hbar$
approximation.  To avoid tedious repetitions,
we shall consider here in some detail this part of the calculation
of the thermodynamic properties, and refer without many additional
comments to the results obtained in this section whenever needed.
We begin with the grand canonical quantities which exhibit
the simplest expressions in terms of the density of states.
In a second subsection we shall consider the canonical ensemble following 
closely the approaches presented in Refs.~\cite{ensemble}.

\subsection{Grand canonical properties}

We begin with the standard definition, Eq.~(\ref{DOS}) of the density
of states
	\[ d(E) = \gs \sum_{\lambda} \delta(E-E_{\lambda}) \ , \]
($\gs = 2$ is the spin degeneracy, $E_{\lambda}$ the eigenenergies)
and its successive energy integrals. They are the energy staircase
	\be
	n(E)  = \int_0^{E} \dif E' \ d(E')  \ ,
	\label{eq:roughn}
	\ee
and the grand potential at zero temperature
	\be
	\omega (E) = - \int_0^E \dif E' \ n(E') \ .
	\label{eq:rougho}
	\ee
These are purely quantum mechanical quantities, depending only on the
eigenstates $E_\lambda$ of the system. 
At finite temperature the corresponding
quantities are obtained by convolution with the derivative
$f'(E-\mu)$ of the Fermi distribution function
	\begin{equation} \label{eq:fermi}
	f(E-\mu) = \frac{1}{(1 + \exp [\beta (E-\mu)])} \ .
	\end{equation}
We then have
   \begin{mathletters}
	\label{allsmooths}
	\begin{eqnarray}
	D (\mu) & = & - \int_0^{\infty} \dif E \ d(E) \ f'(E\!-\!\mu) \ ,
	\label{smootha} \\
	N (\mu) & = & - \int_0^{\infty} \dif E \ n(E) \ f'(E\!-\!\mu) \ ,
	\label{smoothb} \\
	\Omega (\mu) & = & 
	          - \int_0^{\infty} \dif E \ \omega(E) \ f'(E\!-\!\mu)
	\ .
	\label{smoothc}
	\end{eqnarray}
   \end{mathletters}

Integration by parts leads to the standard definition
(\ref{eq:therpot}) of the grand potential and the mean number of
particles in the GCE with a chemical potential $\mu$, {\em i.e.}
	\be
	N(\mu) = \int_0^{\infty} \dif E \ d(E) \ f(E\!-\!\mu) \ .
	\label{eq:mGCE}
	\ee
That means that the thermodynamic properties (\ref{smoothb})--(\ref{smoothc})
are obtained by performing the energy integrations
(\ref{eq:roughn})--(\ref{eq:rougho}) with the Fermi function as a weighting factor.

In the following the separation of the above quantum mechanical and 
thermodynamic expressions into smooth (noted with a ``$\; \bar{~~} \;$'') 
and oscillating (noted with the superscript ``$\; ^{\rm osc} \;$'') parts
is going to play a major role. It has its origin in the well-known decomposition 
of the density of states as
	\begin{equation} \label{eq:dosc1}
	d(E) = \bar d(E) + \dosc(E) \; .
	\end{equation}
This decomposition has a rigorous meaning only in the semiclassical 
($E \rightarrow
\infty$) regime for which the scales of variation of $\bar d$ and $\dosc$
decouple.  To leading order in $\hbar$, the mean component $\bar d(E)$
is the Weyl term reflecting the volume of accessible classical
phase space at energy $E$ (zero-length trajectories), while $\dosc(E)$ is
given as a sum over periodic
orbits (Gutzwiller and Berry-Tabor trace formulas) \cite{gutz_book}.
Generically, it will be expressed as a sum

	\be \label{eq:dosc:gen}
	\dosc(E) = \sum_t d_t(E) \qquad ;  \qquad
	d_t(E) = A_t(E) \, \sin \left(S_t(E)/\hbar + \nu_t \right) \; .
	\ee
running over periodic orbits labeled by $t$ 
where $S_t$ is the action integral along the orbit $t$, $A_t(E)$ 
is a slowly
varying prefactor and $\nu_t$ a constant phase.\footnote{When considering
systems whose integrability is broken by a perturbing magnetic field,
we shall stress the necessity to consider families of recurrent --~but not
periodic~-- orbits of the perturbed system.  This will, however, not
affect the discussion which follows.}

Using the expression~(\ref{eq:dosc:gen}) for $\dosc$ in 
Eqs.~(\ref{eq:roughn}) and (\ref{eq:rougho}), $\nosc$ and $\oosc$ are 
obtained to
leading order in $\hbar$ as 
	\begin{equation} \label{eq:rough_osc:def}
	\nosc(E) = \int^E \dif E' \ \dosc(E')  \qquad ; \qquad
	\oosc(E) = - \int^E \dif E' \ \nosc(E')  \ .
	\end{equation}
The lower bounds are not specified because the constants of
integration are determined by the constraint that $\nosc$ and $\oosc$ 
must have zero
mean values.  (It should be borne in mind that semiclassical 
expressions like
(\ref{eq:dosc1}), and those that will follow, are not applicable at the
bottom of the spectrum.)

In a leading $\hbar$ calculation the integration over energy in
Eq.~(\ref{eq:rough_osc:def}) has to be applied only to
the rapidly oscillating part of each periodic orbit contribution $d_t$.
Noting moreover that if $S_t(E)$ is the action along a periodic orbit, then
$\tau_t(E) \equiv dS_t / dE$ is the period of the orbit, one has in a
leading $\hbar$ approximation
	\be \label{eq:primitive}
	\int^E A_t(E') \, \sin \left( S_t(E')/\hbar +\nu_t \right) dE'
	= \frac{-\hbar}{\tau_t(E)} \,
	A_t(E) \, \cos \left( S_t(E)/\hbar +\nu_t \right)
	\ee
as can be checked by differentiating both sides 
of Eq.~(\ref{eq:primitive}). 
In order to emphasis that the integration over energy merely yields 
a multiplication by
$(- \hbar/\tau)$, we use the notation $(i_\otimes \cdot d_t)$ to assign
the contribution $d_t$ of a periodic orbit after shift of the phase by
$\pi/2$, {\em i.e.}
$(i_\otimes \cdot [B \, \sin(S/\hbar)]) = B \, \cos (S/\hbar)$. We get
	\begin{eqnarray}
	\nosc(E) = \sum_t n_t(E) \qquad & ; & \qquad
	n_t(E) =  \frac{- \hbar}{\tau_t(E)} \, (i_\otimes \cdot d_t(E)) \; ,
	\label{eq:rough_oscn} \\
	\oosc(E) = \sum_t \omega_t(E) \qquad & ;  & \qquad
	\omega_t(E) =  \left( \frac{\hbar}{\tau_t(E)} \right)^2 d_t(E) \; .
	\label{eq:rough_osco}
	\end{eqnarray}

	The thermodynamic functions $\Dosc(\mu)$, $\Nosc(\mu)$ 
	and $\Oosc(\mu)$ are then obtained 
	by application of Eqs.~(\ref{allsmooths}) in which the full
	functions are replaced by their oscillating component.
	The resulting integrals involve the convolution of functions
	($\dosc(E)$, $\nosc(E)$ or $\oosc(E)$) oscillating [locally 
	around $\mu$] with a frequency 
$\tau(\mu) / (2\pi\hbar)$, with the derivative of the Fermi factor
$f'(E-\mu)$ being smooth on the scale of $\beta^{-1} = k_{\rm B} T$.
One can therefore already anticipate that this convolution yields an 
exponential damping of the periodic orbit contribution
whenever $\tau(\mu) \gg \hbar \beta$.
As shown in appendix~\ref{app:convolution}
the temperature smoothing gives rise to
an additional factor for each periodic orbit contribution,
	\begin{equation} \label{sec2:RT}
	R_T(\tau_t) = \frac{\tau_t/\tau_c}{\sinh (\tau_t/\tau_c)}
	\qquad ; \qquad
	\tau_c = \frac{\hbar\beta}{\pi} \; ,
	\end{equation}
in a leading $\hbar$ and $\beta^{-1}$ approximation
(without any assumption concerning the order the limits are taken).
In this way we obtain relations between the following useful
thermodynamic functions and the semiclassical density of states:
    \begin{mathletters} \label{eq:smooth_osc:all}
	\begin{eqnarray}
   	\Dosc(\mu) = \sum_t D_t(\mu) \qquad & ; & \qquad
	D_t(\mu) =  R_T(\tau_t) \, d_t(\mu) \qquad , 
	\label{eq:smooth_oscd} \\
	\Nosc(\mu) = \sum_t N_t(\mu) \qquad & ; & \qquad
	N_t(\mu) =  R_T(\tau_t) \, \left( \frac{-\hbar}{\tau_t} \right)
	(i_\otimes \cdot d_t(\mu)) \; , 	\label{eq:smooth_oscn} \\
	\Oosc(\mu) = \sum_t \Omega_t(\mu) \qquad & ; & \qquad
	\Omega_t(\mu) =  R_T(\tau_t) \,
	\left( \frac{\hbar}{\tau_t} \right)^2 d_t(\mu) \; .
		\label{eq:smooth_osco}
	\end{eqnarray}
    \end{mathletters}
At very low temperature, 
$R_T \simeq 1 - [(\tau_t \pi) /( \hbar \beta)]^2 / 6$ which,
for billiard-like systems where $\tau_t = L_t/v_F$ 
(with $L_t$ being the length of
the orbit and $\vf$ the Fermi velocity),
simply gives the standard Sommerfeld--expansion $R_T \simeq 1 
- [(L_t \pi) /( \hbar \beta \vf)]^2 / 6$.
For long trajectories or high temperature
it yields an exponential suppression and therefore the only
trajectories contributing significantly to the thermodynamic 
functions are those with $\tau_t \leq \tau_c$.
Thus, temperature smoothing has a noticeable effect on the oscillating 
quantities since it effectively suppresses the higher harmonics,
which are associated with long classical orbits
in a semiclassical treatment. On the contrary, for a degenerate electron gas
($\beta \mu \gg 1$), finite  temperature has no effect on the mean
quantities.  Temperature is then the tuning parameter for passing from
$d(E)$ at $T=0$ to
$\bar D(E) = \bar d(E)$ at large temperatures (by the progressive
reduction of $\dosc$). Similar considerations hold for the energy
staircase and the grand potential.

The oscillatory part of the semiclassical
susceptibility in the grand--canonical ensemble is finally obtained
from Eq.~(\ref{eq:susgc}) by replacing $\Omega$ by $\Oosc$.

\subsection{Canonical ensemble}
\label{sec:Canonical}

Let us now consider the
susceptibility in the canonical ensemble, appropriate for systems with
a fixed number of particles. We follow Imry's derivation 
for persistent currents in ensembles of disordered rings \cite{Imry}.
The only important difference
is that we will take averages over the size and the Fermi energy of
ballistic structures instead of averages over impurity realizations.
We will stress the semiclassical interpretation that will be at the 
heart of our work, and highlight some of its subtleties.

As mentioned in the introduction the definition Eq.~(\ref{eq:sus})
of the susceptibility $\chi$ is equivalent to $\cgc$ up to $1/{\bf N}$
({\em i.e.}~$\hbar$ corrections).  Therefore,
in the macroscopic limit of ${\bf N} \rightarrow \infty$ the choice of the
ensemble in which the calculations are done is unimportant. On the 
other hand,
in the mesoscopic regime of small structures (with large but finite
$\bf N$) we have to consider such corrections if we want to take advantage
of the computational simplicity of the Grand Canonical Ensemble (GCE). The
difference between the two definitions is particularly important when
 the GCE
result is zero as it is the case for the
ensemble average of $\cgc$. The evaluation of the corrective terms can be
obtained from the relationship Eq.~(\ref{eq:free}) between the thermodynamic
functions $F({\bf N})$ and $\Omega(\mu)$%
\footnote{In the following we will only write the $\bf N$ dependence of
$F$ and the $\mu$ dependence of $\Omega$, assuming always the $T$ and
$H$ dependence of both functions.} and the relation $N(\mu) = {\bf N}$.  
In the case of finite systems the previous implicit
relation is difficult to invert. However, when $\bf N$ is large we can use
the decomposition of  $N(\mu)$ in a smooth part $\bar N(\mu)$
and a small component $\Nosc(\mu)$ that fluctuates around the secular
part, and we can perturbatively treat the previous implicit relation.
The contribution of a given orbit to $\dosc$ is always of lower order 
in $\hbar$ than $\bar d$ as can be checked for the various examples 
we are going to consider and by inspection of semiclassical trace formulae.
However, since there are infinitely many of such contributions,
we obtain $\dosc$ and $\bar d$ to be of the same order when adding them up.
(This must be the case since the quantum mechanical $d(E)$ is a sum
of $\delta$ peaks.) Thus, we cannot use $\dosc / \bar d$ as a small 
expansion parameter.
On the other hand, finite temperature provides an exponential cutoff
in the length of the trajectories contributing to $\Dosc$
so that only a finite number of them must be  taken into account.
Therefore, $\Dosc$ is of lower order in $\hbar$ than $\bar D$, and
in the semiclassical regime it is possible to expand 
the free energy $F$ with respect to the small parameter $\Dosc / \bar D$.
The use of a temperature smoothed density of states
Eq.~(\ref{smootha}) closely follows the
Balian and Bloch approach \cite{bal69}, where, due to the
exponential proliferation of orbits and the impossibility of
exchanging the infinite time and semiclassical limits, the semiclassical
techniques based on trace formulae are considered meaningful only when applied to
smoothed quantities.
The decomposition of $D(E)$ is depicted in Fig.~\ref{f2}, where
we have taken $\bar D (\simeq \bar d)$ to be energy independent,
corresponding to the two-dimensional (potential free) case.

For a perturbative treatment of the mentioned implicit relation
we define a mean chemical potential $\bar \mu$ by the condition of
accommodating $\bf N$ electrons to the mean number of states
	\be \label{eq:mub_def}
	{\bf N} = N(\mu) = \bar N (\bar \mu) \ .
	\ee
Expanding this relation to first order in $\Dosc/\bar D$, and employing that
$\dif N/\dif \mu = D$, one has
	\be \label{deltamu}
	\Delta \mu \equiv  \mu - \bar \mu
	\simeq - \ \frac{1}{\bar D (\bar \mu)} \ \Nosc (\bar \mu) \; .
	\ee
\nin The physical
interpretation of $\Delta \mu$ is very clear from Fig.~\ref{f2}: The
shaded area represents the number of electrons in the system and it is
equal to the product $\bar{D} \times \bar{\mu}$. 

Expanding the relationship (\ref{eq:free}) to second order
in $\Delta \mu$,
	\be
	F({\bf N}) \simeq (\bar \mu + \Delta \mu) {\bf N}
		+ \Omega (\bar \mu)
		- N (\bar \mu)  \Delta \mu
		- D (\bar \mu) \ \frac{\Delta \mu^2}{2} \ ,
	\label{eq:secor}
	\ee
using the decomposition of $\Omega(\bar \mu)$ and $N(\bar \mu)$
into mean and oscillating parts and
eliminating $\Delta \mu$  (Eq.~(\ref{deltamu})) in the second
order term,
one obtains the expansion of
the free energy to second order in $\Dosc / \bar D$ \cite{Imry,ensemble}
	\begin{equation} \label{eq:fd}
	F({\bf N}) \simeq F^{0} + \Delta F^{(1)} +\Delta F^{(2)} \ ,
	\end{equation}
with
   \begin{mathletters}
   \label{allDF}
	\begin{eqnarray}
	     F^{0} & = & \bar \mu {\bf N} + \bar \Omega(\bar \mu) \; ,
	      \label{DF0}  \\
	     \Delta F^{(1)} & = & \Oosc (\bar \mu)  \; ,
	        \label{DF1} \\
	\displaystyle 
	     \Delta F^{(2)} & = & \frac{1}{2 \bar D (\bar \mu)} \
				\left( \Nosc (\bar \mu) \right)^{2} \ .
	     \label{DF2}
	\end{eqnarray}
   \end{mathletters}
Then $\Delta F^{(1)}$ and $\Delta F^{(2)}$ can be expressed in terms
of the oscillating part of the density of states by means of
Eqs.~(\ref{eq:smooth_oscn}) and (\ref{eq:smooth_osco}).
The first two terms
$F^{0} + \Delta F^{(1)}$ yield the magnetic response
calculated in the GCE with an effective
chemical potential $\bar \mu$. The first ``canonical correction''
$\Delta F^{(2)}$ has a grand canonical form since it is expressed in
terms of a temperature smoothed integral of the density of states
(Eq.~(\ref{eq:mGCE})) for a fixed chemical potential $\bar \mu$.

It is convenient to use the expansion (\ref{eq:fd}) in the calculation of
the magnetic susceptibility of a system with a fixed number of particles
because the leading $\hbar$ contribution to $\bar N (\bar \mu)$ has
no magnetic field dependence, independent of the precise system under consideration.
Therefore, {\em at this level of approximation}, keeping $\bf N$ constant
in Eq.~(\ref{eq:sus}) when taking the derivative with respect to the
magnetic field
amounts to keep $\bar \mu$ constant. Since
$F^{(0)}$ is field independent in a leading order semiclassical expansion
the weak-field susceptibility of a given mesoscopic sample will be dominated
by $\Delta F^{(1)}$. However, when
considering ensembles of mesoscopic devices, with slightly different sizes or
electron fillings, $\Delta F^{(1)}$ (and its associated
contribution to the susceptibility) averages to zero due to its oscillatory
behavior independently of the order in $\hbar$ up to which it is 
calculated.\footnote{In the following, we
shall always calculate $\Delta F^{(1)}$ in a leading order $\hbar$
approximation.  Higher order corrections to $\Delta F^{(1)}$ may
be of the same order as $\Delta F^{(2)}$ but will average to zero 
under ensemble averaging.} Then we must consider the next order 
term $\Delta F^{(2)}$.

As mentioned above, we will essentially work in the semiclassical
regime (to leading order in $\hbar$) where
$F^{0}$ is field independent. However, in the following section we
will examine the next
order $\hbar$ correction to $\bar \Omega (\bar \mu)$ (and to $F^0$),
demonstrating that its field dependence gives rise to the standard
Landau diamagnetism, independent of any confinement.

\newpage

\section{Landau susceptibility}
\label{sec:LanGen}

In the previous section we showed that the various quantum mechanical
({\em i.e.} $d(E)$, $n(E)$, $\omega(E)$) and thermodynamic ({\em i.e.}
$D(\mu)$, $N(\mu)$, $\Omega(\mu)$) properties of a mesoscopic system can
be decomposed into smooth and fluctuating parts. In the semiclassical limit,
where the Fermi wavelength is much smaller than the system size, each of
these quantities allows an asymptotic expansion in powers of $\hbar$. For 
most of the purposes it is sufficient to consider only leading order terms
while higher order corrections must only be added if the former vanish for some reason.
This is the case for the smooth part ${\bar \Omega}(\mu)$
of the grand potential, which is the dominant term at any temperature,
but is magnetic field independent to leading order in $\hbar$. The
present section will be the only part of our work where higher $\hbar$
corrections are considered. We will show that they give rise to the
standard Landau susceptibility.
Our derivation relies neither, on the quantum side, on the existence of
Landau levels, nor, on the classical side, on 
boundary trajectories or the presence of circular
cyclotronic orbits fitting into the confinement potential.
This shows that the Landau
susceptibility is a property of mesoscopic devices as well as 
infinite systems, being the dominant contribution at sufficiently high
temperature\footnote{Analog results have been independently obtained
by S.D.~Prado {\em et al.} \cite{Prado}. The Wigner distribution function
was previously used by R.\ Kubo \cite{Kubo64} in the study of
Landau diamagnetism.}.

We consider a $d$-dimensional ($d=2,3$) system of electrons governed
by the quantum Hamiltonian
	\be \label{eq:quantH}
	{\hat {\cal H}} =  \frac{1}{2m} \ \left(\hat \bp -
	\frac{e}{c} \bA(\hat \bq)\right)^2 \
	   	+ \ V(\hat \bq) \; ,
	\ee
where $\bA$ is the vector potential generating the magnetic field $H$
and $V(\bq)$ is the potential which confines the electrons 
in some region of the space. This region can a priori have any dimension, and
it can be smaller that the cyclotron radius. We will only assume in the
following that $V(\bq)$ is {\em smooth} on the scale of a Fermi
wavelength, so that semiclassical asymptotic results can be used.
In billiards the effect of {\em hard} boundaries on the susceptibility 
is negligible compared to the Landau bulk term \cite{rob86,antoine}, and
therefore the results obtained below apply there, too.

There exist general techniques to compute the semiclassical expansion of
the mean part of the density of states (or of  its integrated versions
Eqs.~(\ref{eq:roughn}), (\ref{eq:rougho})) up to arbitrary
order in $\hbar$.  The most complete approach, which allows one
to take into account the effect of sharp boundaries, can be found in
the work of Seeley \cite{seeley}.  However, assuming the smoothness of 
$V(\bq)$ allows us to follow the standard approach introduced by Wigner 
in 1932 \cite{wig32} which is based on the notion of the Wigner
transform of an operator.
As a starting point we consider the Laplace transform of the level 
density (or heat Kernel),
	\be \label{partition}
	Z(\lambda )  =  \int_0^\infty \dif E \ e^{-\lambda E} \ d(E)
	\ = \ \gs \ {\rm Tr} (e^{-\lambda \hat{{\cal H}}} ) \; ,
	\ee
where $\gs=2$ takes into account the spin degeneracy.
In appendix~\ref{app:wigner} we apply after a brief description the
technique to calculate the first two terms
of the expansion of $Z(\lambda)$ with respect to $\lambda$. They 
yield under the inverse transformation the first two terms 
of the expansion of $d(E)$ in powers of $\hbar$. The
oscillating part $\dosc(E)$ of $d(E)$ is not included in this procedure
since it is associated with exponentially
small terms in $Z(\lambda)$, that is, $Z(\lambda) \simeq \bar Z(\lambda)$
for $\lambda \simeq 0$. This well known property can be easily seen from
the integral treated in appendix~\ref{app:convolution} by identifying
$\beta$ with $\lambda$ and using the exponential form of the distribution
function in the classical limit of high temperatures ($\beta \mu \ll 1)$.

Noting ${\cal H}(\bq,\bp)$ the classical Hamiltonian corresponding to
Eq.~(\ref{eq:quantH}), the leading order [Weyl] contribution to $Z(\lambda)$
is given by Eq.~(\ref{ZW}),
	\be \label{ZWeyl}
	Z_{\rm W}(\lambda)
	 = \frac{\gs}{(2\pi \hbar )^d} \ \int \dif \bq \dif \bp \,
	\exp \left( -\lambda {\cal H}(\bq,\bp)
	\right) \; ,
	\ee
and the inverse Laplace transform yields the familiar result
	\be \label{dWeyl}
	d_{\rm W}(E) =
	\bar d_{\rm W}(E) = \frac{\gs}{(2\pi \hbar )^d} \ \int \dif \bq \dif \bp \,
	\delta \left( E - {\cal H}(\bq,\bp) \right)  \; .
	\ee
In the above integrals, the substitution
	\be \label{change}
	\bp \rightarrow \bp' = \bp - \frac{e}{c} \bA
	\ee
eliminates any field dependence. Therefore
\be \label{eq:defwweyl}
\omega_{\rm W}(E) \ = \
\bar \omega_{\rm W}(E) \ = \ - \int_0^E \dif E' \int_0^{E'}
\dif E'' \
d_{\rm W}(E'') \; ,
\ee
as well as the leading term
$\bar \Omega_{\rm W}(\mu) $ of the grand potential (obtained
in the high temperature limit of Eq.~(\ref{smoothc})),
are field independent. This is the reason for the absence of orbital
magnetism in classical mechanics.
To observe a field dependence, one must consider the first
correcting term of $Z(\lambda)$ which, as shown in
appendix~\ref{app:wigner} (Eq.~(\ref{Z1})), is given by
	\be \label{Z_1}
	Z_1(\lambda,H) = - \lambda^2 \ \frac{\mu_B^2 H^2}{6} \ Z_{\rm W}
		+ Z_1^0 \; .
	\ee
Here, $\mu_B = (e\hbar) / (2mc)$ is the Bohr magneton, and $Z_1^0 =
Z_1({\scriptstyle H\!=\!0})$ is a field independent term that we will drop
from now on since it does not contribute to the susceptibility.

The integrated functions $n(E)$ and $\omega(E)$ can be obtained from
their Laplace
transforms
	\begin{equation} \label{eq:wmu}
	n(\lambda) =   \frac{Z(\lambda)}{\lambda} \; , \qquad
	w(\lambda) = - \frac{Z(\lambda)}{\lambda^2} \; .
	\end{equation}
Then the first correction to the zero-temperature grand potential is 
	\be
	\omega_1(E) \ = \
	\bar \omega_1(E) \ = \frac{\mu_B^2 H^2}{6} \ \bar d_{\rm W}(E) \; .
	\ee
After convolution with the derivative of the Fermi
function (Eq.~(\ref{smoothc})) we obtain the first corrective term of
the grand potential
	\be
	\Omega_1 (\mu) =
	\bar \Omega_1 (\mu) = \frac{\mu_B^2 H^2}{6} \ \bar D_{\rm W}(\mu)
	\; .
	\ee

In the grand canonical ensemble, the above
equation readily gives the leading contribution to the susceptibility
	\be \label{eq:landau}
        \bar \cgc = -\frac{\mu_B^2}{3 A} \
	\bar D_{\rm W} \; ,
	\ee
coming from the mean part of the grand potential. In Eq.~(\ref{eq:landau})
$A$ is the confining volume (area for $d=2$) of the electrons.
Noting that $\bar D_{\rm W} = \dif \bar N_{\rm W} / \dif \mu$,
one recognizes the familiar result of Landau \cite{Land}.
For systems without potential (bulk, or billiard systems), it gives
in the degenerate case $(\beta \mu \gg 1)$ in two, respectively,
three dimensions
	\begin{equation} \label{eq:Lan23}
	\bar \cgc_{2d} = - \frac{\gs e^2}{24 \pi m c^2} \ , \qquad
	\bar \cgc_{3d} = - \frac{\gs e^2 \kf}{24 \pi^2 m c^2} \; .
	\end{equation}
In the non--degenerate limit the susceptibility is
	\be \label{eq:Lannondeg}
        \bar \cgc = -\frac{\mu_B^2}{3 A} \ \frac{{\bf N}}{k_B T} \; .
	\ee
The temperature independence in the degenerate regime and the power--law decay
in the non-degenerate limit cause the dominance of the Landau contribution
at high temperatures since, as mentioned in the previous section (and
demonstrated in appendix~\ref{app:convolution}), the
contributions from $\Delta F^{(1)}$ and $\Delta F^{(2)}$ (Eqs.~(\ref{DF1}) and
(\ref{DF2})) are exponentially damped by temperature.

The Landau diamagnetism is
usually derived for free
electrons or for a quadratic confining potential \cite{LanLip,Peierls}.
We have provided here its
generalization to any confining potential
(including systems smaller than the cyclotron radius).

For a system with fixed number $\bf N$ of electrons, defining a 
Weyl chemical
potential $\mu_{\rm W}$ by
	\be
	{\bf N} = \bar N_{\rm W}  (\mu_{\rm W})
	\ee
and following the same procedure as in Sec.~\ref{sec:Canonical}
one can write
	\be
	F^{(0)}({\bf N}) \simeq F_{\rm W} + \bar \Omega_1(\mu_{\rm W}) \; ,
	\ee
where both $\mu_{\rm W}$ and
	\be
	F_{\rm W} = \mu_{\rm W} {\bf N} +
		\bar \Omega_{\rm W}(\mu_{\rm W})
	\ee
are field independent.  Therefore, the smooth part of the free energy
gives the same contribution than Eq.~(\ref{eq:landau}): We
recover the Landau diamagnetic response in the canonical ensemble, too.

At the end of this section we would like to comment on the
case of free electrons in two dimensions. Since
Eq.~(\ref{DOSLL}) represents an exact formula for the density of states,
$\bar d(E) = (\gs mA)/(2 \pi \hbar^{2})$ can be
interpreted as the exact mean density of states, and
$\dosc(E) =  (\gs mA)/(\pi \hbar^{2}) \sum_{n=1}^{\infty} (-1)^{n} 
\cos{\left((2\pi n E)/(\hbar w)\right)}$ as the exact oscillating part.
However, $\omega(E)$ being obtained by integrating  $d(E)$ twice, has a 
mean value which, in addition to $ -\bar d \, E^2/2$, contains 
the term $(\mu_B^2 H^2/6) \bar d$ yielding the Landau susceptibility.
In the usual derivation, this term  comes from the integration of
$\dosc(E)$, more precisely from the boundary contribution at $E=0$
(i.e., from levels too close to the ground state in order
to properly separate the mean value from oscillating parts).
One should be aware that $\bar \omega(E)$ cannot
be defined by Eq.~(\ref{eq:defwweyl}) as
soon as non leading terms are considered.   For this reason
some care was required for the definitions of the
last section (see the discussion around
Eqs.~(\ref{eq:rough_osc:def})-(\ref{eq:smooth_osc:all})).

\newpage

\section{Systems integrable at arbitrary fields}
\label{sec:integrable}

In the remainder of this work we will provide semiclassical approximations
for the corrective free-energy terms $\Delta F^{(1)}$ and $\Delta F^{(2)}$
(see Eq.~(\ref{eq:fd})) and their associated magnetic responses
for systems that react differently under the influence of an applied 
field. We will be mainly working in the 
weak-field regime (except in section~\ref{sec:highB}), 
where  the magnetic field
acts as a perturbation almost without altering the classical dynamics.
In this regime the nature of the zero-field dynamics ({\em i.e.\ } 
integrable vs.\ chaotic, or more precisely, the organization of periodic
orbits in phase space) becomes the dominant factor determining the 
behavior and magnitude of the magnetic susceptibility. 
For systems which are integrable at zero field
the generic situation is that the magnetic field breaks the integrability
(as any perturbation will do).
It is necessary in that case to develop semiclassical
methods allowing to deal with nearly, but not exactly, integrable systems.
This question will be addressed in sections~\ref{sec:square} and 
\ref{sec:general}.
There exist however ``non generic'' systems where the classical dynamics
remains integrable in the presence of the magnetic field. Due to their
rotational symmetry, circles
and rings (which are the geometries used in many experiments) fall 
into this
category.  In these cases (and similarly for the Bohm-Aharonov flux
\cite{RivO}) the Berry-Tabor semiclassical trace formula
\cite{ber76,ber77} provides the appropriate path to calculate semiclassically the
oscillating part of the density of states $\dosc$, including its field
dependence. Thus, $\Delta F^{(1)}$ and
$\Delta F^{(2)}$, and their respective contributions to the susceptibility,
can be deduced. 
This is the program we perform in this section, treating
specifically the example of circular and ring billiards.

The magnetic susceptibility of the circular billiard can be calculated from
its exact quantum mechanical solution in terms of Bessel functions
\cite{Dingle52,Bog,von_thesis}. The magnetic response of long cylinders 
\cite{Kulik,RiGe} and
narrow rings \cite{RiGe} (the two nontrivial generalizations of 
one-dimensional rings) can be calculated by neglecting the curvature of
the circle and solving the Schr\"{o}dinger equation for a rectangle with periodic
boundary conditions. Our semiclassical derivation provides an intuitive
and unifying approach to the magnetic response of circular billiards 
and rings of any thickness (for individual systems as well as ensembles)
and establishes the range of validity of
previous studies. Moreover,
we present it for completeness since it provides a pedagogical introduction
to the more complicated (``generic") cases of the following sections.

\subsection{Oscillating density of states for weak field}
\label{sec:GenInt}

By definition, a classical Hamiltonian ${\cal H}(\bp,\bq)$ is integrable
if there exist as many constants of motion in involution as degrees of
freedom.  For bounded systems, this implies (see e.g.
\cite{arnold:book}) that all trajectories are trapped 
on torus-like manifolds (invariant tori), each of which 
can be labeled by the action integrals
	\be \label{eq:action}
	I_i = \frac{1}{2\pi} \oint_{{\cal C}_i}  \bp \, \dif \bq
	\qquad (i=1,2) \; ,
	\ee
taken along two independent paths ${\cal C}_1$ and ${\cal C}_2$ on 
the torus.  (We are dealing with two degrees of freedom.)
It is moreover possible to perform a canonical transformation from the
original $(\bp,\bq)$ variables to the action-angle variables $({\bf I}, \phi)$
where ${\bf I} = (I_1,I_2)$ and $\phi = (\varphi_1, \varphi_2)$ with
$\varphi_1,\varphi_2$ in $[0,2\pi]$. Because both, $I_1$ and $I_2$, are
constants of motion, the Hamiltonian ${\cal H}(I_1,I_2)$ expressed in
action-angle variables depends only on the actions.

For a given torus we note $\nu_i = \partial {\cal H} / \partial I_i$ 
$(i=1,2)$ the angular frequencies, and $\alpha \equiv \nu_1/\nu_2$ 
the rotation number.
A torus is said to be ``resonant'' when its rotation number
is rational ($\alpha = u_1/u_2$ where $u_1$ and $u_2$ are coprime
integers).  In that case all the orbits on the torus are periodic, and
the torus itself constitutes a one-parameter family of periodic orbits,
each member of the family having the same period and action.
The families of periodic orbits can be labeled by the two
integers $(M_1,M_2) = (r u_1, r u_2)$ where $(u_1,u_2)$ specifies the
primitive orbits and $r$ is the number of repetitions.
$M_i$ ($i=1,2$) is thus the winding number of $\varphi_i$ 
before the
orbits close themselves. The pair ${\bf M} = (M_1, M_2)$ has been coined
the ``topology'' of the orbits by Berry and Tabor.

For two-dimensional systems, the Berry-Tabor formula can be cast in
the form \cite{ber76,ber77}
	\be \label{BT}
	\dosc(E) = \sum_{\bM \neq (0,0),\epsilon} d_{\bM,\epsilon}(E) \ ,
	\ee
with
	\be \label{BTT}
	d_{\bM,\epsilon}(E) =
	\frac{\gs \ \tau_{\bM}}{\pi \hbar^{3/2} M_2^{3/2} 
	\left| g_E^{''}(I_{1}^{\bM}) 
	\right|^{1/2}} 
        \  \cos{\left(\frac{S_{\bM,\epsilon}}{\hbar}  - \hat\eta_{\bM} \ 
	\frac{\pi}{2} +
	\gamma \ \frac{\pi}{4} \right)} \ .
	\ee
The sum in Eq.\ (\ref{BT}) runs over all families of closed orbits at energy $E$, 
labeled by their topology $\bM$ (in the first quadrant, that is $M_1$ 
and $M_2$ are positive integers), and, except for self-retracing orbits,
by an additional index $\epsilon$ specifying tori related to each other
through time reversal symmetry and therefore having the same topology.
$\gs$ represents the spin degeneracy factor, while
$S_{\bM,\epsilon}$ and $\tau_{\bM}$ are, respectively,
the action integral and the period along the periodic trajectories 
of the family $\bM$. $\hat\eta_{\bM}$ is
the Maslov index which counts the number of caustics 
of the invariant torus encountered by the trajectories.
For billiard systems with Dirichlet boundary conditions, we will
also take into account in $\hat\eta_{\bM}$ the phase $\pi$
acquired at each bounce of the trajectory on the hard walls
(and still refer to $\hat\eta_{\bM}$ as the Maslov index, although
slightly improperly).
The energy surface $E$ in action space whose implicit form is
${\cal H}(I_1,I_2) = E$, is explicitly defined by the function
$I_2 = g_E(I_1)$.
We note ${\bf I}^\bM = (I^\bM_1,I^\bM_2)$ the action variables of the torus
where the periodic orbits of topology $\bM$ live. They are determined
by the resonant-torus condition
	\be \label{eq:resonant}
	\alpha = - \left. \frac{d g_E(I_1)}{d I_1} 
	                       \right|_{I_1=I^\bM_1} \ = 
	\ \frac{M_1}{M_2} \; ,
	\ee
where the first equality arises from the differentiation  of
${\cal H}(I_1,g_E(I_1)) = E$ with respect to $I_1$.
Finally, the last contribution to the phase is given by 
$\gamma = {\rm sgn}(g''_E(I^\bM_1))$.

The [first] derivation of the Berry-Tabor trace formula \cite{ber76} 
follows very
similar lines as the treatment of the density of states performed in the 
introduction for the macroscopic Landau susceptibility. The EBK (Einstein,
Brillouin, Keller) quantization condition is used instead of the exact
form (\ref{LL}) of the Landau levels, followed by the
application of the Poisson summation rule. While in the latter case
this procedure leads to the
exact sum of Eq.~(\ref{DOSLL}), the Berry-Tabor formula is obtained
(similar to the treatment of de Haas~--~van Alphen oscillations
for a non-spherical Fermi surface)  after
a stationary-phase approximation valid in the semiclassical limit where
$S \gg \hbar$ (with a stationary-phase condition according to  
Eq.~(\ref{eq:resonant})).

Given a two-dimensional electron system whose classical Hamiltonian
	\be \label{eq:classH}
	{\cal H}(\bp,\bq) =  \frac{1}{2m} \ 
	\left(\bp - \frac{e}{c} \bA( \bq) \right)^2 
	+ V( \bq) 
	\ee
remains integrable for finite values of the transverse field 
$H {\hat z} = \nabla \times \bA$, the magnetic response can be obtained,
in principle, from the calculation of the various quantities involved 
in the Berry-Tabor formula at finite fields.
However, for weak fields, one can use the fact that the
field dependence of each contribution $d_\bM$ to the oscillating 
part of the density of states is essentially due to the
modification of the classical action, since this latter is multiplied 
by the large factor $1/\hbar$, while the field dependence of the periods 
and the curvatures of the energy manifold can be neglected.
Therefore, in this regime we will use for $\tau_\bM$ and $g_E$ the
values  $\tau^0_\bM$ and $g^0_E$ at zero field
and consider the first order correction $\delta S$ to the unperturbed 
action $S^0_\bM$.  A general result in classical mechanics 
\cite{ozor:book,boh95} states that the change ({\em at constant energy}) 
in the action integral along a closed orbit under the
effect of a parameter $\lambda$ of the Hamiltonian is given by
	\be \label{theorem}
	\left(\frac{\partial S}{\partial \lambda} \right)_E =
	- \oint \dif t \ \frac{\partial {\cal H}}{\partial \lambda} \ ,
	\ee
where the integral is taken along the {\em unperturbed} trajectory.
Therefore, if the Hamiltonian has the form of Eq.~(\ref{eq:classH}),
classical perturbation theory yields for small magnetic fields $H$,
	\be \label{dS}
	\delta S = \frac{e}{c} \ H \A_{\epsilon} \; ,
	\ee
where $\A_{\epsilon}$ is the 
directed area enclosed by the unperturbed orbit. 
This expansion is valid for magnetic fields
low enough, or energies high enough, such that the cyclotron radius of the
electrons is much larger than the typical size of the structure 
($r_c=mcv/eH \gg a$, which is, e.g., the case for electrons at the Fermi 
energy in the
experiments of Refs. \cite{levy93,BenMailly}). In this case we neglect the
change in the classical dynamics and consider the effect of the applied 
field only through the change of the action integral.

For a generic integrable system there is no reason, a priori, that
all the orbits of a given family $\bM$ should enclose the same area.
However, as pointed out above, a characteristic feature of integrable
systems is that the action is a constant for all the periodic orbits
of a given resonant torus.  Therefore, the fact that a system remains
integrable under the effect of a constant magnetic field implies
(because of Eq.~(\ref{dS})) that all the orbits of a
family enclose the same absolute area $\A_{\bM,\epsilon}$.  
Moreover, since the system is time-reversal invariant
at zero field, each closed orbit
($\bM,\epsilon$) enclosing an area $\A_{\bM,\epsilon}$ is associated
with a time-reversed partner having exactly the same
characteristics except for an opposite enclosed area (if the orbit
is its own time reversal, $\A_\bM = 0$). 
Grouping time-reversal trajectories in Eq.~(\ref{BT}) at $H\!=\!0$ we
have
     
	\begin{equation} \label{eq:dmstrhe0}
        d^0_{\bM}(E) = \left\{ \ba{ll} \displaystyle
                d^0_{\bM,\epsilon}(E)  & \qquad \qquad \mbox{for self-retracing orbits}\\
          \displaystyle \sum_{\epsilon=\pm 1} d^0_{\bM,\epsilon}(E) = 2 \ d^0_{\bM,\epsilon}(E)
                   & \qquad \qquad \mbox{for non self-retracing orbits}\\
        \ea \right.\; .
        \end{equation}
For weak fields the contribution of self-retracing orbits is unaltered and therefore
they do not contribute to the magnetic response. For the non
self-retracing ones we have

	\begin{equation} \label{eq:lowBd}
	d_\bM(E,H) = \sum_{\epsilon=\pm 1} d_{\bM,\epsilon}(E,H) = d^0_{\bM}(E) \
	\cos{\left(\frac{eH}{\hbar c} \A_{\bM} \right)} 
 \; \; , \; \; \A_\bM = |\A_{\bM,\epsilon}| \; .
	\end{equation}

\nin This is the basic relation to be used in the examples that follow.

\subsection{Circular billiards}
\label{sec:circle}

We now apply the preceding considerations to a 
two-dimensional gas of electrons
moving in a circular billiard of radius $a$ (where the potential $V(\bq) $
is zero in the region $|\bq| < a$ and infinite outside it). Thus we deal
with vanishing wavefunctions at the boundary (Dirichlet
boundary condition).  

In billiards without magnetic field the magnitude 
$p$ of the momentum is conserved, 
and it is convenient to introduce the wave number,
	\be \label{eq:k}
	k = \frac{p}{\hbar} = \frac{\sqrt{2mE}}{\hbar} 
	\ee

\nin since at $H\!=\!0$ the time-of-flight and the action-integral 
of a given trajectory 
can be simply expressed in terms of its length $L$ as
	\be \label{eq:SandT}
	\tau^0 = \frac{m}{p} \ L \; , \qquad
	\frac{S^0}{\hbar} = k L \;  .
	\ee

Following Keller and Rubinow \cite{keller}, we calculate the action
integrals ${\bf I}=(I_1,I_2)$ by using the independent paths ${\cal C}_1$ 
and ${\cal C}_2$ displayed
in Fig.~\ref{fig:circle}(a). The function $g_E$ is given by
(see \cite{keller} and 
Appendix \ref{app:ring_g})
	\be \label{circle:gE}
	g_E(I_1) = \frac{1}{\pi} \left\{ \left[(pa)^2-I_1^2\right]^{1/2} 
	- \ I_1 \ \arccos\left(\frac{I_1}{pa}\right) \right\} \ ,
	\ee
where $I_1$ is interpreted as the angular momentum and bounded by
$0 \leq I_1 < pa$.

The periodic orbits of the circular billiard are labeled by
the topology $\bM = (M_1,M_2)$, 
where $M_1$ is the number of turns around the circle until
coming to the initial point after $M_2$ bounces. (Obviously 
$M_2 \ge 2 M_1$.)  Elementary geometry yields for the length of 
the topology-$\bM$ trajectories
	\begin{equation} \label{lengthM}
	L_{\bM}=2M_2a \ \sin{\delta} \ ,
	\end{equation}

\nin where $\delta = \pi M_1/M_2$.
The resonant-torus condition, 
Eq.~(\ref{eq:resonant}), allows us to obtain  
${\bf I}^\bM$ as
	\begin{mathletters} \label{allIMs}
	\be \label{IMa}
	I_1^\bM = p a \cos{\delta} \; ,
	\ee
	\be \label{IMb}
	I_2^\bM = \frac{p a}{\pi} \left\{\sin{\delta} 
	\ - \ \delta \
	\cos{\delta} \right\} \; .
	\ee
	\end{mathletters}
The Maslov index of the topology-$\bM$ trajectories is 
$\hat\eta_\bM = 3 M_2$ 
($M_2$ bounces, each of them giving a dephasing of $\pi$, and $M_2$ 
encounters with the caustic per period).
We therefore have all the ingredients necessary to calculate the oscillating
part of the density of states at 
zero field: For the non self-retracing trajectories we obtain
	\begin{equation} \label{dgwgzf}
	d^0_{\bM}(E) = \sqrt{\frac{2}{\pi}} \
	\frac{\gs m L_{ \bM}^{3/2}}{\hbar^2}
	\ \frac{1}{k^{1/2}M_{2}^{2}} \ \cos{\left(k
	L_{\bM}\ + \frac{\pi}{4} - \frac{3\pi}{2} M_{2} \right)} \; .
	\end{equation}
The contribution of a self-retracing orbit is just one half of the contribution
(\ref{dgwgzf}). Its field dependent counterpart is obtained from 
Eq.~(\ref{eq:lowBd}) with the area enclosed by the periodic orbits given by
	\begin{equation} \label{areaMg}
	\A_{\bM} = \frac{M_2 a^2}{2} \sin{2\delta} \ .
	\end{equation}
The bouncing-ball trajectories $M_2= 2 M_1$ (with zero angular momentum) 
are self-retracing and have no
enclosed area; thus they do not contribute to the low field susceptibility.

Using   Eqs.~(\ref{DF1}) and (\ref{eq:smooth_osco}), and  noting
$\kf = k(\bar \mu) = (2/a) \ ({\bar N}({\bar \mu})/\gs)^{1/2}$ the 
Fermi wave vector, we obtain the contribution to the magnetic
susceptibility associated with $\Delta F^{(1)}$:
	\begin{eqnarray} 
	\frac{\chi^{(1)}}{\chi_L} & = &
	\frac{48}{\sqrt{2 \pi}} \ (\kf a)^{3/2} \; \times 
	\label{chicir1}  \\
	& \times &
	\sum_{M_1,M_2 > 2M_1} \frac{(\A_{\bM}/a^2)^2}{(L_{\bM}/a)^{1/2}} \ 
	\frac{1}{M_2^2} 
	\cos{\left(\kf L_{\bM}\ + \frac{\pi}{4} - \frac{3\pi}{2}M_2\right)}
	\cos{\left(\frac{eH}{\hbar c} \A_{\bM} \right)}
	\ R_T(L_{\bM}) \ . \nonumber
	\end{eqnarray}
Since we are working with billiards, the temperature factor $R_T$ is
given in terms of the trajectory length $L_\bM$ by Eq.~(\ref{R_factor2}) 
and the characteristic cut-off length 
$L_c = \hbar \vf \beta/\pi $. For 
$M_2 \gg M_1$ we have $L_\bM \simeq 2 \pi M_1 a$ and 
$\A_\bM \simeq \pi M_1 a^2$, independent of $M_2$.  
Performing the summation over the index $M_2$ (for fixed value of 
$M_1$) by taking the length and area
dependent terms outside the sum we are left with a rapidly convergent 
series (whose general term is $(-1)^{M_2}/M_2^2$). We can therefore 
truncate the series after the first few terms. In Fig.~\ref{fig:chi_circle}
the sum (\ref{chicir1}) is evaluated numerically at zero field
(solid line) for a cut-off length $L_c = 6 a$ which selects only
the first ($M_1=1$) harmonic,
and the beating between the first few periodic orbits is obtained
as a function of wave-vector $\kf$. With only the first two primitive orbits
($M_2=3$ and $4$, dashed line) we give a good account of $\chi^{(1)}$ for
most of the $k$-interval. Taking the first four primitive orbits 
suffices to reproduce the 
whole sum. The short period in $\kf$ corresponds 
approximately to the circle perimeter $L=2\pi a$.
Going to lower temperatures gives an overall increase of
the susceptibility but does not modify the structure of the 
first harmonic contribution
since the length of the whispering-gallery trajectories is bounded
by $L$. However, for larger values of $L_c$ higher harmonics, namely
up to $M_1$ of the order of $L_c/2\pi a$, will be observed. 
The predominance of
the first few trajectories also appears in the beating as a function 
of magnetic field (not shown) that results from the evaluation of 
(\ref{chicir1}) at finite fields.

>From Fig.~\ref{fig:chi_circle} we see that the susceptibility of a circular 
billiard oscillates as a function of the number of electrons (or $\kf$)
taking paramagnetic and diamagnetic values. Its overall magnitude is
much larger than the two-dimensional Landau susceptibility and grows as 
$(\kf a)^{3/2}$. We will later show (Sec.~\ref{sec:general}) 
that this finite-size increase with respect 
to the bulk value is distinctive of systems that are integrable at
zero field. In order to characterize the typical value of the magnetic
susceptibility we define
	\be
	\label{chicir1t}
\chi^{({\rm t})}  \ = \  \left[ \ \overline{(\chi^{(1)})^2} \ \right]^{1/2} 
	\ee

\nin where, as in section \ref{sec:TherFor},
 the average is over a $\kf a$ interval
classically negligible ($\Delta(\kf a) \ll \kf a$)
but quantum mechanically large ($\Delta(\kf a) \gg 2\pi$), so that 
off-diagonal terms $\cos(\kf L_\bM) \cos(\kf L_{\bM'})$ 
with $\bM \neq \bM'$ vanish under averaging. A remark is in order 
here because at fixed $M_1$, $L_\bM$ goes to $2\pi M_1 a$ as 
$M_2$ goes to $\infty$, and $(L_{(M_1,M_2)} - L_{(M_1,M'_2)})$ 
can be made arbitrarily small by increasing $M_2$ and $M'_2$.
Therefore, for any interval of $\kf a$ over which the
average is taken, some non-diagonal terms should remain 
unaffected.  Nevertheless, because of the rapid decay of the
contribution with $M_2$, these non-diagonal terms can be neglected in
practice for the experimentally relevant temperatures.
The typical zero-field susceptibility
of the circular billiard is then given by
       \be
        \label{chicir1td}
	\frac{\chi^{({\rm t})}(H\!=\!0)}{\chi_L} \
	\simeq \ \frac{48}{\sqrt{2 \pi}}  \ (\kf a)^{3/2} \
	\left[ \frac{1}{2} \sum_{M_1,M_2>2M_1} 
	\frac{(\A_{\bM}/a^2)^4}{L_{\bM}/a} \ \frac{R^{2}_T(L_{\bM})}{M_2^4}
	\right]^{1/2} \ .
	\ee
Numerical evaluation of the first harmonic ($M_1\!=\!1$) from (\ref{chicir1t}) on the
$\kf a$ interval of Fig.~\ref{fig:chi_circle} with $L_c = 6a$ gives
$2.20 (\kf a)^{3/2} \chi_L$ (dotted horizontal line),
while Eq.~(\ref{chicir1td}) restricted to $M_2\!\le\!6$
yields $2.16 (\kf a)^{3/2} \chi_L$ 
illustrating the smallness of the off-diagonal and large-$M_2$ terms.

For an ensemble made of circular billiards with a dispersion in size
or in the number of electrons such that
$\Delta (\kf a) > 2 \pi $, the term $\chi^{(1)}$ yields a vanishing 
contribution to the average susceptibility. In such a case it is necessary
to go to the next-order free-energy term $\Delta F^{(2)}$, whose 
associated contribution $\chi^{(2)}$ yields the average susceptibility
by means of Eqs.~(\ref{eq:sus}) and (\ref{DF2}).  For the same
reason as above one can show that only diagonal 
terms of $(\Nosc)^2$ survive the $\kf a$ average, in spite of the degeneracy
of the length of the closed orbits as $M_2$ goes to $\infty$. One therefore has
	\be \label{chicir}
	\frac{\overline{\chi}}{\chi_L}  =
	\frac{48}{\pi} \ \kf a \ \sum_{M_1,M_2 > 2M_1} 
	\frac{(A_\bM/a^2)^2 (L_\bM/a) }{M_2^4} \ 
	\cos{\left(\frac{2eH}{\hbar c} \A_{\bM} \right)}
	 \ R_T^{2}(L_\bM) \ .
	\ee
Again, the terms generally decay rapidly with $M_2$ (as $1/M_2^4$), and
for a cutoff length $L_c$ selecting only the terms with $M_1=1$
the total amplitude at zero field  ($5.2 \kf a$) can be obtained 
from the first few lowest terms. The low field
susceptibility of an ensemble of circular billiards is paramagnetic
and increases linearly with $\kf a$. As for the $\chi^{(1)}$ 
contribution, we will show in the sequel that this behavior does
not necessitate the integrability at finite fields, but rests only
upon the integrability at zero field.

Up to now there have not been measurements of the magnetic response
of electrons in circular billiards (individual or ensembles). Our
typical (Eq.~(\ref{chicir1td})) or average (Eq.~(\ref{chicir})) 
susceptibilities exhibit a large enhancement with respect to the bulk
values (by powers of $\kf a$). Thus it should
be possible to detect experimentally these finite-size effects.

\subsection {Rings}

The magnetic response of small rings can be calculated along the same 
lines as
in the case of the circles. The ring geometry deserves special interest since
it is the preferred configuration for persistent current 
measurements. In a
ring geometry at $H\!=\!0$ we have two types of periodic orbits: 
those which do not touch the inner disk (type-I), 
and those which do hit it (type-II).
(See Fig.~\ref{fig:circle} of Appendix \ref{app:ring_g}; we note
by $a$ and $b$ respectively the outer and inner radius of the ring.)
The function 
$g_{E}(I_1)$ has two branches corresponding to the interval to which the
angular momentum $I_1$ belongs. For $pb<I_1<pa$, (type-I trajectories) 
$g_{E}$ has the same form (\ref{circle:gE}) as for the circle, while for
 $0 \leq I_1 <pb$, (type-II trajectories) we show in
Appendix \ref{app:ring_g} that
	\be \label{ring:gE}
	g_E(I_1) = \frac{1}{\pi} \left\{ \left[(pa)^2-I_1^2\right]^{1/2} -
	\left[(pb)^2-I_1^2\right]^{1/2} \ - \
	I_1 \left[\arccos\left(\frac{I_1}{pa}\right) -
	\arccos\left(\frac{I_1}{pb}\right)\right] \right\} \ .
	\ee

The type-I trajectories are labeled in the same way as for the circle
by the topology $\bM=(M_1,M_2)$ representing the number of turns $M_1$
around the inner circle until returning to the initial point after $M_2$ 
bounces on the outer circle. We therefore obtain the 
resonant-tori condition
Eqs.~(\ref{allIMs}) and the same contribution (\ref{dgwgzf}) to the 
oscillating part of the density of states as in the case of the circle.
The only difference is that in the Berry-Tabor trace formula
(Eq.~(\ref{BT})) the sum corresponding to type-I trajectories is now
restricted to $M_2 \geq \hM_2(M_1) = {\rm Int}[M_1 \pi/\arccos{r}]$. 
We note by
${\rm Int}$ the integer-part function and $r=b/a$.
We stress the fact that the minimum value of $M_2$ is itself a 
function of $M_1$. The previous restriction
can also be expressed as  $\cos{\delta} > r$, with
$\delta = \pi M_1/M_2$. 
Type-II trajectories can be labeled by the
topology $\bM=(M_1,M_2)$, where $M_1$ is the number of turns 
around the inner circle in coming to the initial point after $M_2$
bounces on the {\em outer} circle. We have the same restriction 
$M_2 \geq \hM_2(M_1) $ as for type-I trajectories, and we can use 
$\hat\eta_{\bM}=0$
since there are $2M_2$ bounces with the hard walls and no encounters
with the caustic. From (\ref{ring:gE}) we obtain the  resonant-torus 
condition
        \begin{mathletters} \label{allIMRs}
        \be \label{IMRa}
        I_1^\bM = p b \ 
	\frac{\sin{\delta}}{\sqrt{1+r^2-2r\cos{\delta}}} \; ,
        \ee
        \be \label{IMRb}
        I_2^\bM = \frac{p a}{\pi} \left\{
	\sqrt{1+r^2-2r\cos{\delta}} \ - \ 
	\frac{r \delta \sin{\delta}}{\sqrt{1+r^2-2r\cos{\delta}}} 
        \right\} \; .
        \ee
        \end{mathletters}

The $H\!=\!0$ contribution to the oscillating part of the density of states
from non self-retracing type-II trajectories with topology $\bM$ is given by
        \begin{equation}
        \tilde{d}^0_{\bM}(E) = 4 \sqrt{\frac{2}{\pi}} \
	    \frac{\gs a^2 m}{\hbar^2 } \
	\frac{\left[(1-r\cos{\delta})(r\cos{\delta}-r^2) \right]^{1/2}}
	{\left( k \tilde L_\bM \right)^{1/2}}
        \ \sin{\left(k \tilde{L}_{\bM}\ + \frac{\pi}{4} \right)} \ ,
        \label{dgwgzft1}
        \end{equation}
while its length is
        \begin{equation} \label{lenghtMt1}
        \tilde{L}_{\bM}=2M_2a \ \sqrt{1+r^2-2r\cos{\delta}} \ .
        \end{equation}
The small field dependence follows from Eq.~(\ref{eq:lowBd}) using
the enclosed area
	\begin{equation} \label{areaMt1}
	\tilde A_\bM = 
	M_2 ab \ \sin{\delta} \ .
	\end{equation}

In the case of annular geometries it is customary to characterize the magnetic
moment ${\cal M}$ of the ring by the persistent current 
	\begin{equation} \label{eq:percu}
	I =  \frac{c}{A} \ {\cal M} = 
	- c \left(\frac{\partial F}{\partial \Phi} \right)_{T,N}  \ .
	\end{equation}
In order to pass from the applied magnetic field $H$ to the
flux $\Phi$ we use the area $A$ of the outer circle ($\Phi=A H$, \  
$A=\pi a^2$) as defining area. (For thin rings, all periodic orbits with
the same repetition number $M_1$ enclose approximately the same
flux $M_1 \Phi$.) Applying  Eqs.~(\ref{eq:smooth_osco})--(\ref{allDF}),
and calling $I_0=e \vf/2\pi a$ the typical current of one-dimensional 
electrons at the Fermi energy, the persistent current of a ring billiard 
can be expressed as the sum of two contributions corresponding to  both 
types of trajectories:
	\be \label{I1}
	\frac{I^{(1)}}{I_0}  = 
	\gs \ (\kf a)^{1/2}  \sum_{M_1,M_2 \geq \hM_2}
	\left\{ {\cal I}^{(1)}_{\bM,I} \ \sin{\left(\frac{eH}{\hbar c} \A_{\bM} \right)}
        \ R_T(L_{\bM}) \ + 
	{\cal I}^{(1)}_{\bM,II} \ \sin{\left(\frac{eH}{\hbar c} \tilde \A_\bM \right)}
        R_T(\tilde{L}_{\bM}) \right\} \ ,
	\ee

        \begin{mathletters} \label{allI1Rs}
        \be \label{I1Ra}
        {\cal I}^{(1)}_{\bM,I} = 
	2\sqrt{\frac{2}{\pi}} \ \frac{1}{M_2^2} 
        \ \frac{(\A_\bM/a^2)}{(L_{\bM}/a)^{1/2}}
        \ \cos{\left(\kf L_\bM + \frac{\pi}{4} - \frac{3\pi}{2} M_2\right)}
        \ ,
	\ee
        \be \label{I1Rb}
        {\cal I}^{(1)}_{\bM,II} = 8 \sqrt{\frac{2}{\pi}} 
        \frac{(\tilde \A_\bM/a^2)}{(\tilde{L}_{\bM}/a)^{5/2}}
        \left[(1-r\cos{\delta})(r\cos{\delta}-r^2) \right]^{1/2}
        \sin{\left(k \tilde{L}_{\bM}\ + \frac{\pi}{4} \right)}
        \; .
        \ee
        \end{mathletters}

In Fig.~\ref{fig:chi_ring} we present  the first harmonic
$I_{1}^{(1)}$ of the persistent current for a thin ring and a cut-off
length $L_c=6a$ (solid line). (I.e., we are considering the winding number $M_1=1$.)
The contribution of type-I trajectories (dashed line) is similar as in
the case of the
circle: a rapidly convergent sum showing as a function of $\kf$ the 
beating between the first two trajectories ($\hM_2$ and $\hM_2+1$). On the
other hand, Eq.~(\ref{I1Rb}) shows that the trajectories with low
values of $M_2$ (i.e.~$M_2 \sim \hM_2$) contributing
to ${\cal I}^{(1)}_{\bM,II}$ have negligible weight due to the small
stability prefactor caused by the defocusing effect exerted by the
inner disk ($\cos {\delta} \simeq r$). The sum is dominated by
trajectories with $M_2 > \hM_2$ and therefore we loose the
previous beating structure in the total $I_1^{(1)}$.
The short period in $\kf$ still corresponds to the circle perimeter $L$.

As in the previous subsection, we characterize
the typical value of the magnetic response by averaging
$(I^{(1)})^2$ over a $\kf a$-interval containing many oscillations,
but yet negligible on the classical scale.
        \be  \label{i1t}
        I^{({\rm t})} \ = \ \left[ \ \overline{\left( I^{(1)} \right)^2} 
	   \ \right]^{1/2} \ .
        \ee
In the same way as for the circular billiard, one can in practice consider
that, despite the degeneracy in the length of type-I trajectories for
large $M_2$, only diagonal terms (in both index $\bM$ and 
trajectory-type) survive the averaging for large enough
$\Delta(\kf a)$.  Therefore

\begin{eqnarray} \label{I1td}
\frac{I^{({\rm t})}}{I_0} & \simeq & \gs (\kf a)^{1/2}
\sum_{M_1,M_2 \geq \hM_2}
\left[ \left({\cal I}^{({\rm t})}_{\bM,I}  \right)^2 
\sin^2{\left(\frac{eH}{\hbar c} \A_{\bM} \right)} R_T^2(L_{\bM}) \right. \nonumber + \\  
& + & \left. \left( {\cal I}^{({\rm t})}_{\bM,II} \right)^2 
\sin^2{\left(\frac{eH}{\hbar c} \tilde \A_\bM \right)} R_T^2(\tilde{L}_{\bM})
\right]^{1/2} \ , 
\end{eqnarray}

\nin where  $({\cal I}^{({\rm t})}_{\bM,I})^2$ and 
$({\cal I}^{({\rm t})}_{\bM,II})^2$ are obtained 
from Eqs.~(\ref{allI1Rs}) simply by replacing the average of
$\cos^2(\kf L_\bM + \pi/4 -3M_2\pi/2)$ and 
$\sin^2(\kf \tilde L_\bM + \pi/4)$ by $1/2$.

In Fig.~\ref{fig:chi_ty} we present the typical persistent current and its
two contributions for various ratios $r=b/a$ and cut-off lengths $L_c$ for
the first harmonic ($M_1\!=\!1$).
The contribution ${\cal I}^{({\rm t})}_{\bM,I}$ of type-I trajectories dominates
for small $r$ (where the inner circle is not important and we
recover the magnetic response of the circular billiard) while type-II
trajectories take over for narrow rings. The crossover $r$ depends on
temperature through $L_c$ due to the different dependence of the
trajectory length on $\bM$ (Eqs.~(\ref{lengthM}) and (\ref{lenghtMt1}))
for both types of trajectories.

As in the case of $\chi^{(1)}$ for the circular billiard, $I^{(1)}$ 
 gives a vanishing contribution to
the persistent current of an ensemble of rings with different sizes or
electron fillings as soon as the dispersion in $\kf a$ is of the order
of $2 \pi$. We therefore need to go to the term $\Delta F^{(2)}$ in
the free-energy expansion, which is obtained (see Eq.~(\ref{DF2})) from
	\be \label{noscring}
	\Nosc (\bar \mu) =  \sum_{M_1,M_2 \geq \hM_2}
	\left\{ N_{\bM,I}(\bar \mu)+N_{\bM,II}(\bar \mu) \right\} \ ,
	\ee
where $N_{\bM,I}(\bar \mu)$ and $N_{\bM,II}(\bar \mu)$ are given in terms
of the respective contributions to the field dependent density of 
states through Eq.~(\ref{eq:smooth_oscn}). 
For an ensemble with a large dispersion of
sizes only diagonal terms survive the average and we have (with
$\bar D = \gs m A (1-r^2) / (2\pi\hbar^2)$)
	\be \label{i2av}
	\frac{\overline{I^{(2)}}}{I_0} =
	     \gs \ \sum_{M_1,M_2 \geq \hM_2}
	\left\{ \overline{{\cal I}^{(2)}_{\bM,I}} \ 
	\sin{\left(\frac{2 eH}{\hbar c} \A_{\bM} \right)}
        \ R_T^2(L_{\bM}) + 
	\overline{{\cal I}^{(2)}_{\bM,II}} 
	\ \sin{\left(\frac{2eH}{\hbar c} \tilde \A_\bM \right)}
        \ R_T^2(\tilde{L}_{\bM}) \right\} \ ,
	\ee

   \begin{mathletters} \label{allI2Rs}
        \be \label{I2Ra}
        \overline{{\cal I}^{(2)}_{\bM,I}} = 
	    \frac{2}{\pi} \ \frac{1}{M_2^4}
        \ \left(\frac{L_{\bM}}{a}\right) \
        \left(\frac{\A_\bM}{a^2}\right)\ \frac{1}{1-r^2} \
	\ ,
        \ee
        \be \label{I2Rb}
        \overline{{\cal I}^{(2)}_{\bM,II}} = \frac{32}{\pi} \
	\frac{\left( \tilde \A_\bM/a^2 \right)}{\left( \tilde
					  L_\bM / a \right)^3} 
        \ \frac{(1-r \cos{\delta})(r\cos{\delta}-r^2)}{1-r^2}
        \; .
	\ee
        \end{mathletters}

The $\kf$ dependence of the average persistent current is linear (through $I_0$),
similarly to the case of the average susceptibility of an ensemble of circular
billiards.

\subsubsection*{Thin rings}

In the case of thin rings ($a \simeq b$, \ $r \simeq 1$)
further approximations can be performed on Eqs.~(\ref{allI1Rs}) 
and (\ref{allI2Rs}) using $(1-r)$  as
a small parameter, giving more compact and meaningful expressions
for the typical and average persistent currents.  Since in addition
this is the configuration used in the experiment of 
Ref.~\cite{BenMailly}, we shall consider more closely
this limiting case.  First, we note that $\hat \delta = \arccos{r}
\simeq \sqrt{2(1-r)} \ll 1$.  Thus
	\be \label{mhat}
	\hM_2 = {\rm Int}\left[\frac{\pi M_1}{\hat \delta}\right]
	\simeq \frac{\pi}{\sqrt{2}} \ \frac{M_1}{\sqrt{1-r}}
	\gg M_1 \; ,
	\ee
and for $M_2 \ge \hM_2$, the area and length of contributing orbits 
can be approximated by

        \be \label{ALTRs}
        \A_\bM \simeq \tilde \A_\bM  \simeq M_1 A = M_1 \pi a^2
        \qquad ; \qquad
        L_\bM \simeq M_1L = M_1 2 \pi a  \; .
        \ee
For the length of type-II trajectories we have 
$\tilde{L}_{\bM} \simeq M_1 L$ for $M_2 \simeq \hM_2$, and
${\tilde L}_\bM \simeq 2 M_2(a-b)$ when $ M_2 \gg \hM_2$. 
All trajectories with winding number $M_1$ enclose approximately the same
flux $M_1 \Phi$, and the field dependent terms in Eq.~(\ref{I1})
may be replaced by $\sin{(2 \pi M_1 \Phi/\Phi_0)}$.
There is therefore no difference between the case that we study (where
a uniform magnetic field $H$ is applied) and the ideal case of a flux line
$\Phi$ through the inner circle of the ring.
The length dependent factors $R_T^2$ can also be taken outside the sum
over $M_2$ since the main contribution of type-II trajectories comes from
$M_2 \simeq \pi M_1/[5^{1/6}(1-r)^{2/3}]$.
Even if these $M_2$'s are much larger than $\hM_2$, their associated
${\tilde L}_\bM$ are still
of the order of $M_1 L$ to leading order in $1-r$.
 
Turning now to the typical and ensemble average currents, 
it should be stressed that for narrow rings it is necessary
to go to fairly large energies before an average on a scale
being quantum mechanically large but classically small is possible.
Indeed, one has for both types of trajectories
$\kf(L_{\hM_2+1}-L_{\hM_2}) \simeq \kf(\tilde L_{\hM_2+1} -
\tilde L_{\hM_2}) \simeq (4\sqrt{2}/3) \pi {\cal N} \sqrt{1-r}$,
where ${\cal N}=\kf (a-b)/\pi$ is the number of transverse occupied 
channels.  Therefore, ${\cal N}$ should be much larger than 
$(1-r)^{-1/2}$ if one wants to assume $\Delta (\kf a)$ sufficiently 
large to average out all non-diagonal terms without 
violating the condition $\Delta(\kf a) \ll \kf a$.
Supposing the previous condition is met, and
introducing the typical amplitudes  ${\cal J}^{({\rm t})}_{M_1,I}$ and 
${\cal J}^{({\rm t})}_{M_1,II}$ of each harmonic, we write

	\be 
	\frac{I^{({\rm t})}}{I_0}  = 
	\gs \  \left[ \sum_{M_1} \left\{
	\left( {\cal J}^{({\rm t})}_{M_1,I} \right)^2
	 + \left( {\cal J}^{({\rm t})}_{M_1,II} \right)^2 \right\}
	\sin^2\left(2\pi M_1\frac{\Phi}{\Phi_0}  \right) 
	R^2_T(M_1 L) \right]^{1/2} \ ,
	\ee
    
  \begin{mathletters} \label{allJts}
	\be \label{Jta}
	\left({\cal J}^{({\rm t})}_{M_1,I}\right)^2 = 
	\kf a \ \sum_{M_2 \geq \hM_2} \left({\cal I}^{({\rm t})}_{\bM,I}\right)^2
	=  2 \kf a M_1 \left[ \sum_{M_2 \geq \hM_2} 
	\frac{1}{M_2^4} \right]  \; ,
        \ee
        \be
        \label{Jtb}
        \left({\cal J}^{({\rm t})}_{M_1,II}\right)^2 = 
	\kf a \ \sum_{M_2 \geq \hM_2} \left({\cal I}^{({\rm t})}_{\bM,II}\right)^2 =
	2 \pi \ \kf a M_1^2 \left[\sum_{M_2 \geq \hM_2}
        \frac{(1-r)^2 - \delta^4/4}
        {M_2^5((1-r)^2+\delta^2)^{5/2}} \right] \ .
        \ee
  \end{mathletters}

Since $\hM_2 \gg 1$ we can convert the previous sums into  integrals and obtain

  \begin{mathletters} \label{allJtlos}
	\be \label{Jtloa}
	\left({\cal J}^{({\rm t})}_{M_1,I}\right)^2 \simeq \frac{4 \sqrt{2}}{3 (\pi M_1)^2} \ 
	{\cal N} \ (1-r)^{1/2}  \ .
        \ee
   	\be
	\label{Jtlob}
	\left({\cal J}^{({\rm t})}_{M_1,II}\right)^2 \simeq \frac{4}{3 (\pi M_1)^2}
	{\cal N} \ \left(1 - \sqrt{2} (1-r)^{1/2} \right) \ .
        \ee
  \end{mathletters}

In leading order in $1-r$ the persistent current is dominated by type-II trajectories
(independent of the temperature) and given by
   \be \label{I1TRn5}
        \frac{I^{({\rm t})}}{I_0} = \frac{2}{\pi\sqrt{3}} \ \gs \ 
	\sqrt{\cal N} \left[\sum_{M_1} \frac{1}{M_1^2}
	\ \sin^2{\left(2 \pi M_1\frac{\Phi}{\Phi_0}\right)} \ R_T^2(M_1 L) 
	\right]^{1/2} \ ,
        \ee
consistent with the result of Ref.~\cite{RiGe}. For the next order term
the contribution from type-I trajectories is cancelled by that of type-II
resulting in the relatively flat character of the curves for $I^{({\rm t})}$
in Fig.~\ref{fig:chi_ty}.
	
For the current of an ensemble of thin rings, the calculations are similar 
to those of Eqs.~(\ref{allJtlos}), and in leading order in
$1-r$ we obtain:
     \be
        \frac{\overline{I^{(2)}}}{I_0}  =
        \gs \  \sum_{M_1} \left\{
        \overline{{\cal J}^{({2})}_{M_1,I}}
          + \overline{{\cal J}^{({2})}_{M_1,II}} \right\}
        \sin\left(4\pi M_1\frac{\Phi}{\Phi_0}  \right)
        R^2_T(M_1 L) \ ,
        \ee

  \begin{mathletters} \label{allI3Rs}
        \be \label{I3Ra}
   	\overline{{\cal J}^{({2})}_{M_1,I}} = 
	\sum_{M_2 \geq \hM_2} \overline{{\cal I}^{({2})}_{\bM,I}} =
	\frac{4\sqrt{2}}{3\pi^2}
        \ \sqrt{1-r} \frac{1}{M_1}
	\ ,
        \ee
        \be \label{I3Rb}
        \overline{{\cal J}^{({2})}_{M_1,II}} = 
	\sum_{M_2 \geq \hM_2} \overline{{\cal I}^{({2})}_{\bM,II}} =
	\frac{2}{\pi^2} \ 
        \left( 1 - \frac{2\sqrt{2}}{3} \sqrt{1-r} \right)
	\frac{1}{M_1}
        \; .
	\ee
        \end{mathletters}
\nin Type-II trajectories once again dominate the average magnetic 
response of thin rings and the amplitude for the first harmonic is
$\overline{I_{1}^{(2)}}/I_0 \simeq (2\gs/\pi^2) \sin{(4 \pi \Phi/\Phi_0)} 
R_T^2(L)$,
independently of the number of transverse channels ${\cal N}$. The average 
persistent current shows the halfing of the flux period with respect to 
$I^{(1)}$
characteristic for ensemble results (as found in the disordered case and
consistently with the results for averages in the following sections).

\subsubsection*{Comparison with Experiment}

Persistent currents have been measured by Mailly, Chapelier and Benoit 
\cite{BenMailly} in a thin
semiconductor ring (with effective outer and inner radii $a=1.43 \mu m$ and
$b=1.27 \mu m$) in the ballistic and phase-coherent regime ($l=11 \mu m$ 
and
$L_{\Phi}=25 \mu m$). The Fermi velocity is $\vf=2.6\times 10^7 cm/s$ and
therefore the number of occupied channels is ${\cal N} \simeq 4$. 
The quoted temperature of $T=15 mK$ makes the temperature factor irrelevant
for the first harmonic ($L_c \simeq 30a$, $R_T(L) \simeq 1$). 
The magnetic response exhibits an
$hc/e$ flux periodicity and changes from diamagnetic to paramagnetic 
by changing the microscopic configuration, consistently with 
Eqs.~(\ref{I1})-(\ref{allI1Rs}). Unfortunately, the sensitivity is not high
enough in order to test the signal averaging with these microscopic changes.
The typical persistent current was found to be $4 nA$, while Eq.~(\ref{I1TRn5})
and Ref.~\cite{RiGe} would yield $7 nA$. The difference between the theoretical
and measured values is not significant given the experimental uncertainties as
discussed in Refs.~\cite{BenMailly} and \cite{von_thesis}.
Moreover, as we stressed above,
a very large $\kf a$ interval is needed for the average of $(I^{(1)})^2$
in order to recover $I^{({\rm t})}$; otherwise we expect large
statistical fluctuations. As in the case of the susceptibility of squares
that we analyze in the next section, residual disorder (reducing the magnetic
response without altering the physical picture) and interactions may be 
necessary in order to attempt a detailed comparison with the experiment.
Clearly, new experiments on individual rings of 
various thickness and on ensembles of ballistic rings
would be helpful in order to test the ideas of the present section.

\newpage

%
\section {Simple Regular Geometries: the Square}
\label{sec:square}

The circular and annular billiards studied in section~\ref{sec:integrable}
have the remarkable property that, due to their rotational symmetry,
they remain integrable under the application of a magnetic field.
However, for a generic integrable system (a {\em regular} geometry)
any perturbation breaks the integrability of the dynamics.  Moreover,
the periodic orbits which are playing the central role in the semiclassical
trace formulas are most strongly affected by the perturbation.  Indeed,
the Poincar\'e-Birkhoff theorem \cite{arnold:book} states that as soon as the
magnetic field is turned on,  all resonant tori (i.e.~all
families of periodic orbits) are instantaneously broken, leaving only two
isolated periodic orbits (one stable and one unstable). It is therefore no longer
possible to use the Berry-Tabor semiclassical trace formula to
calculate the oscillating part of the density of states for finite field
since it is based on a sum over resonant tori, which do not exist any further.
One has therefore to devise a semiclassical technique allowing to
calculate $\dosc(E)$ for nearly, but not completely, integrable systems.

To achieve this, it is necessary to go back to the basic equations from
which the standard semiclassical trace formulae (Gutzwiller \cite{gut71},
Balian-Bloch \cite{bal69}, Berry-Tabor \cite{ber77}) are derived.
The density of states $d(E)$, Eq.~(\ref{DOS}), is related to the trace  of the
energy dependent Green function $G({\bf q},{\bf q'};E)$ by
	\be  \label{eq:traceG}
	d(E) = - \frac{\gs}{\pi} \ {\rm Im} \ \G(E) \; ,
	\qquad \qquad
	\G(E) = \int \dif {\bf q} \,
	G ({\bf q},{\bf q};E)  \; ,
	\ee
where again the factor $\gs=2$ comes from the spin degeneracy.
$G({\bf q},{\bf q'};E)$ has a singularity (logarithmic in two dimensions)
when $\br \rightarrow \br'$ which just gives the smooth [Weyl] part $\bar d(E)$
of the density of states in a leading order semiclassical expansion. However,
in order to consider only the oscillating part of $d(E)$
one can use the semiclassical approximation of the Green function
\cite{gutz_book}
	\be \label{eq:green}
	\Gsc({\bf q},{\bf q'};E) =
            \frac{1}{i \hbar} \frac{1}{\sqrt{ 2 i \pi \hbar}}
	       \sum_t D_t  \exp{\left[\frac{i}{\hbar} S_t - i
                            \eta_t\frac{\pi}{2}\right]}
	\ee
where the sum runs over all classical  trajectories $t$ joining
${\bf q}$ and ${\bf q'}$ {\em at energy $E$}.
$S_t$ is the action along the trajectory $t$,
$D_t$ a determinant involving second derivatives of the action
(the general expression of which is given in appendix~\ref{app:D_M})
and $\eta_t$ is the Maslov index of the trajectory, i.e. the number of
focal points encountered  when  traveling from $\bq$ to $\bq'$.
As in section~\ref{sec:integrable}, we shall also take into account
in $\eta_t$ the phase $\pi$ acquired at
each reflection at the wall of a billiard with Dirichlet boundary conditions.

By taking the trace (\ref{eq:traceG}) the sum in Eq.~(\ref{eq:green})
becomes a sum over all orbits closed in configuration (i.e.~$\bq$) space,
to which we will refer in the following as {\em recurrent} orbits.
The standard route to obtain $\dosc$ is to
evaluate this integral by stationary-phase approximation.
This selects the trajectories which are not only closed in configuration
space $({\bf r}'\!=\!{\bf r})$, but also closed in phase space
($\bf p'\! =\! \bf p$), i.e. {\em periodic} orbits.
When these latter are [well] isolated the Gutzwiller Trace Formula \cite{gut71}
is obtained. For integrable systems, all recurrent orbits are in fact periodic
since the action variables are constants of motion.

Periodic orbits appear in continuous
families associated with resonant tori. All orbits of a family have the same
action and period, and one can calculate the density of states
using the Berry-Tabor Formula as described in
the previous section.  For systems such as the square billiard, the
physical effect which generates the susceptibility comes along with
the breaking of the
rational tori, so that just ignoring this, i.e.~using the Berry-Tabor Formula,
is certainly inadequate. On the other hand, for $H \rightarrow 0$ the
remaining orbits are not sufficiently well isolated to apply the Gutzwiller
Trace Formula. Therefore, as stated before, we need a uniform treatment of the perturbing
field, in which not only the orbits being closed in phase space are taken
in account, but also the orbits closed in configuration space  which can be
traced back to a periodic orbit when $H \rightarrow 0$.

In this section we show how this can be performed in the particular case of
a square billiard. Because of the simplicity of its geometry,
the integrals involved in the trace Eq.~(\ref{eq:traceG})
can be performed exactly for weak magnetic fields.
Moreover, the square geometry deserves special interest since it was the first
microstructure experimentally realized to measure the magnetic response
in the ballistic regime.
We present here a semiclassical approach addressing the
physical explanation of the experimental findings of Ref.~\cite{levy93},
which have pointed the way
for the ongoing research.  In order to obtain semiclassical expressions
for the susceptibility of individual and ensembles of squares
we will proceed as outlined in section~\ref{sec:TherFor}: 
We will calculate the density
of states and use the decomposition of the susceptibility according to
Eq.~(\ref{eq:fd}) into contributions corresponding to $\Delta F^{(1)}$
and $\Delta F^{(2)}$.  In section~\ref{sec:general} we present the
theory for a generic  integrable system perturbed by a magnetic field,
generalizing the results of this section.

\subsection{Oscillating density of states for weak field}
\label{seq:square1}

To start with, we consider a square billiard (of side $a$) in
the absence of a field. Each family of periodic orbits can be labeled by
the topology $\bM = (M_x, M_y)$ where $M_x$ and $M_y$ are the
number of bounces occurring on the bottom and left side
of the billiard (see Fig.~\ref{fig:fam11}).
The length of the periodic orbits for all members of a family is
	\be \label{square:length}
	L_\bM = 2 a \sqrt{M_x^2 + M_y^2} \ .
	\ee

\nin The unperturbed action along
the trajectory is, as for any billiard system,
$S^0_\bM/\hbar  = k L_\bM $ where $k$ is the wavenumber.
The Maslov indices are $\eta_\bM = 4 (M_x + M_y)$,
and we will omit them from now on since they only yield a dephasing of
a multiple of $2\pi$.  Finally the unperturbed determinant reduces to
	\be \label{eq:square:DM}
	D_\bM = \frac{m}{\sqrt{\hbar k L_\bM}} \; .
	\ee
One way to obtain this result is to use the method of images
(see Fig.~\ref{fig:image})
and express the exact Green function $G(\bq,\bq';E)$ in terms of the
free Green function $G^0(\bq,\bq';E)$ as \cite{bal69,gut71}
	\begin{equation} \label{image}
	G(\bq,\bq';E) = G^0(\bq,\bq';E)
		+ \sum_{\bq_i} \epsilon_i G^0(\bq_i,\bq') \ ,
	\end{equation}
where the $\bq_i$ represent all the mirror images of $\bq$ by any combination
of symmetry across a side of the square, and $\epsilon_i = +1$ or $-1$
depending on whether one needs an even or odd number of symmetries to map
$\bq$ on $\bq_i$.  $G^0(\bq,\bq';E)$ gives the above mentioned
logarithmic singularity of $G$ when $\bq' \rightarrow \bq$, but the
long range asymptotic behavior of the two-dimensional free Green function
	\be
	G^0(\bq_i,\bq') \simeq
		\frac{1}{i \hbar} \frac{m}{\sqrt{ 2 i \pi \hbar}}
		\frac{ \exp(ik|\bq'-\bq_i|)}{ \sqrt{\hbar k |\bq'-\bq_i|}}
	\ee
can be used for all other terms (images).

For sufficiently weak magnetic fields,  one may follow the same
approach as in the previous section, keeping in Eq.~(\ref{eq:green}) the
zero'th order approximation for the prefactor $D_\bM$,
and using the first-order correction $\delta S$ to the action which, as
expressed by Eq.~(\ref{dS}), is proportional to the area enclosed
by the unperturbed trajectory.  Here however, as is the generic case (and
contrary to circular or annular geometries) the area enclosed by an orbit
varies within a family.

Let us consider the contribution to the density of states
of the family of recurrent trajectories which for $H\! \rightarrow \!0$ tends
to the family of shortest periodic orbits with non-zero enclosed area,
that plays a crucial role in determining the magnetic response,
as already recognized in Ref.~\cite{levy93}.
For $H\!=\!0$, this family consists in the set of orbits which, say, start
with an angle of 45 degrees with respect to the boundary on
the bottom side of the billiard at a distance $x_0$ ($0 \leq x_0 \leq a$)
from its left corner, bounce once on each side of the square before returning
to their initial position (family $\bM = (1,1)$, see Fig.~\ref{fig:fam11}(a)).
It is convenient to use as configuration space coordinates $x_0$ which labels the
trajectory, the distance $s$ along the trajectory, and the index
$\epsilon = \pm 1$ which specifies the direction in which the trajectory is
traversed. In this way, each point $\bq$ is counted four times
corresponding to the four sheets of the invariant torus.
The enclosed area $\A_{\epsilon}(x_0,s)$ obviously does not depend on $s$ and is
given by
	\be  \label{area}
	\A_{\epsilon}(x_0) = \epsilon \ 2 \ x_0 \ (a-x_0) \; .
	\ee
Periodic orbits are those paths for which the action is extremal
(${\bf \nabla} S = {\bf p'} - {\bf p} = 0$).  Therefore Eqs.~(\ref{dS}) 
and (\ref{area}) illustrate the contents of the Poincar\'e-Birkhoff
theorem, that for any non-zero field only the two trajectories
corresponding to $x_0 = a/2$ remain periodic (one stable, one unstable
according to the two possible directions of traversal).
The contribution of the family (1,1) to $\dosc(E)$ is $d_{11}(E) = - (\gs/\pi)
\ {\rm Im} \ \G_{11}(E)$. Inserting Eqs.~(\ref{area}) and
(\ref{dS}) into the integral of Eq.~(\ref{eq:traceG}) we have
	\be \label{dg11}
	\G_{11}(H) =
	    \frac{1}{i \hbar} \frac{1}{\sqrt{ 2 i \pi \hbar}}
            \int_0^{L_{11}} d s \left(\frac{ d y}{d s} \right)
	    \int_0^{a} d x_0   \sum_{\epsilon=\pm 1} D_{11}
	    \exp{\left[i k L_{11} + i \frac{2e\epsilon}{\hbar c}
            H x_0(a-x_0) \right]} \; .
	\ee
The contribution to the density of states of the family (1,1) factorizes
into an unperturbed (Berry-Tabor-like) term and a field dependent
factor
	\be \label{dosc11}
	    d_{11}(E,H) = d^0_{11}(E) \ {\cal C}(H)
	\ee
with
	\be \label{dosc11nf}
	   d^0_{11} \equiv d_{11}(H\!=\!0) =
	   \frac{4 \gs}{\pi} \ \frac{ m a^2}{\hbar^2 (2\pi k L_{11})^{1/2}}
	    \ \sin{\left(kL_{11}\!+\!\frac{\pi}{4}\right)} \ ,
	\ee
and
	\begin{equation}
	\label{Csimple}
	{\cal C}(H) = \frac{1}{a}
	  \int_0^{a} {\rm d} x_0  \cos \left( \frac{2e}{\hbar c} H x_0 (a-x_0)
	  \right)
 	=
	  \frac{1}{\sqrt{2 \varphi}}
	  \left[ \cos(\pi \varphi) {\text C}(\sqrt{\pi \varphi}) +
	         \sin(\pi \varphi) {\text S}(\sqrt{\pi \varphi}) \right]
    	\; .
	\end{equation}
$\text C$ and $\text S$ respectively denote the cosine and sine
Fresnel integrals \cite{Gradshtein}, and
\be \label{eq:varphi}
        \varphi = \frac{Ha^2}{\Phi_0}  
\ee
is the total flux through the square measured in units of the flux quantum
($\Phi_0 = hc/e$).
For the circular and annular geometries, the field dependence
of the density of states, and therefore the susceptibility, was related to
the dephasing between time reversal families of orbits.  Here,
Eq.~(\ref{Csimple}) expresses that the dependence of $\dosc$ on the 
field is also
determined by the field induced decoherence of different orbits
{\em within} a given family.

As soon as $\varphi$ reaches a value close to one, the Fresnel
integrals can be replaced by their asymptotic value $1/2$,
which amounts to evaluate $\C(\varphi)$ by stationary phase, i.e.
\be \label{eq:statphas}
{\cal C}^{\rm S} (\varphi) = \frac{\cos(\pi \varphi \! - \! \pi/4)}{
                              \sqrt{4\varphi}} \; .
\ee
This means that for $\varphi>1$ the dominant contribution
to ${\cal C}(\varphi)$ comes from the neighborhood
of the two surviving periodic orbits ($x_0 = a/2, \epsilon = \pm 1$),
and the oscillations of  ${\cal C}(\varphi)$ are related to the successive
dephasing and rephasing of these orbits. In fact, one would have obtained  just
$d_{11}^{\rm S} = d_{11}^0 {\cal C}^{\rm S} (\varphi)$ by evaluating the
contribution to the density of states of the two surviving periodic orbits
using the Gutzwiller trace formula
with a first-order classical perturbative evaluation of
the actions and stability matrices.
${\cal C}^{\rm S} (\varphi)$ however diverges when $H \! \rightarrow \! 0$,
while the full expression Eq.~(\ref{Csimple}) simply gives ${\cal C}(0)=1$.

To compute the contribution $d_\bM$ of longer trajectories, it
is worthwhile to write $(M_x, M_y)$ as $(r u_x, r u_y)$, where $u_x$ and $u_y$
are coprime integers labeling the primitive orbits and $r$ is the number
of repetitions.  As illustrated in Fig.~\ref{fig:fam11}(b), for any orbit
of the family the square can be decomposed into $u_x \times u_y$ cells, such
that the algebraic area enclosed by the trajectory inside two adjacent
cells exactly compensate.  Therefore, keeping $x_0$ as a label of the
orbit (with  $x_0 \in [0, a/u_x]$ to avoid double counting),
the total area enclosed by the trajectory $(ru_x,ru_y)$ is
	\begin{equation} \label{eq:Anxny}
	\A_\bM = \left\{ \ba{cl} \displaystyle
		0  & \qquad \qquad \mbox{$u_x$ or $u_y$ even}\\
	        \displaystyle r \frac{\A_{\epsilon}(u_x x_0)}{u_x u_y}
		   & \qquad \qquad \mbox{$u_x$ and $u_y$ odd}\\
	\ea \right. \; ,
	\end{equation}
where $\A_{\epsilon}(x_0)$ is given by Eq.~(\ref{area}).  From the above
equation, and proceeding in the same way as for the orbit $(1,1)$
Eq.~(\ref{dosc11}) can be generalized to
  	\be \label{square:GM}
	d_\bM(E,H) =
	\left\{ \ba{ll} \displaystyle
	    d^0_\bM(E) \ & \qquad \qquad \mbox{$u_x$ or $u_y$ even}\\
		\displaystyle
	    d^0_\bM(E) \ {\cal C}\left(\frac{r\varphi}{u_xu_y}\right)
		   & \qquad \qquad \mbox{$u_x$ and $u_y$ odd}\\
	\ea \right. \; ,
	\end{equation}
where $\C(\varphi)$ is given by Eq.~(\ref{Csimple}) and
$d^0_\bM \equiv d_\bM(H\!=\!0) $ is the zero-field contribution of the 
family $\bM$
	\be \label{square:G0M}
	    d^0_\bM =
	    \frac{4 \gs}{\pi} \ \frac{ m a^2}{\hbar^2 (2\pi k L_\bM)^{1/2}}
	     \ \sin{\left(kL_\bM\!+\!\frac{\pi}{4}\right)} \ .
	\ee

\subsection{The susceptibility: individual samples vs. ensemble averages}
\label{sec:chisemicl}

For clarity of the presentation we will calculate in a first stage the
susceptibility contribution of the family (1,1) of the shortest flux enclosing
orbits only. This corresponds to the temperature regime of the experiment
Ref.~\cite{levy93} where the characteristic
length $L_c$ given by Eq.~(\ref{R_factor2}) is of the order of
$L_{11}$, the length
of the shortest orbits, and contributions of all longer orbits are eliminated
due to temperature damping. In the next subsection we will state the results
valid at arbitrary temperature by taking into account the contribution of 
longer orbits.

>From the expressions (\ref{dosc11}) and (\ref{dosc11nf}) of the
contributions of the family $(1,1)$ to $\dosc(E,H)$ one obtains
the corresponding contribution to $\Delta F^{(1)}$
(Eqs.~(\ref{eq:smooth_osco}) and (\ref{DF1})) as
	\begin{equation}
	\label{square:F11}
	\Delta F^{(1)}_{11}(H) = \frac{\gs\hbar^2}{m}
	 \left( \frac{2^3 a}{\pi^3 L^5_{11}} \right)^{1/2} (\kf a)^{3/2} \
	\sin{\left(\kf L_{11}\!+\!\frac{\pi}{4}\right)} \
	{\cal C}(H) R_T(L_{11}) \; .
	\end{equation}
$R_T(L_{11})$ is the temperature dependent
reduction factor Eq.~(\ref{R_factor2}), valid for billiard systems.
The field-dependent factor ${\cal C}(\varphi)$ is given by
Eq.~(\ref{Csimple}).
Taking the derivatives with respect to the
magnetic field, we have [for $L_c \simeq L_{11}$]
	\be \label{square:chi1}
	\frac{\chi^{(1)}}{\cl}  = -\frac{3}{(\sqrt{2}\pi)^{5/2}} \
	(\kf a)^{3/2} \ \sin{\left(\kf L_{11}+
	\frac{\pi}{4}\right)}\frac{{\rm d}^2 \ {\cal C}}{{\rm d} \varphi^2}
	\ R_T(L_{11}) \ .
	\ee
The susceptibility of a given square oscillates as a function
of the Fermi energy and can be paramagnetic or
diamagnetic (see Fig.~\ref{fig:chi1}(a)). 
Since we are considering only one kind of trajectory the  typical 
susceptibility $\chi^{({\rm t})}$ (with the definition (\ref{chicir1t}))
is simply proportional to the prefactor of $\chi^{(1)}$. Therefore, it
is of the order of $(k_F a)^{3/2}$, which is much larger than
the Landau susceptibility $\chi_{\scriptscriptstyle L}$.
As shown in Fig.~\ref{fig:chi1}b (solid line) 
$\chi^{(1)}$ exhibits also (by means of
$\partial^2 {\cal C}/\partial \varphi^2$) oscillations as a function of the
flux at a given number of electrons in the square.
The divergent susceptibility
obtained from ${\cal C}^{\rm S}$ (dashed line) provides a good
description of $\chi^{(1)}$ for $\varphi \stackrel{>}{\sim} 1$.

For a measurement made on an ensemble of squares of different sizes $a$,
$\chi^{(1)}$ vanishes under averaging if the dispersion of
$k_{\scriptscriptstyle F} L_{11}$
across the ensemble is larger than $2\pi$.  In that case
the average susceptibility is given by the contribution to $\Delta F^{(2)}$
arising from the $(1,1)$ family (Eq.~(\ref{DF2})).
Proceeding in a similar way as for the first--order
term, the contribution of the family $(1,1)$ to the integrated density 
$\Nosc$ is given by Eq.~(\ref{eq:smooth_oscn}) as
	\begin{equation} \label{square:N11}
	N_{11}(\bar \mu,H) = - \gs
	 \left( \frac{2^3 a^3}{\pi^3 L^3_{11}} \right)^{1/2} 
	(\kf a)^{1/2} \
	\cos{\left(\kf L_{11}\!+\!\frac{\pi}{4}\right)} \
	{\cal C}(H) R_T(L_{11}) \; .
	\end{equation}
To calculate $\chi^{(2)}$ we have to consider
$\Delta F^{(2)} = (\Nosc)^2/2\bar{D}$ (with 
$\bar D = (\gs m a^2)/(2\pi\hbar^2)$), and in particular the term
	\begin{equation}
	\frac{(N_{11}(\bar \mu,H))^2}{2\bar D}  =
	\frac{\gs\hbar^2}{(\sqrt{2})^3 \pi^2 m a^2} \ \kf a
	\  \cos^{2}{\left(\kf L_{11}+\frac{\pi}{4}\right)}
	\ {\cal C}^2 (\varphi) \ R^2_T(L_{11}) \ .
	\end{equation}
This contribution is of lower order in $\kf a$ than that of
$\Delta F_{11}^{(1)}$,
but its sign does not change as a function of the phase $\kf L_{11}$.
Therefore the squared cosine survives the ensemble average\footnote{
Beside the orbits (1,1) the orbits (1,0) and (0,1) which are even shorter
contribute to $\Delta F^{(1)}$ in the limit $L_c \sim L_{11}$. Since
they do not enclose any flux the second derivative of $\Delta F_{10}^{(1)}$
with respect to $H$, i.e. $\chi_{10}^{(1)}$ can be neglected for small
fields. However, they enter into $\chi^{(2)}$ by means of the cross 
products
$(N_{10}+N_{01})N_{11}$ in $(\Nosc)^2$. Nevertheless, they play no role for
the averaged $\overline{\chi^{(2)}}$ because $N_{10}$ and $N_{11}$
do not oscillate with the same frequency and therefore their product
averages out.}
and we obtain, performing the derivatives with respect to $\varphi$
(still in the regime $L_c \simeq L_{11}$),
	\begin{equation} \label{square:chi}
	\frac{\overline{\chi^{(2)}}}{\chi_{\scriptscriptstyle L}}  =
	- \frac{3}{(\sqrt{2}\pi)^3} \ k_{\scriptscriptstyle F} a
	\ \frac{{\rm d}^2 {\cal C}^2}{{\rm d} \varphi^2} \ R_T^2(L_{11})
	\; .\end{equation}
The total averaged susceptibility is therefore
	\[ \overline{\chi}  = - \chi_{\scriptscriptstyle L}
		+ \overline{\chi^{(2)}} \; ,\]
since, as seen in section~\ref{sec:LanGen}, one has also to include
the diamagnetic (bulk) ``Landau term'' $-\chi_{\scriptscriptstyle L}$ 
arising
from $\hbar$ corrections to $F^0$.  In the regime $L \simeq L_c$ we are
considering here, $\chi_{\scriptscriptstyle L}$ is negligible 
with respect to $\overline{\chi^{(2)}}$ as $\hbar \rightarrow 0$,
and one can use $\overline{\chi} \simeq \overline{\chi^{(2)}}$.
Note however that when $L_c \ll L$, Eqs.~(\ref{square:chi1}) and
(\ref{square:chi}) remain valid but $\chi^{(1)}$ as well
as $\chi^{(2)}$ is exponentially suppressed.  In this ``trivial'' regime
$\chi$ (and thus $\overline{\chi}$) reduces to the Landau susceptibility,
and becomes independent of the underlying classical dynamics. The linear
dependence of the average susceptibility on $\kf$ is shown in Fig.~\ref{fig:chi2}a.

Since ${\cal C}$ has its absolute maximum at $\varphi\!=\!0$, the average
zero-field susceptibility is paramagnetic and attains a maximum value
of \cite{URJ95,vO95}
	\begin{equation} \label{square:chizf}
	\overline{\chi^{(2)}}({H\!=\!0}) =
	\frac{4\sqrt{2}}{5\pi} \ \kf a \ \chi_{\scriptscriptstyle L} \ R^2_T(L_{11}) \; .
	\end{equation}
For small fields the average susceptibility (thin solid line,
 Fig.~\ref{fig:chi2}b)
has an overall decay as $1/\varphi$ and oscillates in sign on the scale
of one flux quantum through the sample. As in the disordered case \cite{BM}
the period of the field oscillations of the average is half of that of the
individual systems (see Fig.~\ref{fig:chi1}(b)).
In our case the difference can be traced to the
$ \C^2$ dependence that appears in Eq.~(\ref{square:chi}) in contrast to
the simple $\C$ dependence of Eq.~(\ref{square:chi1}). 

For an ensemble with a wide distribution of lengths 
(as in Ref.~\cite{levy93}) an average $\langle \cdots \rangle$ 
on a classical scale (i.e.~$\Delta a /a \not \ll 1$) rather 
than on a quantum scale ($\Delta (\kf a) \simeq 2\pi$) needs 
to be performed, and the dependence of ${\cal C}$
on $a$ (through $\varphi$) has to be considered. Since the scale 
of variation of ${\cal C}$ with $a$ is much slower than that of
$\sin^2{(k_{\scriptscriptstyle F}L_{11})}$
we can effectively separate the two averages and obtain the total
mean by averaging the local mean:

     \be \label{clasav}
        \langle \chi \rangle =
            \int d a  \ \overline{\chi} \ P(a) \; ,
        \ee

\nin where the quantum average $\overline{\chi}$ is given by Eq.~(\ref{square:chi}) and
$P(a)$ is the probability distribution of sizes $a$. Taking for $P(a)$ a Gaussian
distribution with a $30\%$ dispersion we obtain the thick solid line of Fig.~\ref{fig:chi2}b.
The low-field oscillations with respect to
$\varphi$ are suppressed under the second average, while the zero-field
behavior remains unchanged.

The expected value for the susceptibility measured in an ensemble of $n$ squares is
$n\langle \chi \rangle \propto n \kf a$, with a {\em large} statistical dispersion of 
$\sqrt{n} \chi^{\rm (t)} \propto \sqrt{n} (\kf a)^{3/2}$. However, for experiments 
like the one of Ref.~\cite{levy93} where $n \simeq 10^5 \gg \kf a \simeq 10^2$, it
is not possible to obtain a diamagnetic response by a statistical fluctuation.

\subsection{Contribution of longer orbits}
\label{sec:longorbit}

In the zero temperature limit\footnote{It should be kept in mind however
that the expansion in Eq.~(\ref{eq:fd}) is a priori not valid when
$T\rightarrow 0$.} or more generally if one is interested in results
valid at any temperature, it is
necessary to take also into account the contribution of longer trajectories.
This can be done following exactly the same lines as for the contribution
of the family (1,1).  From Eqs.~(\ref{square:GM}) and (\ref{square:G0M})
one obtains the contribution of the family
$\bM = (M_x,M_y) = (r u_x, r u_y)$, (where $ u_x$ and $u_y$ are coprime)
to $\Delta F^{(1)}$
	\begin{equation}
	\label{square:FM}
	\Delta F^{(1)}_{\bM}(H) = \frac{\gs\hbar^2}{m}
	 \left( \frac{2^3 a}{\pi^3 L^5_{\bM}} \right)^{1/2} (\kf a)^{3/2} \
	\sin{\left(\kf L_{\bM}\!+\!\frac{\pi}{4}\right)} \
	{\cal C}_\bM(\varphi) R_T(L_{\bM}) \; ,
	\end{equation}
where
  	\be \label{square:CbM}
	{\cal C}_\bM(\varphi) =
	\left\{ \ba{ll} \displaystyle
	    1 \ & \qquad \qquad \mbox{$u_x$ or $u_y$ even}\\
		\displaystyle
	    {\cal C}\left(\frac{r\varphi}{u_xu_y}\right)
		   & \qquad \qquad \mbox{$u_x$ and $u_y$ odd}\\
	\ea \right. \; .
	\end{equation}
$L_\bM$ and the function ${\cal C} (\varphi)$ are given respectively by
Eqs.~(\ref{square:length}) and (\ref{Csimple}). In order to get
$\chi^{(1)}$ we have to take the second derivative of ${\cal C}_\bM$
with respect to the magnetic field. This yields zero if either $u_x$ 
or $u_y$ is even and a factor $r^2/(u_x u_y)^2$, if both are odd.  
We therefore obtain
	\be \label{square:chi1_tot}
	\frac{\chi^{(1)}}{\cl}  = -\frac{3}{\pi^{5/2}} \ (\kf a)^{3/2}
	\sum_r \! \! \sum_{\ba{c} \scriptstyle u_x, \; u_y \\
			    \scriptstyle {\rm odd} \ea}
	\frac{1}{r^{1/2}(u_x^2+u_y^2)^{5/4} (u_x u_y)^2}
	\sin{\left(\kf L_{\bM}+\frac{\pi}{4}\right)}
	\ {\cal C}'' \! \left(\frac{r \varphi}{u_x u_y} \right)
	\ R_T(L_{\bM}) \ ,
	\ee
valid at any temperature.

The low temperature result for $\chi^{(2)}$ follows in essentially the same
way, but taking the average is made rather intricate in the case of a square
(as compared for instance to a rectangle) because of the degeneracies in
the lengths of the particular orbits of this system.  
Indeed, there are infinitely many integers
which can be decomposed in at least two different ways into sums of two squares.
For instance, $ 11^2 + 7^2 = 13^2 + 1^2 = 170$.  As a consequence,
$L_{11,7} = L_{13,1}$, and
$ \overline{ N_{11,7} N_{13,1} } \neq 0$. An explicit formula
for $\overline{\chi^{(2)}}$ therefore
requires to handle correctly all the non--diagonal terms containing orbits of degenerated
lengths which do not average to zero.
This leads to a number theoretical problem (i.e.\ characterizing all numbers
which decomposition as the sum of two squares is not unique), with which we
do not deal and which moreover will be seen
to be of no practical relevance.   Therefore, instead of considering
a square, we will give the expression for $\overline{\chi^{(2)}}$ for
a rectangle of area $a^2$ and of horizontal and vertical lengths
$a\cdot e$ and $a\cdot e^{-1}$.  In that case,
all the formulae given in section~\ref{seq:square1} remain valid.
As the only difference one has now
	\[ L_\bM = 2 a \sqrt{(M_x/e)^2 + (M_y e)^2} \]
instead of Eq.~(\ref{square:length}), which does not give rise to length degeneracies
if, as we will suppose, $e^4$ is irrational.
Noting that the prefactor of
$ N_\bM^2$ depends as $L_\bM^{-3}$ on the length of the orbit
(instead of $L_\bM^{-5/2}$ for $\Delta F^{(1)}_\bM$),
one obtains for the canonical correction to the susceptibility
	\begin{equation} \label{square:chi_tot}
	\frac{\overline{\chi^{(2)}}}{\cl}  =
	- \frac{3}{\pi^3} \ k_{\scriptscriptstyle F} a
	\sum_r \! \! \sum_{\ba{c} \scriptstyle u_x, u_y \\
			    \scriptstyle {\rm odd} \ea}
	\frac{1}{r\left((u_x/e)^2+(u_y e)^2\right)^{3/2} (u_x u_y)^2}
	\ \  ({\cal C}^2)'' \! \left(\frac{r\varphi}{u_x u_y}\right)
	\ R_T^2(L_{\bM}) \; .
	\end{equation}

The equations (\ref{square:chi1_tot}) and (\ref{square:chi_tot}) show that
even at zero temperature the strong
flux cancelation typical for the square (or rectangular) geometry
generates a very small prefactor
$1 / (r^{1/2} (u_x^2+u_y^2)^{5/4} (u_x u_y)^2)$ for $\chi^{(1)}$ (square
geometry) and $ 1 / (r \left((u_x/e)^2+(u_y e)^2\right)^{3/2} (u_x u_y)^2 )$
for  $\chi^{(2)}$ (rectangular geometry).  For the second shortest contributing
primitive orbit, $\bM = (1,3)$, this yields for instance for
$\chi^{(1)}$ a damping of $1/(9 \times 10^{5/4}) \simeq 0.0062$.
For  $\overline{\chi^{(2)}}$ the multiplicative factor is even smaller.
In practice only the repetitions $(r,r)$ of the family (1,1)
will contribute significantly to the susceptibility, and
one can use Eqs.~(\ref{square:chi1_tot}) and (\ref{square:chi_tot})
keeping only the term $u_x = u_y = 1$ of the second summation.
As a consequence, all the complications due to the
degeneracies in the length of the orbits for the square are
of no practical importance (Eq.~(\ref{square:chi_tot}) restricted
to $u_x = u_y = 1$ can be used for the square with $e=1$),
showing why their detailed treatment was not necessary.
As illustrated in Fig.~\ref{fig:repetition} for $\overline{\chi^{(2)}}$,
the repetitions of the orbit (1,1) are yielding a diverging susceptibility
at zero field when the temperature goes to zero, but barely affect the
result even as $T \rightarrow 0$ for finite $H$, where the contributions of the
repetitions do no longer add coherently.

\subsection{Numerical calculations}
\label{sec:numeric}

As a check of our semiclassical results we calculated quantum mechanically 
the orbital susceptibility of spinless particles in
a square potential well $[-a/2,a/2]$ in an homogeneous magnetic
field. Within the symmetric gauge ${\bf A} = H (-y/2, x/2, 0)$ the
corresponding
Hamiltonian in scaled units $\tilde x=x/a$ and
$\tilde E=(m a^2/\hbar^2) E$ reads
	\begin{equation}
	\tilde {\cal H} = -\frac{1}{2} 
	  \left(\frac{\partial^2}{\partial \tilde x^2}
          + \frac{\partial^2}{\partial \tilde y^2}\right)
	      - i\pi \, \varphi 
	\left(\tilde y \frac{\partial}{\partial \tilde x}
	       - \tilde x \frac{\partial}{\partial \tilde y}\right)
          + \frac{\pi^2}{2} \varphi^2 ({\tilde x}^2 + {\tilde y}^2) \ ,
\label{eq:hamilton}
\end{equation}
with the normalized flux $\varphi$ defined as in Eq.~(\ref{eq:varphi}).
Taking into account the invariance of the Hamiltonian (\ref{eq:hamilton})
with respect to rotations by $\pi, \pi/2$ we use linear
combinations of plane-waves which 
are eigenfunctions of the parity operators
${\bf P}_\pi$, ${\bf P}_{\pi/2}$, respectively. Omitting the tilde in 
order to simplify the notation, they read
\begin{eqnarray}
 & & \sqrt{2} [S_n(x) C_m(y) \pm i C_m(x) S_n(y)]
     \hspace{5mm} ; \hspace{5mm} (P_\pi = -1)    \; , \\
 & & \begin{array}{l}
         \sqrt{2} [C_n(x) C_m(y) \pm  C_m(x) C_n(y)]  \\
                   \sqrt{2}i\, [S_n(x) S_m(y) \pm  S_m(x) S_n(y) ]
     \end{array}
     \hspace{5mm} ; \hspace{5mm} (P_\pi = +1)
\end{eqnarray}
with $S_n(u) = \sin(n \pi u )$, $n$ even, and $C_m(u) = \cos(m \pi u )$,
$m$ odd, obeying Dirichlet boundary conditions. In this representation
the resulting matrix equation
is real symmetric and decomposes into four blocks representing the
different symmetry classes. By diagonalization we calculated the first
3000 eigenenergies taking into account up to 2500 basis functions
for each symmetry class.  A typical energy level diagram 
of the symmetry class $(P_\pi,P_{\pi/2})=(1,1)$ as a function of 
the magnetic field is shown in Fig.~\ref{f1}(a). In between the two
separable limiting cases $\varphi = 0$ and $\varphi \longrightarrow \infty$
the spectrum exhibits a complex structure typical for a non--integrable
system which classical dynamics is at least partly chaotic.

We calculate numerically the {\em grand-canonical} susceptibility 
(Eq.~(\ref{eq:susgc}), Fig.~\ref{f1}b) from
\begin{equation}
\cgc(\mu) = -\frac{\gs}{a^2}\frac{\partial^2}{\partial H^2}\,
\sum_{i=1}^\infty \, \frac{\epsilon_i }{1+\exp[\beta(\epsilon_i-\mu)]}
\end{equation}
where $\gs$ accounts for the spin degeneracy and $\epsilon_i$ denotes the
single particle energies.

However, in order to address the experiment of Ref.~\cite{levy93} and
to compare with the semiclassical approach of the preceeding subsection
we have to work in the canonical ensemble.
At $T=0$ the free energy $F$ reduces to the total energy and the
canonical susceptibility (Eq.~(\ref{eq:sus})) is
given as the sum
	\begin{equation}
	\chi(\!T=\!0) = -\frac{\gs}{a^2} \sum_{i=1}^N \,
	\frac{\partial^2\, \epsilon_i  }{\partial H^2}
	\end{equation}
over the curvatures of the $N$ single particle energies
$\epsilon_i$. The susceptibility is therefore dominated by large
paramagnetic singularities each time the highest occupied state undergoes a
level crossing with a state of a different symmetry class or a narrow avoided
crossing with a state of the same symmetry.
This makes $T=0$ susceptibility spectra of quasi--integrable billiards (with
nearly exact level crossings) or systems with spectra composed of energy levels
from different symmetry classes (as it is the case for the square)
looking much more erratic than those of chaotic systems with stronger level
repulsion \cite{NakTho88}.

The $T=0$ peaks are compensated
once the next higher state at a (quasi) crossing is considered, and therefore
disappear at finite temperature when the occupation of nearly degenerated
states becomes almost the same. Thus finite temperature regularizes the
singular behavior of $\chi$ at $T=0$ and of course describes
the physical situation. We obtain the canonical susceptibility at finite $T$
from
\begin{equation}
\chi = \frac{\gs}{a^2 \beta}\,\frac{\partial^2}{\partial H^2} \ln Z_N(\beta) \ .
\label{chiqm}
\end{equation}
The canonical partition function $Z_N(\beta)$ is given by
\begin{equation}
Z_N(\beta) = \sum_{\{\alpha\}} \; \exp[-\beta E_\alpha(N)]
\label{Zcanonical}
\end{equation}
with
\begin{equation}
E_\alpha(N)  = \sum_{i=1}^\infty \, \epsilon_i \; n_i^\alpha \hspace{5mm} ,
\hspace{5mm} N =  \sum_{i=1}^\infty \; n_i^\alpha \; .
\label{Ealpha}
\end{equation}
The $n_i^\alpha \in \{0,1\}$ describe the occupation
of the single particle energy-levels. A direct numerical computation of
the canonical partition function becomes extremely
time consuming at finite temperature.  We approximate the sum in
Eq.~(\ref{Zcanonical}) which runs over all (infinitely many)
occupation distributions $\{\alpha\}$ for $N$
electrons by a finite sum $Z_N(M;\beta)$ over all possibilities to distribute
$N$ particles over the first $M$ levels with $M \ge N$ sufficiently large.
Following Brack et al.~\cite{bra91} we calculate $Z_N(M;\beta)$ recursively
using
	\begin{equation}
	Z_N(M;\beta) = Z_N(M-1;\beta) +
		Z_{N-1}(M-1;\beta) \exp(-\beta \epsilon_M)
	\label{recrelation}
	\end{equation}
with initial conditions
	\begin{equation}
	Z_0(M;\beta) \equiv 1 \hspace{3mm} , \hspace{3mm} Z_N(N-1;\beta)
	 \equiv  0
	\end{equation}
and increase $M$ until convergence of $Z_N(M,\beta)$, i.e.
the difference between $Z_N(M;\beta)$ and $Z_N(M-1;\beta)$ is negligible.
This recursive algorithm reduces the number of algebraic operations to
calculate $Z_N$ drastically and is fast and accurate even if $\kb T$ is of
the order of 10 or 20 times the mean level spacing, i.e., in a regime 
where a direct calculation of $Z_N$ is not feasible.

\subsection{Comparison between numerical and semiclassical results}
\label{sec:num}

Our numerical results for the susceptibility of individual and ensembles
of squares are displayed as the dashed lines in Figs.~\ref{fig:chi1} and
\ref{fig:chi2} and are in excellent agreement with the semiclassical
predictions of Sec.~\ref{sec:chisemicl}. Fig.~\ref{fig:chi1}a shows
the numerical result for the canonical susceptibility and the
semiclassical leading order contribution $\chi^{(1)}_{11}$ at zero field
as a function of $k_F a$ ($\sqrt{4\pi N/\gs}$ in terms of the number of
electrons). The temperature $\kb T$ is equal to five times the mean level
spacing $\Delta$ of the single particle spectrum. The quantum result oscillates
with a period $\pi/\sqrt{2}$ as semiclassically expected (Eq.~(\ref{square:chi1}))
indicating the dominant effect of the fundamental orbits of length $L_{11} =
2\sqrt{2}a$. The semiclassical amplitudes (solid line)
are slightly smaller than the numerics because only the shortest orbits
are included.

Fig.~\ref{fig:chi1}b shows the flux dependence of $\chi$ for a fixed number of electrons
$N \approx 1100 \gs$. 
The semiclassical prediction (Eq.~(\ref{square:chi1}), solid curve)
is again in considerable agreement with the quantum result while the
analytical result (Eq.~(\ref{eq:statphas}), dashed line)
from stationary phase integration yields an (unphysical)
divergence for $\varphi \rightarrow 0$ as discussed in Sec.~\ref{sec:chisemicl}.

For the numerical calculations we can perform the ensemble average directly and 
we obtain the averages on the quantum scale (thin dashed line, Fig.~\ref{fig:chi2}b) 
or classical scale (thick dashed line) by taking a Gaussian distribution 
of sizes with respectively a small or large $\Delta a/a$ dispersion.
Fig.~\ref{fig:chi2}(a) depicts the $k_F a$ dependence of
$\overline{ \chi }$ assuming a Gaussian distribution of lengths $a$ with a 
standard deviation $\Delta a/a \approx 0.1$ for each of the
three temperatures $\kb T/\Delta = 2,3,5$. The dashed curves are the ensemble
averages of the quantum mechanically calculated {\em entire} canonical
susceptibility $\overline{ \chi}$. The dotted lines are the
{\em exact} (numerical) results for the averaged term  $\overline{
\chi^{(2)}_{\rm qm}} = (\overline{ \Nosc_{\rm qm})^2}/2\Delta $.
They are nearly indistinguishable (on the scale of the figure) from  
the {\em semiclassical} approximation of Eq.~(\ref{square:chi}) (solid line).
Although a small flux $\varphi \approx 0.15$
has been chosen (here the contribution from the next longer orbits
$(2,2)$ nearly vanishes) the precision of the semiclassical approximation
based on the fundamental orbits (1,1) is striking. The difference between
the results for $\overline{ \chi }$ and
$\overline{ \chi^{(2)} }$ gives an estimate
for the precision of the thermodynamic expansion Eq.~(\ref{eq:fd}).
The convoluted semiclassical result has been shifted
additionally by $-\cl$ to account for the diamagnetic Landau
contribution and is again in close
agreement with the numerical result of the averaged susceptibility
$\bar{ \chi }$. 

\subsection {Comparison with the experiment}
\label{sec:experiment}

In a recent experiment, L\'evy {\em et al.} \cite{levy93} measured
the magnetic response of an {\em ensemble} of $10^5$ microscopic
billiards of square geometry lithographically defined on a high mobility
GaAs heterojunction. The size of the squares is on average $a=4.5 \mu m$,
but has a large variation (estimated between 10 and 30\%) between the center
and the border of the array. Each square can be considered as phase-coherent
and ballistic since the phase-coherence length and elastic mean free path are 
estimated, respectively, to be between 15 and 40 $\mu m$ and between
5 and 10 $\mu m$.

Therefore, it is worthwhile to compare the observed magnetic response
with the prediction of our clean model of non-interacting electrons,
to see whether this simple picture contains  the main physical input
to understand the experimental observations, although one should control in addition
that the residual impurities do not alter fundamentally the magnetic response
of the system. This is the subject of a forthcoming article \cite{rod2000}. Ongoing
calculations including (weak) disorder indeed indicate that the underlying
physical picture remains correct.

At a qualitative level, a large paramagnetic peak at zero field
has been observed in Ref.~\cite{levy93}, two orders of magnitude larger 
than the Landau susceptibility, decreasing on a scale of approximately 
one flux quantum through each square. Since there is a large dispersion
of sizes we do not observe the field oscillations of the quantum average
(\ref{square:chi}), but the comparison has to be established with the 
classical average results Eq.~(\ref{clasav}). The corresponding
results from our semiclassical calculations 
(Eq.~(\ref{square:chi},\ref{clasav}))
and the full quantum calculations are shown in Fig.~\ref{fig:chi2}b) as
the thick full, respectively dashed, lines (denoted by $\langle \chi\rangle$ in the
figure). The offset in the semiclassical curve with respect to the
quantum mechanical curve is due to the Landau
susceptibility $\cl$ and additional effects from bouncing--ball orbits
(see section \ref{sec:highB} A) not included in the semiclassical trace.
Our theoretical results for the flux dependence of the average $\langle \chi \rangle$
with respect to a wide distribution in the size of the squares 
agree on the whole with the experiment. However, the
diamagnetic response for $\langle \chi\rangle$ 
that we obtain for $\varphi \approx 0.5$ is not
observed experimentally, indicating that there may be a more important 
size-dispersion than estimated.  As will be discussed in more detail in 
section~\ref{sec:general}, a very large distribution of lengths
enhances the effect of the breaking of time reversal invariance due
to the magnetic field, yielding a vanishing average response at 
{\em finite field} and a paramagnetic susceptibility at {\em zero
field} decaying on a field scale $\Phi_0$
by the dephasing of the contribution of time reversal symmetric orbits
to the density of states. 

More quantitatively, the experiment of Ref.~\cite{levy93}
yielded a paramagnetic susceptibility at $H\!=\!0$ with a
value of approximately 100 (with an uncertainty of a factor of 4) 
in units of $\chi_{\scriptscriptstyle L}$. The two electron densities 
considered in the experiment are $10^{11}$ and 
$3 \! \times \! 10^{11} {\rm cm}^{-2}$ corresponding
to approximately $10^4$ occupied levels per square. Therefore our
semiclassical approximation is well justified. For a temperature of $40mK$
the factor $4\sqrt{2}/(5\pi) k_{\scriptscriptstyle F} a R_T^2(L_{11})$ from
Eq.~(\ref{square:chizf}) gives zero field susceptibility values of 60 
and 170, respectively,  in reasonable agreement with the measurements. 
In order to attempt a more
detailed comparison with the measurements we need to incorporate
the suppression of the clean susceptibility by the residual
disorder, which depends on the strength and correlation length
of the impurity potential \cite{rod2000}.
The field scale for the decrease of 
$\langle \chi(\varphi) \rangle$ is of the order of one flux quantum
through each square, in agreement with our theoretical findings.
The temperature dependence experimentally observed seems however less
drastic than the theoretical prediction.

\newpage

\section{Generic integrable and chaotic systems}
\label{sec:general}

In sections~\ref{sec:integrable} and \ref{sec:square} we have studied
in detail specific geometries of conceptual as well as experimental relevance.
In particular, we have demonstrated the degree 
of accuracy of our semiclassical approach by a careful comparison with 
exact quantum results.  The aim of the present section is to take
a broader point of view and to give more general
semiclassical implications concerning the  magnetic properties of ballistic
quantum dots.  We shall first consider the weak--field behavior of 
generic integrable systems, generalizing the results of the previous 
section.  We focus on weak fields because only this regime
is affected by the integrability of the dynamics at zero
field. The case of systems which remain integrable at 
arbitrary field strength was discussed in section~\ref{sec:integrable}.
In the second stage we shall turn to chaotic systems
(at weak as well as finite fields) and finally finish the section by
discussing the similarity and differences of the magnetic response
for the various cases of classical stability.

\subsection{Generic integrable systems}
\label{sec:gen_int}

We consider the generic magnetic response
of two--dimensional integrable systems perturbed by a weak magnetic field
breaking the integrability.
The Eqs.~(\ref{eq:smooth_osc:all}) and (\ref{allDF}), which relate
the thermodynamic functions $\Delta F^{(1)}$ and $\Delta F^{(2)}$
to the oscillating part $\dosc(E)$ of the density of states, are general
relations which apply in particular here. The main difficulty
is therefore to obtain semiclassical uniform approximations 
for $\dosc(E)$ interpolating between
the zero field regime, for which the Berry-Tabor Formula \cite{ber76,ber77}
(suitable for integrable systems) applies, and higher fields 
(still classically perturbative however),
for which the periodic orbits which have survived under the perturbation
are sufficiently well isolated in order to use the Gutzwiller trace formula
\cite{gut71}.
This problem of computing for a generic system the oscillating part of the
density of states in the nearly but not exactly integrable regime
has been addressed by Ozorio de Almeida \cite{ozor86,ozor:book}. 
We are going to
follow this approach for the case of a perturbation by a magnetic field.
However, for the sake of completeness and in order to define their regime 
of validity, we will give a brief derivation of the basic results needed.
This is the subject of section~\ref{sec:6A1}.
In section~\ref{sec:6A2} we then deduce the grand-canonical
and canonical contributions to the susceptibility.

\subsubsection{Perturbation theory for magnetic fields}
\label{sec:6A1}

Let $\Hch (\hat {\bf p},\hat{\bf q})$ be a quantum Hamiltonian which
classical analogue can be expressed as
	\be \label{perturbed_H}
	{\cal H}({\bf p},{\bf q}) =
	{\cal H}^0\left({\bf p} \! - \! \frac{e}{c}{\bf A},{\bf q}\right) \; .
	\ee
${\cal H}^0(\bp,\bq)$ is the Hamiltonian describing the motion in
the absence of a magnetic field and ${\bf A}$ is the vector potential 
generating
a uniform magnetic field $H$.  ${\cal H}^0$ is supposed to be integrable
which permits to define action-angle coordinates $({\bf I},{\bf \varphi})$,
$\varphi_1, \varphi_2 \in [0,2\pi]$ such that at zero field
the Hamiltonian ${\cal H}^0(I_1,I_2)$ depends only on the actions.

To compute $\dosc(E)$ we start from the same basic equations as for the
square geometry.  In the weak--field regime which we are considering, the only
recurrent trajectories of the sum Eq.~(\ref{eq:green}) which
contribute noticeably to the trace Eq.~(\ref{eq:traceG}) are those which
merge into  periodic orbits of the unperturbed Hamiltonian as 
$H \rightarrow 0$.
Considering only these contributions, which we can label by the topology
$\bM$ of the unperturbed periodic orbits, and dropping the Weyl
part of the trace $\G(E)$ of the Green function we can write
	\be \label{contribution}
	\G(E) \simeq \sum_{\bsM} \G_{\bsM} \; ,
	\qquad
	\G_{\bsM}(E) = \frac{1}{i \hbar} \frac{1}{\sqrt{ 2 i \pi \hbar}}
	\int \dif q_1 \dif q_2 \, D_{\bsM}
	\exp{\left[\frac{i}{\hbar} S_{\bsM} -
                            i\eta_{\bsM}\frac{\pi}{2} \right]} \ .
	\ee

Let us now focus on the contribution $\G_{\bsM}$ of the family of
closed orbits $\bf M$.  For sufficiently low fields we will employ (as
in sections~\ref{sec:integrable} and \ref{sec:square}) 
that the change in the semiclassical 
Green function by changing  $H$ is essentially given by the
modification of the phase, $S_M/\hbar$ being large in the semiclassical
limit. The variation in the determinant $D_\bsM$ can usually be neglected.
Therefore, in the evaluation of the integral in Eq.~(\ref{contribution})
one should keep the (unperturbed) zero'th order approximation for $D_\bsM$
and evaluate the action up to the first order correction.
For the action this yields
	\be
	S_\bsM(\bq,\bq) = S^0_\bsM + \delta S_\bsM(\bq,\bq)
	\ee
with
	\be \label{S0}
	S^0_{\bsM} = \oint_{\rm orbit} {\bf p} \cdot d {\bf q}
                  = \oint_{\rm orbit} {\bf I} \cdot d {\bf \varphi}
	          = 2\pi I_{\bsM} \cdot {\bf M} \; ,
	\ee
noting $\bf I_\bsM$ the action coordinates of the periodic
orbit family $\bM$ at $H=0$.
The contribution $\delta S_\bsM$ is expressed in terms of the 
area enclosed by the
{\em unperturbed} orbit by means of Eq.~(\ref{dS}).
$S^0_\bsM$ is constant for all members of the family, but
$\delta S$ generically depends on the trajectory on which
the point $\bq$ lies.  However,
the area enclosed by the orbit and thus $\delta S_\bsM$ does not change when 
varying $\bf q$ along the orbits. It is therefore convenient to use a
coordinate system such that one coordinate is constant along
 the unperturbed trajectory.  
Writing $\bM = (r u_1,r u_2)$ where $u_1$ and $u_2$ are 
coprime integers, this is provided explicitly by the standard canonical
transformation $({\bf I},{\bf \varphi}) \rightarrow ({\bf J},{\bf \theta})$
generated by $F_2 ({\bf J},{\bf \varphi}) = 
(u_2 \varphi_1 -u_1 \varphi_2)J_1 + \varphi_2 J_2$~:
	\be  \label{canonical_transform}
	\ba{ll}
	\theta_1 = u_2 \varphi_1 - u_1 \varphi_2 \ 
	\ \ & J_1 = I_1 / u_2 \\
	\theta_2 = \varphi_2     &  J_2 = I_2 + (u_1/u_2) I_1
	\ea \; ,
	\ee
for which $\theta_1$ is constant along a trajectory {\em on the torus 
$\bf I_{\bsM}$}. Then $\theta_1$ specifies
the trajectory and $\theta_2$ the position on the trajectory.
For a square geometry, $\theta_1$ and $\theta_2$ are up to a dilatation,
respectively, the variables $x_0$ and $s$ introduced in 
section~\ref{sec:square}.
$\theta_2$ should be taken in the range $[0,2\pi u_2]$ (rather 
than $[0,2\pi]$) to ensure that the transformation
Eq.~(\ref{canonical_transform}) constitutes a one to one correspondence.

After substituting ${\bf q}$ by ${\bf \theta}$ in the
integral of Eq.~(\ref{contribution}), $\delta S$ depends only 
on $\theta_1$, but no longer on $\theta_2$.  One can moreover 
show (see appendix~\ref{app:D_M})
the following relation for the zero field approximation of the 
determinant $D_\bsM$:
	\be  \label{eq:DM_relation}
	D_\bsM \cdot
	\left| \left( \frac{\partial \bq}{\partial \theta} \right) \right|
	= \frac{1}{\dot{\theta}_2}
	  \frac{1}{ \left| 2 \pi r u_2^3 g_E^{''} \right|^{1/2}}  \; ,
	\ee
where $I_2 = g_E(I_1)$ is the function introduced in
section~\ref{sec:integrable} to describe the energy surface $E$. From
Eq.~(\ref{contribution}) and (\ref{eq:DM_relation}) one gets
	\be  \label{BT_green}
	\G_{\bsM}(E) = \frac{1}{i \hbar} \frac{1}{\sqrt{ 2 i \pi \hbar}}
	\frac{1}{\left| 2\pi r u_2^3 g_E^{''} \right|^{1/2}}
	\exp{\left[\frac{i}{\hbar} S^0_{\bsM} -
                            i \eta_{\bsM}\frac{\pi}{2} \right]}
	\int_0^{2\pi u_2}  \frac{\dif \theta_2 }{\dot{\theta}_2}
	\int_0^{2\pi} \dif \theta_1
	\exp{\left[\frac{i}{\hbar} \delta S (\theta_1) \right]} \; .
	\ee
The integral over $\theta_2$ is the period $\tau_\bsM / r$ of the primitive
periodic orbit.  In the absence of a field the integral over $\theta_1$
is simply $2\pi$ which gives
	\be \label{BT_GM}
	\G^0_\bsM(E) = -
	   \frac{i \tau_\bsM}{\hbar^{3/2} M_2^{3/2} 
	   \left| g_E^{''} \right|^{1/2}}
	   \exp i \left[ \frac{S^0_\bsM}{\hbar}
		- \eta_\bsM \frac{\pi}{2} - \frac{\pi}{4} \right] \; .
	\ee
$d^0_\bsM(E)$, the zero field contribution of the orbits of 
topology $\bM$ to the oscillating part of the density of states,
is obtained from Eq.~(\ref{BT_GM}) as 
$d^0_\bsM(E) = -(\gs/\pi) {\rm Im} \G^0_\bsM(E)$.  
Therefore, except for the evaluation of the Maslov indices that we have 
disregarded here, one recovers in this way for the integrable limit
the Berry-Tabor formula Eq.~(\ref{BTT}) of a two-dimensional 
system (as we have used in section~\ref{sec:integrable}).

Inspection of Eq.~(\ref{BT_green}) for weak magnetic fields shows that,
upon perturbation, $\G_\bsM$ is just given by the product of 
the unperturbed result $\G^0_\bsM$ and a factor
	\be \label{reduction1}
	\tilde \C_\bsM(H) = \frac{1}{2\pi} \int_0^{2\pi} \dif \theta_1 \,
       \exp \left[ 2 i \pi \frac{H \A_\bsM(\theta_1) }{ \Phi_0} \right] \;.
	\ee
This accounts for the small dephasing between different
  closed (in configuration space) orbits of topology $\bf M$ 
due to the fact that the
resonant torus on which they are living is slightly broken by the
perturbation. (An orbit of topology $\bf M$ closed in configuration
space is then generally not periodic, i.e.\ closed in phase space.)
Supposing the unperturbed motion to be time reversal invariant, 
it can be seen
moreover that only the real part of $\tilde \C_\bsM(H)$ has
 to be considered:
The function $\A_\bsM (\theta_1)$ is defined for the unperturbed
system.  Therefore, the time reversed of a trajectory labeled by $\theta_1$
is a periodic orbit of the unperturbed system which encloses an area
$-\A_\bsM(\theta_1)$.  Its contribution cancels the imaginary part of
$\exp [ 2 i \pi H \A_\bsM(\theta_1) / \Phi_0 ]$, and one can use
	\be \label{reduction2}
	\C_\bsM(H) = \frac{1}{2\pi} \int_0^{2\pi} \dif \theta_1 \,
	  \cos \left[ 2 \pi \frac{H \A_\bsM(\theta_1) }{ \Phi_0} \right] \;
	\ee
instead of $\tilde \C_\bsM(H)$. Since  $\C_\bsM(H)$ is real, one obtains
from Eq.~(\ref{eq:traceG})
	\be \label{uniform_d}
        \dosc(E) = \sum_{\bsM \neq 0} \C_\bsM(H) d^0_\bsM(E)  \; ,
	\ee
where $d^0_\bsM(E)$ is the zero--field contribution given by the 
Berry-Tabor expression of Eq.~(\ref{BTT}). At zero field we obviously 
have $\C_\bsM(0) = 1$.  At sufficiently large field, the integral
(\ref{reduction1}) can be evaluated using stationary phase
approximation.\footnote{To be precise the ratio $HA/\Phi_0$ rather 
than the  field must be large. Formally, one has to
consider not the $H\rightarrow \infty$ limit, which is incompatible
with the classical perturbation scheme, but an
$\hbar \mbox{ ({\em i.e.}~$\Phi_0$) } \rightarrow 0$ 
limiting process, which
does not change the classical mechanics. In practice this means that the 
fluxes considered are large on a quantum scale, but still small on the 
classical scale. This is achieved at high enough energies.}
$\C_\bsM$ can be expressed as a sum over all extrema of
$\A_\bsM(\theta_1)$
({\em i.e.}~of $\delta S$). These are all the periodic orbits which
survive under the perturbation.  It can be seen \cite{gri95} that,
in this approximation, Eq.~(\ref{uniform_d}) yields exactly the
Gutzwiller trace formula for which the actions, periods
and stabilities of the periodic orbits are evaluated using classical
perturbation theory. Eq.~(\ref{uniform_d}) thus provides
an interpolation between the Berry-Tabor and Gutzwiller formulae.

The functions $\A_\bsM (\theta_1)$, and therefore $\C_\bsM (H)$, are system
and trajectory dependent. One can, however, gain some further understanding of
the perturbative regime by following again Ozorio de Almeida and
writing $\A_\bsM (\theta_1)$ in term of its Fourier series
	\be \label{fourier_s}
	\A_\bsM = \sum_{n=0}^\infty \A^{(n)}_\bsM 
	          \sin(n \theta_1 - \gamma^{(n)}) \ .
   	\ee
If $\A_\bsM$ is a smooth function of $\theta_1$, the coefficients 
$\A_\bsM^{(n)}$ are usually rapidly decaying functions of $n$.
For systems where one can neglect all harmonics higher than the first
one, the integral Eq.~(\ref{reduction2}) can be performed, and
it is possible to distinguish two types
of functions $\C_\bsM (H)$, depending on the symmetry properties of the
unperturbed family of orbits under time reversal.

Indeed, one may encounter
two different situations depending on whether the torus ${\bf I_\bsM}$ is
time reversal invariant (e.g.~square geometry) or 
has a distinct partner ${\bf I_\bsM}^*$ 
in phase space which is its counterpart under time reversal (e.g.~circular
geometry).  In the former case, the origin of the angles $\theta_1$ can
be chosen such that $\A_\bsM (\theta_1)$ is an antisymmetric 
function, while
in the latter case it can be in principle any real function of $\theta_1$.%
\footnote{Note in the former case $\tilde \C_\bsM = \C_\bsM$, while in the
latter $\tilde \C_\bsM \neq \C_\bsM$ but $\G_\bsM + \G_{\bsM^*}
= \G^0_\bsM \C_\bsM(H) + \G^0_{\bsM^*} \C_{\bsM^*}(H)$.}

If $I_\bsM$ is time reversal invariant, 
$\A_\bsM (-\theta_1) = - \A_\bsM (\theta_1)$ implies that 
$\A_\bsM^{(0)} = 0$ (as well as all the phases $\gamma^{(n)}$).  
In this case
	\be \label{square_like}
	\C(H) \simeq J_0(2\pi H \A_\bsM^{(1)}/\Phi_0) \; .
	\ee
It is interesting to compare the approximation of $\C(H)$ given by the
above Bessel function with the exact integral Eq.~(\ref{Csimple}) obtained
in section~\ref{sec:square} for the shortest family (${\bf M} = (1,1)$)
of the square geometry.
Noting that $\theta_1 = \epsilon \pi x_0 / a$ and using Eq.~(\ref{area}),
the Fourier coefficients $\A_{11}^{(n)}$ of $\A_{11} (\theta_1)$ 
are given by
	\begin{equation}
	\A_{11}^{(n)}  = \left\{ \begin{array}{l l} 
	\displaystyle \frac{16}{(n\pi)^3} a^2  &
		\qquad \mbox{$n$ odd} \; ,\\
              0 &  \qquad \mbox{$n$ even}  \; .
	\end{array} \right.
        \end{equation}
Keeping only the first harmonic of $\A_{11} (\theta_1)$ amounts 
to approximate
the function $\C(\varphi)$ of Eq.~(\ref{Csimple}) by
$J_0 (32 \varphi / \pi^2)$ which, as seen in
Fig.~\ref{FresnelvsBessel}, is an excellent approximation.

If the torus $\bf I_\bsM$ is not its own time reversal, 
$\A_\bsM (\theta_1)$ is not constrained to be an antisymmetric 
function, and in particular $\A_\bsM^{(0)}$ is usually non zero.  
Neglecting, as above,
all harmonics of $\A_\bsM (\theta_1)$
except the first gives
	\be \label{circle_like}
	\C_\bsM(H) =
	\cos \left( 2\pi \frac{H \A_\bsM^{(0)}}{\Phi_0} \right)
        J_0 \left( 2\pi \frac{H \A_\bsM^{(1)}}{\Phi_0} \right)  \; .
	\ee
If moreover $\A_\bsM^{(1)} \ll \A_\bsM^{(0)}$, then the field 
oscillation frequency is
essentially given by the mean area $\A_\bsM^{(0)}$ enclosed by the orbits
of the family while the overall decrease is determined by the
first harmonic coefficient $\A_\bsM^{(1)}$.  The circular billiard
 can be regarded as a particular case where $\A_\bsM^{(0)}$ is 
non zero while $\A_\bsM^{(1)}$ as well as all other coefficients vanish.

\subsubsection{Magnetic susceptibility for a generic integrable system}
\label{sec:6A2}

>From the expression (\ref{uniform_d}) of the oscillating part of the
density of states the contributions $\chi^{(1)}$ and $\chi^{(2)}$ to the
susceptibility are obtained by the application of 
Eqs.~(\ref{eq:smooth_osc:all})
and (\ref{allDF}), which express $\Delta F^{(1)}$ and $\Delta F^{(2)}$
in terms of $\dosc(E,H)$. Taking twice the field derivative
according to Eq.~(\ref{eq:sus}) and introducing the dimensionless quantities
	\[ \C''_\bsM(H) \equiv \left( \frac{\Phi_0}{2\pi A} \right)^2
		\frac{d^2 \C_\bsM}{d H^2}
	\qquad ; \qquad
	(\C^2)''_\bsM(H) \equiv \left( \frac{\Phi_0}{2\pi A} \right)^2
		\frac{d^2 \C^2_\bsM}{d H^2} \; , \]
($A$ is the total area of the system) one obtains for the grand 
canonical contribution to the susceptibility
	\be \label{gen:chi1}
	\frac{\chi^{(1)}}{\cl} =- 24 \pi m A
	\sum_{\bM} \frac{R_T(\tau_\bsM) }{\tau_\bsM^2}
	\, \frac{d^0_\bsM(\mu)}{\gs}\, \C''_\bsM(H) \; .
	\end{equation}
If one assumes moreover that there are no degeneracies in the length
of the orbits, one has for the averaged canonical correction
	\begin{eqnarray} \label{eq:chi2_int}
	\frac{\overline{ \chi^{(2)} }}{\cl} & = &
	- 24 \pi^2 \hbar^2
	\, \sum_{\bM} \frac{R^2_T(\tau_\bsM) }{\tau_\bsM^2}
	\, \frac{\overline{ (d^0_\bsM(E))^2 }}{\gs^2}
	 \, (\C^2)''_\bsM(H) \label{gen:chi2} \\
	& = & - \frac{12}{\hbar} \sum_\bM
	\frac{R^2_T(\tau_\bsM) }{M^3_2 |g''_\mu({\bf I}_\bM)|} 
	\, (\C^2)''_\bsM(H)  
	\nonumber \; .
	\end{eqnarray}
The field--dependent component of $\overline{\chi^{(2)} }$ for weak fields
is given by
	\[
	(\C^2)''_\bsM(H\!=\!0) = 
	- \frac{1}{2\pi A^2} \int_0^{2\pi} d\theta_1 \, A_\bsM^2(\theta_1) \; ,
	  \]
which is always negative.  Therefore, for an ensemble of integrable
structures the magnetic response is always paramagnetic at zero
field.  We shall come back to this point in the last part of this
section.

\subsection{Generic chaotic systems}
\label{sec:gen_chaos}

Let us now consider generic chaotic systems, more generally, systems where
all the periodic orbits  are sufficiently 
isolated that the trace of the semiclassical Green function 
Eq.~(\ref{eq:traceG}) can be evaluated within stationary phase
approximation. In this case the Gutzwiller trace formula
provides the appropriate path to calculate the oscillating part
of the density of states (with or without magnetic field).
The Gutzwiller trace formula expresses the oscillating part of 
the density of states as a sum over all [here isolated] periodic
orbits $t$ as \cite{gutz_book}
	\be \label{gutz}
	\dosc(E,H) =  \sum_t  d_t \qquad ; \qquad
	d_t(E,H) = \frac{1}{\pi \hbar} 
	\frac{\tau_t}{r_t |{\rm det}(M_t-I)|^{1/2}}
	\cos (\frac{S_t}{\hbar} - \sigma_t  \frac{\pi}{2}) \; .
	\ee
$S_t$ is the action along the orbit $t$, $\tau_t$ the period
of the orbit, $M_t$ the stability matrix,  $\sigma_t$
its Maslov index, and $r_t$ the number of repetitions of the
full trajectory along the primitive orbit. 
All these classical quantities generally depend on 
energy and magnetic field.  If, as considered above for the integrable 
case, one is interested in the magnetic response to weak field, 
 one can express $d_t(E,H)$ in terms of the characteristics of 
the orbits at zero field by taking into
account the field dependence only in the actions.  Proceeding in 
exactly the same way as in section~\ref{sec:GenInt}, i.e.~grouping
together the contributions of time--reverse symmetrical orbits,
one obtains the same relation as Eq.~(\ref{eq:lowBd}) \cite{aga94,Prado}:
	\be \label{chaotic_d}
         d_t(E,H) =  d^0_{t}
	\cos \left[ 2 \pi \frac{H \A^0_{t}}{ \Phi_0} \right] \; .
	\ee
$d^0_{t}$ is the zero--field contribution of the orbit, obtained
from Eq.~(\ref{gutz}) at $H=0$, and $\A^0_{t}$ is the enclosed area
of the {\em unperturbed} orbit. In the case of a generic integrable 
system, the zero field regime played a peculiar role: except for 
the circular 
and annular geometries which remain integrable at all fields, a generic 
integrable system looses its integrability under the effect of a perturbing
magneti field. For chaotic geometries on the contrary, the zero field 
behavior is not substantially different from that at finite fields 
(as far as 
the stability of the dynamics is concerned). Since we are discussing the
general semiclassical formalism of chaotic systems without referring
to specific examples we do not need to restrict ourselves to weak
fields. Within this generic framework the chaotic geometries have 
the same conceptual simplicity as 
the systems which remain integrable at arbitrary field studied in 
section~\ref{sec:integrable}.  Namely Eq.~(\ref{gutz}) applies
independently of the field, and for derivatives with respect to the
field one can use 
	\be \label{eq:dSdH}
	 \frac{\partial S_t(H)}{\partial H} = \frac{e}{c}
	\A_t(H) \ ,
	\ee
 where $\A_t(H)$ is the area enclosed by the trajectory $t$ at the 
considered field. Therefore the computation of the contribution
$\chi^{(1)}$ and $\chi^{(2)}$ to the susceptibility follows essentially along
the same lines as described in
section~\ref{sec:integrable}: $\Delta F^{(1)}$ and $\Delta F^{(2)}$ 
are given by Eqs.~(\ref{eq:smooth_osc:all}) and
(\ref{allDF}), and to leading order in $\hbar$ the derivatives with
respect to the field should be applied only to the rapidly
varying term.  As a consequence, taking twice the derivative 
of the contribution of the orbit $t$ to $\Delta F^{(1)}$ merely amounts to 
a multiplication by a factor $(e\A_t)^2/(c\hbar)^2$, yielding 
	\be \label{eq:chi1_c}
	\frac{\chi^{(1)}}{\cl}  =  24\pi m A \sum_t
	   \frac{R_T(\tau_t)}{\tau_t^2}  \left( \frac{\A_t}{A} \right)^2
	   \frac{d_t(\mu)}{\gs}    \; ,
	\ee
where $d_t$ is given by Eq.~(\ref{gutz}).  Note that  
Eq.~(\ref{eq:chi1_c}) applies also to systems which remain 
integrable at all fields provided the Berry-Tabor formula 
Eq.~(\ref{BTT}) is used instead of the Gutzwiller one. For chaotic as well
as for integrable systems, $\chi^{(1)}$ can be paramagnetic or diamagnetic
with equal probability. The response of an ensemble of structures is given 
by $\Delta F^{(2)}$, which can be calculated  as a double 
sum over all pairs of orbits
	\begin{eqnarray}
        \label{eq:chichi}
	\frac{\chi^{(2)}}{\cl}
	 & = & 24 \sum_{tt'} \frac{R_T(\tau_t) R_T(\tau_{t'})}
	{r_tr'_t|{\rm det}(M_t-I)\,{\rm det}(M_{t'}-I)|^{1/2}} 
	        \left[ \left( \frac{\A_t - \A_{t'}}{A} \right)^2
	        \cos\left(\frac{S_t -S_{t'}}{\hbar} - 
	        (\sigma_t-\sigma_{t'})\frac{\pi}{2} \right)
			- \right.  \label{eq:chi2_c}  \nonumber \\
	           &   &  - \qquad \qquad \left.
	       \left( \frac{\A_t + \A_{t'}}{A} \right)^2
	        \cos\left(\frac{S_t + S_{t'}}{\hbar} - 
	        (\sigma_t+\sigma_{t'})\frac{\pi}{2} \right) \right]
	  \; .
	\end{eqnarray}
Here some remarks are in order. Due to the exponential proliferation
of closed orbits in chaotic systems off--diagonal terms should be 
considered at low temperatures since near--degeneracies in the actions 
of long orbits may appear, so that their contributions do not average out.
However, at sufficiently high temperatures where only short
periodic orbits are relevant, off--diagonal terms (of orbits not related
by time reversal symmetry) are eliminated
upon averaging. At finite field where time--reversal symmetry is broken
(more precisely, when no anti-unitary symmetry is preserved) only the 
terms with  $t'=t$ survive the averaging process, and 
(at the order of $\hbar$ considered)
$\overline{ \chi^{(2)} }$ vanishes since then 
$\A_t = \A_{t'}$.   The origin of the weak--field response 
for an ensemble is a consequence of time--reversal symmetry since
non--diagonal terms involving an orbit
and its time reversal have an action sufficiently close to
survive the average process but an area of opposite sign.
Indeed, assuming (in the weak--field regime) an ensemble average such 
that only diagonal and time reversal related terms are not affected,
Eq.~(\ref{eq:chichi}) reduces to
	\begin{equation}
        \label{eq:chichi_zero}
	\frac{\overline{\chi^{(2)}}_D}{\cl}
	  =  24 \sum_{t} \frac{R^2_T(\tau^0_t)}{r_t^2|{\rm det}(M^0_t-I)|} 
	        \left( \frac{2\A^0_t}{A} \right)^2
	        \cos \left(\frac{4\pi A^0_t H}{\Phi_0} \right)
	\; . \end{equation}
At zero field the cosine of the surviving terms in 
Eq.~(\ref{eq:chichi_zero}) is one and their prefactors positive.
This merely reflects that the dephasing of time reversal
orbits due to the perturbing magnetic field necessarily induces
on average a decrease of the amplitude of $\Nosc$, and therefore
by means of Eq.~(\ref{DF2}) a {\em paramagnetic} susceptibility.
For extremely large distributions in systems size, such as those
discussed in section~\ref{sec:experiment}, even the oscillating patterns
of Eq.~(\ref{eq:chichi_zero}) due to the subsequent rephasing
and dephasing of the time reversal orbits contributions vanish upon
 smoothing.
In this case, only the paramagnetic response related to the original
dephasing is observed, and the average susceptibility reaches zero
 as soon as $4\pi A^0_t H / \Phi_0$ is of the order of $2\pi$ 
for all trajectories.

\subsubsection*{Magnetization line-shape for chaotic systems}

The expressions we have obtained up to now in
this subsection do not require the system to be actually chaotic,
but only that periodic orbits are isolated.  They should therefore
be valid also for the contribution of isolated
orbits in mixed systems, where the phase space contains both
regular and chaotic regions.  This includes for instance 
the contributions of elliptic, i.e.\  stable orbits,
provided they are not close to any bifurcation and the surrounding
island of stability is large enough.

For geometries being actually chaotic it is however possible
to proceed further and to derive  a general expression for 
the line-shape of the field dependent susceptibility, 
if the temperature is low enough. 
For temperatures such that the cutoff time $\tau_c$ of 
the damping factor $R_T(\tau_t)$ is of the order of
the period of the fundamental periodic orbits, the average susceptibility
will be dominated by the shortest orbits, whose characteristics
are largely system dependent. However, for higher $\tau_c$ a large
number of trajectories will contribute to $\overline{ \chi^{(2)} }_D$,
and a statistical treatment of the sum on the r.h.s.\ of 
Eq.~(\ref{eq:chichi_zero}) is possible, yielding an {\em universal}
line-shape for the average susceptibility. 
  For sake of clarity, we discuss
here only the case of billiard like structures, but the following 
developments can be generalized in a straightforward way to any
kind of potentials.

Two basic ingredients are required here in addition to 
Eq.~(\ref{eq:chichi_zero}) to obtain the magnetization peak
line-shape.  The first
one is the semiclassical sum rule derived by Hannay and
Ozorio de Almeida \cite{han84}, which states that in sums like 
Eq.~(\ref{eq:chichi_zero}) the two effects of an exponential decrease
in the prefactors on the one hand and the exponential
proliferation of orbits on the other hand cancel each other
yielding
	\begin{equation} \label{eq:sumrule}
	\overline{ \sum_t \frac{\delta(\tau_t - \tau)}
		{|{\rm det}(M_t-I)|} } = \frac{1}{\tau} \; .
	\end{equation}
(Note, that in the above sum the contributions of orbits with 
number of repetitions $r_t > 1$ are neglected.)
To be valid, this equation requires that the periodic
orbits are uniformly distributed in phase space which will
only be achieved for sufficiently large $\tau$.  For
billiards the periods are given, up to a 
multiplication by the Fermi speed, by the length of the 
orbits and the periods $\tau$ in Eq.~(\ref{eq:sumrule}) can be replaced
by the lengths $L$.  We call $L_1^*$ the characteristic length for which
periodic orbits can be taken as uniformly distributed in phase 
space.  Typically, $L_1^*$ is  not much larger than the shortest 
period of the system.

The second ingredient is the distribution of area enclosed
by the trajectories. For chaotic systems, this distribution
has a generic form \cite{Chaost,dor91}.
Namely the probability $P_N(\Theta)$ for a trajectory to
enclose an algebraic area $\Theta$ after $N$ bounces on the
boundaries of the billiard is given by
	\begin{equation} \label{eq:Adist1}
	P_N(\Theta) = \frac{1}{\sqrt{2\pi N \sigma_N}}
	             \exp\left(-\frac{\Theta^2}{2 N
\sigma_N}\right) \; .
	\end{equation}
This result actually follows from a general argument 
\cite{dor91}
which in our case can be stated as
follows:   With a proper choice of the origin, the area swept by the 
ray vector for a given bounce
is characterized by a distribution, with zero mean  value
and a width  $\sigma_N$  which define the parameter
of the distribution Eq.~(\ref{eq:Adist1}).  For a strongly
chaotic system, successive bounces can be taken as independent 
events, which by means of the central limit theorem yield the distribution
Eq.~(\ref{eq:Adist1}).  Denoting $\bar{ L }$ the average 
distance between two successive reflections and 
$\sigma_L=\sigma_N/\bar{ L }$, this is equivalent to
	\begin{equation} \label{eq:Adist2}
	P_L(\Theta) = \frac{1}{\sqrt{2\pi L \sigma_L}}
	             \exp\left(-\frac{\Theta^2}{2 L \sigma_L}\right) \; .
	\end{equation}
Now $P_L(\Theta)$ is the distribution of enclosed areas
 for trajectories of length $L$, and the above equation is valid
for $L$ larger than a characteristic value $L^*_2$, which again
is of the order of the shortest closed orbit's length.

For temperature  sufficiently low so that $L_c > L^*_1,L^*_2$, 
Eqs~(\ref{eq:sumrule}) and (\ref{eq:Adist2}) can be used to 
replace the sum over periodic orbits Eq.~(\ref{eq:chichi_zero})
by the integral
	\begin{equation}
	\frac{ \overline{ \chi^{(2)} }_D }{\cl} = 
	24 \int_0^\infty \frac{dL}{L} \int_{-\infty}^{+\infty} d\Theta
	P_L(\Theta) \,
	R^2_T(L) \left( \frac{4\Theta^2}{A^2} \right) \!
	\cos \left( \frac{4\pi \Theta H}{\Phi_0} \right) \; .
	\end{equation}
Performing the Gaussian integral over $\Theta$, and introducing the
dimensionless factor $\xi = 2\pi H \sqrt{\sigma_L L_c}/\Phi_0$, one 
obtains the average susceptibility as
	\begin{equation} \label{eq:lineshape}
	\frac{ \overline{ \chi^{(2)} }_D }{\cl} = 
	96 \left(\frac{\sigma_L L_c}{A^2}\right) \,
	{\sf F}(\xi) \,
	\end{equation}
where the function ${\sf F}(\xi)$ is defined as 
	\begin{equation} \label{eq:Fzeta}
	{\sf F}(\xi) = \int_0^\infty \left(\frac{x}{\sinh{x}} \right)^2
	(1 - 4 \xi^2 x^2) \exp(-2 \xi^2 x)  \, dx\;\; ; \hspace{1cm}  x = L/L_c  \; .
       \end{equation}
The quadrature cannot be performed analytically (in a closed expression) for
arbitrary $\xi$\footnote{Using for $R_T(L)$ the asymptotic expression 
$R_T(L) = 2 (L/L_c) \exp(-L/L_c)$, valid for $L>L_c=\hbar \beta
v_F/\pi$, yields ${\sf F}(\xi) = (1-5\xi^2)/(1+\xi^2)^4$, but the 
contribution of the range $L \leq L_c$ is of the same order.},
but it can easily be calculated numerically.  As seen in
Fig.~\ref{fig:Fzeta}, ${\sf F}(\xi)$ has a  maximum at $\xi=0$ with a 
half-width $\Delta \xi \simeq 0.252$. Expansion of ${\sf F}(\xi)$ for small 
$\xi$ yields ${\sf F}(\xi) \approx \pi^2/6  -( 3 \zeta(3) +   2 \pi^4/15) \xi^2$
(where $\zeta(x)$ is the Zeta--function).
Denoting $\Lambda = \sigma_L L_c / A^2$, 
 the susceptibility at zero field is thus given by
	\begin{equation} \label{eq:hight}
	\frac{ \overline{ \chi^{(2)} }_D }{\cl} (H\!=\!0) =
	16 \pi^2 \Lambda \; ,
	\end{equation}
and the value half-width $\Delta \Phi$ by
	\begin{equation} \label{eq:width}
	\frac{\Delta\Phi}{\Phi_0} = \frac{\Delta\xi}{2\pi} \Lambda^{-1/2}
	\; .
	\end{equation}

The experimental observation of Eq.~(\ref{eq:lineshape}) would
be a very stringent confirmation for the applicability of the whole 
semiclassical picture developed here. However, two remarks are in order: 

(i)  
it is experimentally  usually rather difficult to make a clear cut
distinction between the function ${\sf F}(\xi)$ we obtained and, 
say, a Lorentzian shape.  Therefore, 
the temperature dependence (through $L_c$) of both the height
and, more surprisingly,  the
width of the magnetization peak should be observable rather than
the precise functional form of Eq.~(\ref{eq:lineshape}).

The physical picture underlying these
results is that at a given temperature, the cutoff length $L_c$
determines the length of the orbits providing the main contribution
to the susceptibility.  The smaller the temperature, the larger
$L_c$ and the longer the contributing orbits.  The typical areas
enclosed by these orbits thus increase, making them more sensitive to
the magnetic field and yielding  a larger susceptibility at zero
field and a smaller width since time reversal invariance is more 
rapidly destroyed.  The precise temperature dependence of the height
and the width (and their relationship, which might be useful when
$\sigma_L$ is unknown) is given by Eqs.~(\ref{eq:hight}) and
(\ref{eq:width}).

(ii)
It should be
borne in mind that Eq.~(\ref{eq:lineshape}) gives only the contribution
of the diagonal part of $\overline{ \chi^{(2)} }$, but does not
take into account the contribution of pairs of orbits which are not
related by time reversal symmetry.  Moreover, the statistical approach
used implies that fairly long orbits are contributing to the susceptibility,
which because of the exponential proliferation of such orbits should
yield an increasing number of quasidegeneracies in their length.
Therefore, to smooth out these non-diagonal term, one should a priori
require that the smoothing is taken on a very large range of $(\kf a)$.
In practice however, and as will be discussed in more detail in 
\cite{rod2000}, the smooth disorder characteristic of the GaAs/AlGaAs
heterostructures for which this kind of experiments are
done will actually be responsible for the cancelation of the
non-diagonal terms {\em without affecting} (for small enough disorder)
{\em the contribution we have calculated}.%
\footnote{  Without entering into
any details, the reason for this is the following.  For smooth
disorder, one should distinguish between an ``elastic mean free path''
$l$, and a transport mean free path $l_T$ which is much larger than
$l$.  For small disorder, $l_T$ can be assumed  infinite,
but long orbits will usually be longer than $l$.  
As a consequence, the action of each orbit is going to acquire a random
phase  from sample to sample, which is decorrelated for different orbits,
but is the same for time reversal symmetric orbits. Thus the diagonal
contribution we have calculated will not be affected, but non-diagonal
terms will be strongly suppressed.}
The effects of non-diagonal  terms should therefore be noticeably
less important in actual systems that it might appear in a clean
model.

\subsection{Integrable vs. chaotic geometries}
\label{sec:gen_comp}

The magnetic responses of chaotic and integrable systems have 
similarities and differences with respect to their treatment as well as to the resulting
susceptibility. The most remarkable similarity is the paramagnetic character of the
average susceptibility, while the magnitude of this response greatly differs for
both types of geometries. Concerning their treatment the differences
arise form the lack of structural stability of integrable systems
under a perturbing magnetic field.  Indeed, for non generic integrable
systems 
such as the  ring or circular billiards which remain integrable at 
all fields, the structure of the obtained equations are, except
for the use of the Berry-Tabor trace formula instead of the Gutzwiller
trace formula, the same as those for the chaotic systems.  
For generic integrable systems however, the breaking of invariant
tori requires a more careful treatment yielding slightly less
transparent, though essentially similar expressions.

\subsubsection{Paramagnetic character of the average susceptibility}

Because of this formal similarity, the qualitative behavior  of the
magnetic response is also quite the same for generic chaotic and integrable 
systems.  The susceptibility of a single structure can be paramagnetic or
diamagnetic and changes sign with a periodicity in $\kf a$  of the
order of $2\pi$. On the other hand, the average 
susceptibility for an ensemble of microstructure is, as expressed by 
Eqs.~(\ref{eq:chi2_int}) and (\ref{eq:chichi_zero}),
paramagnetic at zero field independent of the kind of dynamics
considered.  Indeed Eq.~(\ref{DF2}) states that 
$\overline{\Delta F^{(2)}}$ is, up to a multiplicative factor,
 the variance of the [temperature
smoothed]  number of states for a given chemical potential $\mu$.
In integrable and chaotic systems
the basic mechanism involved is that the magnetic field reduces
the degree of symmetry of the system, which as a general result
lowers this variance.  Therefore the $\overline{\Delta F^{(2)}}$ 
necessarily decreases when the magnetic field is applied and the average
susceptibility is paramagnetic at zero field.

There are some differences worth being considered. First, for 
chaotic systems the only symmetry existing at zero field
is the time reversal invariance, while for integrable
systems the breaking of time reversal invariance {\em and}
the breaking of invariant tori together reduces
the amplitude of $\Nosc(E)$. For chaotic systems
{\em the paramagnetic character of the ensemble susceptibility arises
as naturally as the negative sign of the magnetoresistance in
coherent microstructures}. The situation is similar to a random matrix point
of view, where the ensembles modeling the fluctuations of time reversal
invariant systems are known to be less rigid (in the sense that
the fluctuation of the number of states in any given stretch of energy
is larger) compared to the case where time reversal invariance is broken.  
The transition from one symmetry class to the other can be understood by the
introduction of generalized ensembles whose validity can be justified
semiclassically \cite{boh95}. It is however important to recognize that 
even for the chaotic case we do not have the standard GOE-GUE
transition \cite{RevBoh} since (\ref{DF2}) involves the integration over
a large energy interval. We are therefore not in the universal,
but in the ``saturation" regime where $(\Nosc(E))^2$ is given by the
shortest periodic orbits.

Secondly, for chaotic systems and for temperatures sufficiently
low that a large number of orbits contribute to the susceptibility, 
it is possible --- similar as in the weak localization effect in
electric transport \cite{Chaost} --- to derive a universal shape
of the magnetization peak.  This is 
not possible for integrable systems, which do not 
naturally lend themselves to a statistical treatment.

\subsubsection{Typical magnitude of the magnetic susceptibility}

Even if there are some analogies between the magnetic response of chaotic
and integrable systems (especially when the latter remain integrable at
finite fields), the {\em  magnitude} of the susceptibility exhibits
significant differences.
The contribution of an orbit to the Gutzwiller formula 
for two--dimensional systems is half an order in $\hbar$ smaller than a 
term in the Berry-Tabor formula for the integrable case.
More generally, in the case of $f$ degrees of freedom,
the $\hbar$ dependence of the Berry--Tabor formula is 
$\hbar^{-(1+f)/2}$ being the same as in the semiclassical Green function.
The Gutzwiller formula is obtained by performing the trace integral of the
Green function by stationary phase in $f-1$ directions,
each of which yielding  a factor $\hbar^{1/2}$.  This results in an entire 
$\hbar^{-1}$ behavior independent of $f$ for a chaotic system.  
Important consequences therefore arise for the case of two--dimensional
billiards of typical size $a$ at temperatures such
that only the first few shortest orbits are significantly 
contributing to the
free energy, and gives rise to a different
parametrical $\kf a$ characteristic of integrable and chaotic systems.
The $\kf a$ behavior of the density of states and susceptibility
for individual systems as well as ensemble averages is displayed in
Table \ref{tab:I}. While the magnetic response of chaotic systems
results from {\em isolated} periodic orbits, it is the existence of
{\em families} of flux enclosing orbits in quasi-- or partly integrable
systems which is reflected in a parametrically different dependence
of their magnetization and susceptibility on $\kf a$ (or $\sqrt{N}$ in terms
of the number of electrons). The difference is especially drastic for
ensemble averages where we expect a $\kf a$ independent 
response $\bar{ \chi}$
for a chaotic system while the averaged susceptibility for integrable
systems, e.g. the ensemble of square potential wells in the experiment
discussed in section \ref{sec:square}, increases linearly in $\kf a$. Under
the conditions of that measurement \cite{levy93} the enhancement 
should be of
the order of 100 compared to an ensemble of chaotic quantum dots.
We therefore suggested \cite{URJ95} to use the different parametrical
behavior of the magnetic response as a tool in order to unambiguously
distinguish (experimentally) chaotic and integrable dynamics in
 quantum dots.
We stress that this criterion is not based on the long time behavior of the
chaotic dynamics  but on 
short time properties, namely the existence of families of orbits 
contributing in phase to the trace of the Green function of 
integrable systems.

\newpage

\section{Non--perturbative fields:\ bouncing--ball-- and
de Haas--van Alphen--oscillations}
\label{sec:highB}

Up to now we have essentially focused on mesoscopic effects in the weak
magnetic field regime where the classical cyclotron radius 
$r_c$ is large compared to the typical size $a$ of the 
system, i.e.
	\begin{equation}
	\frac{r_c}{a} = \frac{c \hbar\, k}{e \, H \, a} \gg 1 \; .
	\label{classcond}
	\end{equation}
Then, electron trajectories can be considered as straight lines 
between bounces and the
dominant effect of the magnetic field enters as a semiclassical phase 
in terms of the enclosed flux. Nevertheless, 
as shown in Fig.~\ref{f1} in the introduction (for the case of a
square) the low--field oscillations of $\chi$ are accurately described
by {\em classical} perturbation theory in terms of the family (11) 
of unperturbed orbits (left inset in Fig.~\ref{f1}(b)). They persist 
up to field strengths $\varphi \approx 10$ which is by orders of magnitude larger 
 than the typical flux scale which describes the breakdown
of first order {\em quantum} perturbation theory, i.e., magnetic fluxes
where the first avoided level crossings appear. Due to condition
(\ref{classcond}) the relevant classical ``small'' parameter is $H/\kf$.
The  semiclassical ``weak--field'' regime increases with
increasing Fermi energy.

In this section we will go beyond this (classically) perturbative
regime and discuss microstructures under larger fields, where the magnetic
response reflects the interplay between the scale of the confining
energy and the scale of the magnetic field energy $\hbar\omega_c$ on the
quantum level. Classically, non--perturbative fields affect
the motion not only through a change of the actions (by means of the
enclosed flux), but additionally due to the bending of the trajectories.
A priori, the semiclassical approach we used for weak magnetic fields applies
also to this case without any difference: Oscillating components
of the single--particle density of states can be related to periodic
(or nearly periodic) orbits by taking the trace of the semiclassical Green
function. The magnetic response is then obtained from integration over the energy 
and taking the derivatives with respect to the magnetic field.
These operations correspond to the multiplication
by the inverse of the period of the orbit, by the damping factor $R_T$
and by the area enclosed by the orbit. Three field regimes
(weak ($a \ll r_c$), intermediate ($a \simeq r_c$), and high
($a \geq 2 r_c$) fields) can be clearly distinguished as is illustrated in
Fig.~\ref{f1}(b) for the square geometry.  The distinction of the 
three regimes appears not because they deserve a fundamentally different
semiclassical treatment, but simply because of some salient features
of the classical dynamics associated to each of these regimes.

In the high--field regime, most of the orbits simply follow
a cyclotron motion.  In that case, the system behaves essentially as an
infinite system, and one recovers the well known de Haas-van Alphen oscillations
for $\chi^{(1)}$.  We shall moreover see below that within our semiclassical
approach, the destruction of some of the cyclotronic orbits due to reflections at
the boundaries can be taken into account, allowing to handle
correctly the crossover regime where $a \geq 2 r_c$ but $r_c$ is not yet
negligible with respect to $a$.

While the high field ($a \gg r_c$) classical dynamics is generally
(quasi) integrable the dynamics in the intermediate field regime is always
mixed (in the sense that chaotic and regular motion coexists in phase space)
except for particular cases of systems with rotational symmetry 
which remain integrable independent of the magnetic field.
In contrast to that, systems in the small field regime can exhibit
any degree of chaoticity  {\em in the zero field limit}.
Indeed, there is a large variety of geometries for which the
motion of the electrons in the absence of a magnetic field is either
integrable, or completely chaotic. Therefore, increasing the field starting
from an integrable (respectively chaotic) configuration at $H=0$, the
intermediate field regime will be characterized by an increase 
(respectively
a decrease) of the degree of chaos of the classical dynamics, which
will noticeably affect the magnetic response of the system.  However,
if the zero--field configuration already shows a mixed dynamics (which is
generically the case), the only noticeable difference
between the weak and intermediate field regime will consist in the
complete lost of time reversal symmetry and naturally its consequences
on $\overline{ \chi^{(2)} }$ as discussed in section~\ref{sec:general}.

In addition, for some particular geometries, namely those for which the
boundary contains some pieces of parallel straight lines, the intermediate
field susceptibility will be characterized by the dominating influence
of {\em bouncing--ball orbits}, periodic electron motion due to reflection
between opposite boundaries.  Fig.~\ref{f1}(b)
depicts a whole scan of the magnetic susceptibility of
a square from zero flux up to flux $\varphi = 55$ $(3r_c \approx a$). 
We can see there, and we will discuss in detail below, that there are
--- besides the small--field oscillations due to orbits (11)
--- two well separated regimes
of susceptibility oscillations: The intermediate field regime
($2r_c > a$) reflects quantized {\em bouncing--ball periodic orbits}
(second  inset) and the oscillations in the strong
field regime ($2r_c < a$) which, as mentioned above,
are related to {\em cyclotron orbits} (right inset).
Although the results to be reported are of quite general nature
we will discuss them quantitatively for the case of 
square microstructures.
We study individual squares and perform our analysis within the grand
canonical formalism.

\subsection{Intermediate fields: Bouncing-ball magnetism}
\label{sec:interfield}

The full line in Fig.~\ref{fig:chibb}(a) shows the quantum mechanically
calculated (see section~\ref{sec:square}.D)
grand canonical susceptibility for small and
intermediate fluxes at a Fermi energy corresponding to $\sim$2100 enclosed electrons
in a square at a temperature such that $k_BT/\Delta = 8$. 
The semiclassical result $\chi^{(1)}_{(11)}$
from the family (11) (Eq.~(\ref{square:chi1})) shown as the dashed--dotted line
(with negative offset) in Fig.~\ref{fig:chibb}(a) exhibits
the onset of deviations from the quantum result
with respect to phase and amplitude starting at $\varphi \approx 8$ 
($r_c \approx 2a$)
indicating the breakdown of the family (11) of straight line
orbits. With increasing flux we enter into a regime where the
non--integrability of the system manifests itself in
a complex structured energy level diagram (see Fig.~\ref{f1}(a)) on the
quantum level and in a mixed classical phase space \cite{rob86b} of
co--existing regular and chaotic motion. However, besides the variety
of isolated stable and unstable periodic orbits there remains a family
of orbits with specular reflections only on opposite sides of the square.
We will denote these periodic orbits shown in Fig.~\ref{fig:bbschema} which are
known as ``bouncing--ball'' orbits in billiards without magnetic
field by $(M_x,0)$ and $(0,M_y)$ according to the labeling introduced
in section~\ref{sec:square}.A. ($M_x$ and $M_y$ are the number of bounces at the
bottom and left side of the square.) These orbits form families
which can be parameterized, e.g., for the case $(M_x,0)$ in terms of the
point of reflection $x_0$ at the bottom of the square. We thus expect
--- as in the case of the families $(M_x,M_y)$ in section~\ref{sec:square} ---
in the semiclassical limit a parametrical dependence on $\kf a$
of the related susceptibilities which should strongly dominate the contributions
of the co--existing isolated periodic orbits.

We present our semiclassical calculation of the susceptibility contribution
related to bouncing--ball orbits for the primitive periodic
orbits, i.e., $(M_x,0) = (1,0)$ and generalize  our results at the end
to the case of arbitrary repetitions. We proceed as in section~\ref{sec:square}
for the derivation of $\chi^{(1)}_{11}$. However, while those
calculations were performed in the limit of a small magnetic field (assuming
$H$--independent classical amplitudes and shapes of the trajectories (11))
we now have to consider explicitly the field dependence of the
classical motion.  The contribution to the diagonal part of the Green
function of a recurring path starting at a point
${\bf q}$ on a bouncing--ball orbit reads
	\begin{equation}
	{\cal G}_{10} ({\bf q, q'=q};E,H) = 
	\frac{1}{i\hbar \sqrt{2\pi i\hbar}} \, D_{10} \,
	\exp\left[i\left(\frac{S_{10}}{\hbar} - 
	\eta_{10}\frac{\pi}{2}\right) \right]
	\ . \label{eq:Gbb}
	\end{equation}
Simple geometry yields for its length, enclosed area, and action
	\begin{equation}
	L_{10}(H) = \frac{2a \zeta}{\sin\zeta}    \quad; \qquad
        A_{10}(H) = -(2\zeta-\sin 2\zeta) \, r_c^2 \quad; \qquad
        \frac{S_{10}}{\hbar}= 
	k \left(L_{10} +  \frac{A_{10}(H)}{r_c(H)}\right)  \; ;
	\label{eq:LAS10}
	\end{equation}
where $\zeta$, the angle between the tangent to a bouncing--ball trajectory
at the  point of reflection and the normal to the side, is given by (see
Fig.~\ref{fig:bbschema})
	\begin{equation}
	\sin\zeta = \frac{a}{2 r_c} \; .
	\label{eq:beta}
	\end{equation}
The Maslov index $\eta_{10}$ is 4 and will be therefore omitted from now on.

As in section~\ref{sec:square}, we will use as configuration space coordinates
the couple $\bq=(x_0,s)$, where $x_0$ labels the abscissa of the last
intersection of the trajectory with the lower side of the square (see Fig.\
\ref{fig:bbschema}) and
$s$ is the distance along the trajectory.  This choice has the advantage that
$D_{10} (x_0,s)$ is constant, and therefore taking the trace of the Green
function merely amounts to a multiplication
by the size of the integration domain.  As discussed in more detail
in Appendix~\ref{app:D_M}, the semiclassical amplitude $D_{10}$ is
given by \cite{gutz_book}
	\begin{equation}
	D_{10}({\bf q, q'=q}) = \frac{1}{|\dot{s}|} \, \left|
	\frac{\partial x_0'}{\partial p_{x_0}}
	\right|^{-\frac{1}{2}}_{{x_0}' = x_0} \; ,
	\label{eq:D10:def}
	\end{equation}
where $ (x_0,p_{x_0}) \rightarrow (x_0', p'_{x_0} )$ is the Poincar\'e map
between two successive reflections on the lower side of the billiard.
Noting $u_{x_0} = (p_{x_0} - e A_x/c)/(\hbar k)$ ($u_{x_0}$ is the
projection of the unit vector parallel to the initial velocity
 on the $x$ axis) one obtains from simple geometrical considerations
	\begin{eqnarray}
	p_{x_0}' & = & p_{x_0}    \nonumber \\
	x_0'     & = & x_0 + 2 r_c
	   \left( \sqrt{1 - ( u_{x_0} \! - \! {a}/{r_c} )^2} -
	   \sqrt{1 - (u_{x_0})^2} \right)   \; .
	\label{eq:bb:map}
	\end{eqnarray}
For the periodic orbits, $x_0'= x_0$ implies that $u_{x_0} = a/2 r_c =
\sin \zeta$, and therefore
	\begin{equation}
	D_{10}({\bf q, q'=q})
	 = \frac{1}{|\dot{s}|} \,  \sqrt{\frac{\hbar k \cos\zeta}{2 a}}
	\label{eq:D10}
	\end{equation}
which reduces to Eq.~(\ref{eq:square:DM}) in the limit $H=0$ ($\zeta=0$).
For the contribution of the whole family (1,0) we must perform the trace
integral Eq.~(\ref{eq:traceG}).
The integral over $s$ gives as usual a multiplication by the period
	\[ \tau_{10} = \frac{L_{10}}{\hbar k/m} \]
of the orbit.  Moreover, since neither the actions $S_{10}$, nor the
amplitude $D_{10}$ depend on $x_0$, the $x_0$-component of the trace
integral simply yields a length factor
	\begin{equation}
	l(H) = a\left(1 - \tan\frac{\zeta}{2} \right)
	\label{eq:l}
	\end{equation}
(see Fig.~\ref{fig:bbschema}) which describes the magnetic field dependent
effective range for the lower reflection points of bouncing--ball
trajectories (1,0).
$l(H)$ vanishes for magnetic fields corresponding to $2 r_c = a$.
We therefore obtain for the
bouncing--ball contribution $d_{10}=-(\gs / \pi) $Im${\cal G}_{10}$
to the density of states
	\be
	d_{10}(E,H) = - \frac{2\gs }{(2\pi \hbar)^{3/2}}
	l(H) L_{10} D_{10}
	\sin\left(\frac{S_{10}}{\hbar}+ \frac{\pi}{4}\right) \; .
	\label{eq:d10}
	\ee
In order to compute the contribution $\chi^{(1)}_{10}$ to the
(grand canonical) susceptibility we first have to calculate
$\Delta F_{10}^{(1)}$ by performing the energy integral
Eq.~(\ref{eq:smooth_osco}), and then
to take twice the derivative with respect to the magnetic field. In a leading $\hbar$
calculation, integrals and derivative should again be applied only on
the rapidly oscillating part of $d_{10}$.  Noting moreover that
Eq.~(\ref{dS}) is not restricted to perturbation around $H=0$,
i.e.~that at any field
	  \[ \frac{\partial S_{10}}{\partial H} = \frac{e}{c} A_{10} \; , \]
we therefore obtain in the same way as we did for Eq.~(\ref{eq:chi1_c})
	\begin{equation} \label{eq:chi110}
	\chi^{(1)}_{10} = \frac{1}{a^2}
	\left( \frac{e A_{10}}{c \tau_{10}} \right)^2
	d_{10}(\mu,H)  R_T (L_{10}) \; .
	\label{eq:Chi10}
     	\end{equation}
Inserting the expressions Eqs.~(\ref{eq:LAS10}), (\ref{eq:l})
and (\ref{eq:D10}) into Eqs.~(\ref{eq:d10}) and (\ref{eq:Chi10}), 
we finally have $\chi^{(1)}_{10}$ explicitly in terms of $\zeta$ as
	\begin{eqnarray}
	\frac{\chi^{(1)}_{10}}{\chi_L} & = &
	\frac{3 }{8 \pi^{1/2}} (\kf a)^\frac{3}{2} \,
	\frac{\sqrt{\cos \zeta} (\sin \zeta+\cos \zeta -1)}{\zeta}
	\frac{ (2 \zeta - \sin (2 \zeta) )^2}{\sin^4\zeta} \times
        \label{eq:chibb} \\
	& & \qquad \qquad \qquad \times
	\sin\left(\frac{S_{10}}{\hbar}+ \frac{\pi}{4}\right)
	\, R_T (L_{10}) \; .  \nonumber
\end{eqnarray}

The entire bouncing--ball susceptibility $(\chi^{(1)}_{10}+\chi^{(1)}_{01})/
\chi_L = 2 \chi^{(1)}_{10}/\chi_L$ according to Eq.~(\ref{eq:chibb})
is shown in Fig.~\ref{fig:chibb}(a) as the dashed line. At fluxes up to
$\varphi \approx 15$ it just explains the low frequency shift in the
oscillations of the quantum result indicating that the overall small field
susceptibility is well approximated by $\chi_{11} + \chi_{10}+ \chi_{01}$.
For fluxes between $\varphi \approx 15$ ($r_c=1.2 a$) up to $\varphi\approx 37$
(the limit where $r_c = a/2$, i.e., the last bouncing--ball orbits vanish)
the magnetic response is entirely governed by bouncing--ball periodic motion
and the agreement between the semiclassical prediction and the
full quantum result is excellent.

The flux dependence of the actions $S_{10}$ (see Eq.~(\ref{eq:LAS10}))
is rather complicated. However, an expansion for
$a/r_c = 2\pi \varphi/(\kf a) \ll 1$ yields a quadratic dependence 
on $\varphi$
	\begin{equation}
	\frac{S_{10}}{\hbar} \simeq 2\, \kf \, a \left[1 -
        \frac{1}{24} \left(\frac{2\pi\varphi}{\kf a} \right)^2 \right] \; .
	\label{eq:S10exp}
	\end{equation}
The susceptibility from Eq.~(\ref{eq:chibb}) with $S_{10}$ according to
Eq.~(\ref{eq:S10exp}) is shown as dotted curve in Fig.~\ref{fig:chibb}(a).
It agrees well at moderate fields and runs out of phase at a flux
corresponding to $a/r_c > 1$. While the period of the $\chi_{11}$ small field
oscillations is nearly constant with respect to $\varphi$ we find a
quadratic $\varphi$ characteristic for the oscillations in the
intermediate regime which turns
into a $1/\varphi$ behavior in the strong field regime (see next subsection).

To show that the agreement between the semiclassical (dashed) curve and the
quantum result is not an artefact of the particular number of electrons
chosen, Fig.~\ref{fig:chibb}(b) depicts semiclassical and quantum 
bouncing--ball oscillations for $k_B T/\Delta=7$ and
at a different Fermi energy corresponding to $\sim$1400 electrons.
With decreasing Fermi energy the upper limit $r_c=a/2$ (or 
$\kf a/(2\pi\varphi)=1/2$)
of the bouncing--ball oscillations is shifted towards smaller fluxes
($\varphi \approx 30$ in Fig.~\ref{fig:chibb}(b)) and the
number of oscillations shrinks. The oscillations for $\varphi > 30$
belong already to the strong field regime discussed in the next subsection.

Up to know we discussed the magnetic response of the family of primitive
orbits (1,0) and (0,1) which completely describes the intermediate 
field regime at rather high temperatures corresponding to a 
temperature cutoff length in the order of the system size. 
At low temperatures we have to include contributions
from higher repetitions $(r,0)$, $(0,r)$ along bouncing--ball paths.
$L_{r0}$ and $A_{r0}$ have a linear $r$-dependence, and from
the Poincar\'e map Eq.~(\ref{eq:bb:map}), one obtains that
$D_{r0}=r^{-1/2} D_{10}$.  Therefore
	\begin{eqnarray}
	\frac{\chi^{(1)}}{\chi_L} & = &
	\frac{1}{\chi_L} \, \sum_{r=1}^\infty \,
	(\chi^{(1)}_{r0} +\chi^{(1)}_{0r}) \nonumber \\
	& = &
	\frac{3 }{4 \pi^{1/2}} (\kf a)^\frac{3}{2} \,
	\frac{\sqrt{\cos \zeta} (\sin \zeta+\cos \zeta -1)}{\zeta}
	\frac{ (2 \zeta - \sin (2 \zeta) )^2}{\sin^4 \zeta} \times
	\label{eq:chibbr} \\
	& & \qquad \qquad \times
	\sum_{r=1}^\infty \, r^{-{1}/{2}} \,
	  \sin\left(r\, \frac{S_{10}}{\hbar}+ \frac{\pi}{4}\right)  \,
	  R_T (r\, L_{10})\; .   \nonumber
	\end{eqnarray}
Fig.~\ref{fig:chibb}(c) shows the susceptibility at the same Fermi energy as in
Fig.~\ref{fig:chibb}(b) but at a significantly lower temperature 
$k_B T/\Delta = 2$.
The bouncing--ball peaks are much higher and new peaks related to long
periodic orbits differing from the bouncing--ball ones appear.
However, the bouncing--ball peak heights and even their shape
(which is no longer sinusoidal and symmetrical with respect 
to $\chi = 0$) is
well reproduced by the analytical sum Eq.~(\ref{eq:chibbr}) showing the
correct temperature characteristic of the semiclassical theory.

The $\kf a$ behavior of the bouncing--ball susceptibility at a fixed flux
is not as simple as in the case of the weak--field oscillations (where
$\chi^{(1)}_{11} \sim (\kf a)^{3/2}$) since the angle $\zeta$ occurring 
in the
prefactor in Eq.~(\ref{eq:chibb}) depends on $\kf a$ and the action is
non--linear in $\kf a$. Nevertheless, the overall oscillatory behavior
is similar as for example in Fig.~\ref{fig:chi1}(a).
However, at a given non--zero magnetic field the
classically relevant parameter Eq.~(\ref{classcond}) changes with energy.
Therefore, by increasing the Fermi energy beginning at the ground state
one generally passes from the strong field regime (at small energies 
or high field strengths, see next section) to
the bouncing--ball regime and will finally reach the regime of oscillations
related
to the family (11). A unique behavior of periodic orbit oscillations 
is only expected by changing magnetic field and Fermi energy 
simultaneously in order to keep the classical parameter 
Eq.~(\ref{classcond}) which determines the classical phase
space of the microstructure constant. Such a technique is 
known as {\em scaled energy spectroscopy} in the context of atomic 
spectra \cite{ERWS88}.

Bouncing--ball oscillations are expected to exist in general in
microstructures with parts of their opposite boundaries
being parallel and in spherical symmetrical microstructures as the disk
discussed in section \ref{sec:integrable}. (In the latter case the oscillations
should be even stronger than in the square since the effective length $l(H)$
(Eq.~(\ref{eq:l})) is not reduced with increasing magnetic field.)
An investigation of rectangular billiards for instance shows a splitting
of the frequencies of oscillations related to orbits $(M_x,0)$ and $(0,M_y)$
due to the different lengths of the orbits in $x$ and $y$ direction.

\subsection{Strong field regime}

At large magnetic field strengths or small energy the spectrum of a square
potential well exhibits the Landau fan corresponding to bulk--like
Landau states being almost unaffected by the system boundaries, while
surface affected states fill the gaps between the Landau levels and
condensate successively into the Landau channels with increasing
magnetic field (see, e.g., Fig.~\ref{f1}(a)). This spectral characteristic
corresponds to susceptibility oscillations which emerge with increasing
amplitude for fluxes corresponding to $r_c < a/2$, for instance for
$\varphi > 40$ in Fig.~\ref{f1}(b). They are shown in more detail in
Fig.~\ref{fig:dhva} where the full line depicts the numerical quantum result.
These susceptibility oscillations
exhibit the same period $\sim 1/H$ as de Haas--van Alphen bulk
oscillations but differ in amplitude, because here the cyclotron radius is not
negligible compared to the system size.

For the bulk or in the extreme high field regime $r_c \ll a$, where
quantum mechanically the influence of the boundaries of
the microstructure on the position of the quantum levels can
be neglected, an expression for the susceptibility
is most easily obtained by Poisson summation of the quantum density of states
as was briefly sketched in the introduction following standard textbooks
\cite{LanLip}. One then obtains the bulk magnetism  as given
by Eq.~(\ref{intro:susdHvA}).   It may be interesting to note however
that a semiclassical interpretation of this equation follows naturally
from an analysis similar to the one we followed throughout this paper.
In this case only one type of primitive periodic orbits exists,
the cyclotron orbits with length, enclosed area, and action given by
	\begin{equation}
	L_0(H) = 2 \pi r_c \quad; \qquad
        A_0(H) = -\pi r_c^2 \quad; \qquad
        \frac{S_0}{\hbar}= k L_0 + \frac{e}{c\hbar} H A_0 = k \pi r_c  \; .
	\label{eq:LASdHvA}
	\end{equation}
Moreover, the trajectory passes through a focal point after each half
traversal along the cyclotron orbit. Therefore, using $\eta_n = 2n$ for
the Maslov indices and omitting the Weyl part of $G$,  one obtains from
Eq.~(\ref{eq:green}) a semiclassical expression for the diagonal part
of the Green function
	\begin{equation} \label{sec7:G_dHvA}
	G(\br , \br' \! = \! \br) =
	\frac{1}{i\hbar \sqrt{2\pi i\hbar}} \sum_n (-1)^n D_n \,
	\exp(i n \pi k r_c  ) \; ,
	\end{equation}
in which the main structure of Eq.~(\ref{intro:susdHvA}) is already apparent.
A direct evaluation of the amplitude $D_n$ in configuration space is
however complicated here by the fact that all trajectories
starting at some point $\br$ refocus precisely at $\br$
(focal point).  Therefore, an expression like Eq.~(\ref{eq:D10:def})
for $D_n$ is divergent and cannot be used.
A method to overcome this problem by working with a Green function
$\tilde{G} (x,y;p_x',y')$ in momentum representation
for the $x'$ direction instead of $G(x,y;x',y')$ is described in
appendix~\ref{app:highB}.  It yields (see Eq.~(\ref{app:G_dHvA}))
	\begin{equation}
	\frac{D_n}{i\hbar \sqrt{2\pi i\hbar}} =
	\frac{m}{i \hbar^2} \; .
	\end{equation}
Inserting this expression in Eq.~(\ref{sec7:G_dHvA}) we obtain
the oscillating part of density of states
	\begin{equation}
	\dosc(E;H) = \sum_n d_n(E,H) = \frac{\gs A m}{\pi \hbar^2}
	\sum_n (-1)^n  	\cos(n \pi k r_c  ) \; ,
     	\end{equation}
from which the de Haas--van Alphen susceptibility Eq.~(\ref{intro:susdHvA})
is  obtained by using 
	\begin{equation}
	\chi^{(1)} = \frac{1}{A}
	\left( \frac{e A_0}{c \tau_0} \right)^2
	\sum_n d_n(\mu,H)  R_T (n L_0) \; .
     	\end{equation}
(with $\tau_0 = L_0/ \vf$) which applies for the
same reasons as Eq.~(\ref{eq:chi110}).

For an infinite system, this direct semiclassical approach to
the susceptibility therefore yields the same result as
the Poisson summation.
For billiard systems, it allows moreover to take correctly into
account the fact that the trajectories too close to the boundary
do not follow a cyclotron motion.  Indeed, as seen in
appendix~\ref{app:highB}, the contribution of cyclotron orbits to
the susceptibility Eq.~(\ref{intro:susdHvA}) has to be modified when
$r_c$ is not negligible compared to $a$ by the introduction of a
multiplicative factor $s(H)$. It accounts for the effect that the family
of periodic cyclotron orbits (not affected by the boundaries)
which can be parameterized by the positions of the orbit centers is
diminished with decreasing field since the minimal distance between orbit
center and boundary must be at least $r_c$.
One therefore obtains for a billiard like quantum dot
	\begin{equation}
	\frac{\chi^{GC}_{cyc}}{\chi_L} = - 6 s(H) \,
	(\kf r_c)^2
	\sum_{n=1}^\infty \, (-1)^n \, R_T(2\pi n r_c) \,
	\cos \left(n \pi \kf r_c \right) \; ,
	\label{sec7:susdHvA}
	\end{equation}
where $s(H)$ is given by Eq.~(\ref{app:s(H)}).  In the case of the square
we find for the area reduction factor
	\begin{equation} \label{eq:s(H)}
	s(H) = \left(1-2\frac{r_c}{a}\right)^2 \,
	\Theta\left(1-2\frac{r_c}{a}\right) \; ,
	\end{equation}
$\Theta$ being the Heavyside step function. The last cyclotron orbit
disappears at a field where $r_c = a/2$, i.e. $s(\varphi)=0$ which happens near
$\varphi \approx 38$ in Fig.~\ref{fig:dhva}. There the dashed line
showing the semiclassical expression (\ref{sec7:susdHvA}) is in good
agreement with our numerical results and reproduces the decrease 
in the amplitudes of the de Haas--van--Alphen oscillations when 
approaching $\varphi(r_c=a/2)$ from the strong field limit. This 
behavior is specific for quantum dots and does not occur in the 
two--dimensional bulk. Corresponding bulk de Haas--van Alphen 
oscillations under the same conditions as for the curves in 
Fig.~\ref{fig:dhva} have (nearly constant) amplitudes in the order
of $\chi/\chi_L \approx 3000$.

The semiclassical curve which only reflects
the contribution from unperturbed cyclotron orbits agrees with the
numerical curve (representing the complete system)
even in spectral regions which show a complex variety
of levels between the Landau manifolds (see Fig.~\ref{f1}). Due to temperature
cutoff and since angular momentum is not
conserved in the square the corresponding edge or whispering gallery orbits
are mostly chaotic and do not show up in the magnetic response.
The strong de Haas--van Alphen like oscillations manifest the dominant
influence of
the family of cyclotron orbits. In related work on the
magnetization of a (angular momentum conserving) circular disk in the
quantum Hall effect regime Sivan and Imry \cite{SivanImry} observed
additional high frequency oscillations related to
whispering gallery orbits superimposed on the de Haas--van Alphen oscillations.

\newpage

\section{Conclusion} 
\label{sec:concl}

In this work we have studied orbital magnetism and persistent currents
of small mesoscopic samples in the ballistic regime. 
Within a model of non-interacting electrons we have provided
a comprehensive semiclassical description  of these phenomena
based on the semiclassical trace formalism initiated by Gutzwiller,
Balian, and Bloch.  We have moreover 
treated in detail a few examples of experimental
relevance such as the square, circle and ring geometries.

 The global picture that emerges from our study can be  summarized
as follows.  The magnetic
response is obtained from the variation of the thermodynamic 
potential (or the free energy) under an  applied magnetic field
and therefore, in a non-interacting model, from the knowledge of the
single-particle  density of states. 
The semiclassical formalism naturally leads to a separate 
treatment of the smooth (in energy) component of the density of 
states (or its integrated versions) and of its rapidly 
oscillating part. The former is related to the local properties 
of the energy manifold, while the latter is associated with the dynamical
properties of the system, more precisely to its periodic 
(or nearly periodic) orbits.  For the smooth component
we have shown that, despite the leading (Weyl) term 
 in an $\hbar$ expansion is independent of the field, 
higher order terms can be computed and give rise to the 
standard Landau diamagnetism for any confined electron system 
at arbitrary magnetic fields.  In the high temperature regime,
where the rapidly oscillating component of the density of states 
is suppressed  by the rounding of the Fermi surface, 
the magnetic response reduces to the Landau diamagnetism.
On the other hand, for the temperatures of experimental relevance the 
contribution coming from the oscillating part of the density
of states is much larger than the Landau term and dominates
the magnetic response.  Similarly to the case of diffusive
systems, the susceptibility of a ballistic sample in contact with 
a particle reservoir with chemical potential $\mu$  can be paramagnetic or
diamagnetic (depending on $\mu$) with equal probability. The fact
that the samples are isolated (with respect to electron transfer)
forces us to work in the canonical ensemble.  Because of the
breaking of time reversal invariance occurring when the field is 
turned on, this results, for essentially the same reason as in 
the diffusive regime, in a small paramagnetic asymmetry for the 
probability distribution of the susceptibility of a given sample.
For generic integrable systems, this effect is reinforced by the 
breaking of invariant tori, which acts concurrently with the lost
of time reversal invariance.
The asymmetry disappears for a flux $\Delta \Phi$ inside the system
 which is of the order of the quantum flux $\Phi_0$  at a 
temperature selecting only the first few shortest orbits contributions, 
but may be smaller for lower  temperature.
Measuring the magnetic response of an ensemble of 
structures with a large dispersion in the size or the number of 
electrons
magnifies this asymmetry and yields a total  response   [per
structure] which is  paramagnetic and much smaller than the 
typical susceptibility for a flux smaller than $\Delta \Phi$, and zero
for larger flux.  For ensembles with only microscopic differences 
between the individual structures (i.e.~$\Delta(\kf a) 
\geq 2 \pi$, but still $\Delta a/a \ll 1$ and $\Delta {\bf N}/{\bf
N} \ll 1$) further oscillating patterns in the average susceptibility
should be observed for larger fields.

Since the oscillating part of the density of states is semiclassically
related to the classical periodic orbits, the nature of the classical
dynamics quite naturally plays a major role in the determination
of the amplitude of the magnetic response.  Indeed, for a system
in which continuous families of periodic orbits are present, these
orbits contribute in phase to the density of states, yielding
much larger fluctuations of the density of states than for systems
possessing only isolated orbits, and therefore much larger magnetic 
response.  Families of periodic orbits are characteristic for
integrable systems, while for chaotic systems the periodic orbits are usually
isolated.  This different behavior can therefore be referred to
as the hallmark for the distinction between integrable and chaotic 
systems.  It should be borne in mind however that this difference 
is due to short-time properties, namely the existence or absence of 
families of orbits, rather than to long-time properties such as 
exponential divergence of orbits.   In this respect, some atypical
chaotic systems, such as the Sinai billiard for instance, may show 
a magnetic response typical for an integrable system because of the 
existence of marginally stable families of orbits.

The importance of classical mechanics can be illustrated in the 
[experimentally
relevant] case of two-dimensional billiard-like quantum dots in the 
weak-field regime.  If the system is chaotic, more precisely
if the periodic trajectories are  isolated, the typical susceptibility
 scales as $(\kf a) \cl$, where $\kf$ is the Fermi wave number and
 $a$ the typical size of the dot.  By comparison, the typical 
susceptibility of an integrable
system scales with $(\kf a )^{3/2} \cl$.   This characteristic behavior of
integrable systems is found in the generic case (like the square) where
the magnetic field breaks the integrability as well as in the non-generic
case (like the disk) where the system remains integrable at finite
fields. The difference due to the nature of the classical mechanics is 
even stronger for measurements on ensembles of structures since
one obtains a $(\kf a) \cl$ dependence for integrable systems and
no dependence on $(\kf a)$ for the chaotic ones.
 The same parametric dependences are 
obtained for the persistent currents in integrable and chaotic
multiply-connected geometries. 
Therefore, the nature
of the dynamics yields an order-of-magnitude difference in the 
magnetic response of integrable and chaotic systems, which should be 
easy to observe experimentally (especially for ensemble measurements).
Finally, for systems with mixed dynamics, for which the phase-space
is characterized  by the coexistence of regular and chaotic
motion, the magnetic response should be dominated by the nearly integrable 
regions of phase-space.  This gives rise to a  $(\kf a )^{3/2} \cl$ 
dependence for the typical susceptibility as long as some families
of periodic orbits remain sufficiently unperturbed. The precise calculation
of the prefactor may however present some complications that we have not 
considered here (the general semiclassical treatment of 
mixed systems remains an open problem) and should 
depend on the fraction of phase-space being integrable.

The semiclassical approach we are using not only allows a global
understanding of the magnetic response of ballistic devices,
but also provides precise predictions when specific systems 
are considered. 
The detailed comparison between exact quantum calculations
and semiclassical results for the square geometry 
demonstrates indeed that the semiclassical predictions are
extremely accurate.  This has been shown in section~\ref{sec:square} 
for weak fields, such that the trajectories are essentially unaffected
by the magnetic field, and also in section~\ref{sec:highB}
for fields large enough to yield a cyclotron radius of the
order of the typical size of the structure (where the bending of the
classical trajectories has to be taken into account).  For intermediate
fields we have identified a new regime where the magnetic susceptibility
is dominated by bouncing-ball trajectories that alternate between 
opposite sides of the structure (enclosing flux due to their bending).
For high fields the electrons move on cyclotron orbits and we have
recovered the de Haas~-~van Alphen oscillations (with finite-size 
corrections that we calculated semiclassically).

In order to understand the success of the semiclassical approach, it 
should be kept in mind
that the lack of translational invariance characteristic for
the ballistic regime, where the shape of the device 
plays an important role, complicates the application of other
approximation schemes as, e.g., diagrammatic expansions.  Therefore,
except for very specific cases where exact quantum calculations
are possible, and unless one is satisfied by direct numerical 
calculations,
 some semiclassical ideas have to be implemented to deal with such
problems. Moreover, from a more practical point of view, the 
semiclassical trace 
formalism we have used appears perfectly adapted to deal
with thermodynamic quantities such as the Grand
Potential $\Omega(\mu)$ or its first and second derivatives
$N(\mu)$ and $D(\mu)$.  Indeed, the beauty of this approach is
that the oscillating part of the density of states
is directly  expressed in terms of Fourier-like components, each of which
is associated with a periodic (or nearly periodic) orbit.  The 
thermodynamic properties are obtained from their  purely quantal 
(or zero temperature) analogues $\omega$, $n$ and $d$ by temperature 
smoothing, which merely amounts to  multiply each oscillating 
component by a temperature-dependent damping factor. For all fields 
(high, intermediate, or weak),  this factor depends
only on the ratio of the {\em period} $\tau$ of the 
corresponding orbit and the temperature-dependent 
cutoff time $\tau_c = \beta \hbar / \pi$
and suppresses exponentially the contribution of orbits with period
longer than $\tau_c$.
As a consequence, not only the effect of temperature is taken into
account in an intuitive transparent way, but in addition only the shortest
periodic orbits have to be considered in the semiclassical
expansion.  All the problems concerning the convergence of trace formulae and
the validity of semiclassical propagation of the wave function
for very long times are of no importance here.   One therefore avoids
most of the problems which plague the field of quantum chaos
when semiclassical trace formulae are used to resolve the spectrum
on a mean-spacing scale. Mesoscopic physics is usually
concerned with the properties of the spectrum on an energy-scale large
compared to the mean-spacing.  In the spirit of
the work of Balian and Bloch \cite{bal69}, this is the situation
for which the semiclassical trace formalism is especially appropriate. 

Having stressed the success of the semiclassical approach in dealing
with our model of non-interacting electrons evolving in a clean 
medium, it is worthwhile to consider in more detail
how the above picture should be modified when going closer to the real
world, and incorporating the effects of residual disorder, 
electron-electron or electron-phonon interactions.
As stressed in the introduction, the first of these points is 
relatively harmless because of finite temperature smoothing.
The restriction to short periodic orbits actually justifies an approach to
the ballistic regime using a model for clean systems since long diffusive 
trajectories do not contribute to the finite-temperature susceptibility.
Indeed, careful numerical and semiclassical studies of the effect of small
residual disorder \cite{rod2000} show that, except for a possible
reduction of the magnetic response, the above description of the orbital
magnetism of ballistic systems remains essentially unaltered. In
particular, the mechanism proposed by Gefen et al.\ \cite{gef94}
is not borne out by the numerical simulations at the temperatures of
experimental relevance.
For smooth disorder, such as  presumably prevails in the 
systems of Refs.~\cite{levy93} and \cite{BenMailly}, 
the magnetic response is decreased by the
dephasing of nearby trajectories in a way that depends on its strength
and the ratio between the correlation length and the size of the
structure \cite{rod2000}, but diffusive trajectories can be seen to
be absolutely irrelevant if the elastic mean free path is larger than
the size of the structure.  The precise knowledge of this reduction is
however needed in order to make a decisive comparison with the experimental
results of Ref.~\cite{levy93}.

At the low temperatures of the experiments the inelastic mean free path
of the electrons is much larger than the system size since electron-phonon
interactions are suppressed. On the other hand,  
the effect of electron-electron 
interactions on the magnetic response is a much more controversial
point. In particular, it has been invoked to be the necessary mechanism 
to obtain the measured values \cite{inter}  for the problem of persistent 
currents in disorder metals. In a first approximation to the
experimental conditions that we investigated in this work we would 
infer that 
electron-electron interactions are not crucial since the screening
length is much smaller than the size of the samples and since the
2-d renormalization of the effective mass at these electron densities
is only about 11\% \cite{JDS}. Clearly the two previous criteria 
will not be satisfied in smaller structures, and the possibility
that electron-electron interactions express themselves through
a mechanism for which these estimates are not relevant remains
open even in the experimental realizations we consider.

Contrarily to the effect of disorder, which can be implemented
within a semiclassical framework without essential difficulties,
a semiclassical treatment of the electron-electron interaction
still remains an open problem.  However, the genuine effects 
that we have found within our semiclassical approach for the
clean model of non--interacting electrons should prevail in more
sophisticated theories. We think that the rich variety of
possible experimental configurations for ballistic devices 
(the shape and the size can nowadays be chosen at will)
provides an ideal testing ground for these more complete approaches.
We hope that the work presented here will stimulate 
experimental and theoretical activity addressing the magnetic 
response of ballistic microstructures.

\section*{Acknowledgments}

We acknowledge helpful discussions with H.~Baranger, O.~Bohigas, Y.~Gefen,
M.~Gutzwiller, L.~L\'evy, F.~von~Oppen, N.~Pavloff, B.~Shapiro, and
H.~Weidenm\"{u}ller. We are particularly indebted to H.~Baranger for
continuous support and a careful reading of the manuscript, and to
O.~Bohigas for forcing us not to stop until getting to the
bones of the problem. We thank B.~Mehlig for communicating us Ref. \cite{Kubo64}.
 KR and RAJ acknowledge support from the 
``Coop\'eration CNRS/DFG" (EB/EUR-94/41).
KR thanks the A. von Humboldt foundation for financial support.
The Division de Physique Th\'eorique is ``Unit\'e de recherche des
Universit\'es Paris~11 et Paris~6 associ\'ee au C.N.R.S.''.

\newpage

\appendix

\section{Convolution of a rapidly oscillating function with the
derivative of the Fermi function}
\label{app:convolution}

When considering thermodynamic quantities related to the oscillating part
of the density of states at finite temperature $T$, one has to evaluate 
integrals of the form
	\begin{equation} \label{convolution}
	I(T) = \int_0^\infty \dif E \,
		A(E) \exp \left[ \frac{i}{\hbar} S(E) \right]
		f' (E-\mu) \; ,
	\end{equation}
where $f'(E-\mu)$ is the derivative of the Fermi function
	\[ f(E-\mu) = \frac{1}{1+\exp\beta(E-\mu)} \ , \]
and $\beta=1/k_BT$. 
The rapidely oscillating function
$A(E) \exp \left[ \frac{i}{\hbar} S(E) \right]$ usually originates from the
contribution of a classical orbit (or a family of orbits) to the oscillating
part of the density of states.  In this case $S(E)$ is the action integral
along the orbit, and its derivative $\dif S / \dif E \equiv  \tau(E)$ is
the period of the orbit.

At zero temperature $f'= - \delta(E-\mu)$ giving for
$I_0 \equiv I({T\!=\!0})$
	\begin{equation} \label{I_0}
	I_0 = - A(\mu) 	\exp \left[ \frac{i}{\hbar} S(\mu) \right] \; .
	\end{equation}
In this appendix we show that, to leading order in $\hbar$ and in
$\beta^{-1}$ (but without making any assumption concerning their relative
value), the integral of Eq.~(\ref{convolution}) is given by
	\begin{equation} \label{I_beta}
	I(T) = I_0 R_T(\tau)
	\end{equation}
with the temperature dependence
	\begin{equation} \label{R_factor1}
	R_T(\tau) = \frac{\tau(\mu)/\tau_c}{\sinh (\tau(\mu)/\tau_c)}
	\qquad \qquad 	\tau_c = \frac{\beta\hbar}{\pi} \; .
	\end{equation}
For systems without potential, i.e.\ free particles confined
in a box (billiards), the period of the trajectory is related to its length $L$
by $\tau(\mu) = L / \vf$, where $\vf = \hbar \kf / m$ is the Fermi 
velocity.   $R_T$ can then be written as
	\begin{equation} \label{R_factor2}
	R_T(L) = \frac{ L/L_c}{\sinh (L/L_c)}
	\qquad ;  \qquad 	
	L_c = \frac{\hbar \vf \beta}{\pi}  \; .
	\end{equation}
In the case of unconfined free particles, the formulae
(\ref{R_factor1}) and (\ref{R_factor2}) are 
equivalent to the usual form of the temperature
dependence of the de Haas--van Alphen effect Eq~(\ref{R:dHvA}) given in
the introduction.  Below we present 
a slight variation of a standard calculation (see
e.g.~\cite{LanLip}) of the
temperature dependence of the de Haas--van Alphen effect, which
generalizes it to any type of dynamics, once we caste it in the
form of Eq.~(\ref{R_factor1}).

Performing the integral (\ref{convolution}) along the contour shown in
Fig.~\ref{contour} and noting
that the singularities of the derivative of the Fermi function are
double poles located at $E_k = \mu + i(2k+1)\pi/\beta$ $\quad
(k= 0,\pm 1, \pm 2, \ldots)$ with a coefficient $1/\beta$, one finds the
following relation 

        \begin{eqnarray}
        &&\int_0^\infty \dif E \,
                A(E) \exp \left[ \frac{i}{\hbar} S(E) \right]
                f' (E-\mu)
        -
        \int_{0 + i 2\pi/\beta}^{\infty + i 2\pi/\beta} \dif E \,
                A(E) \exp \left[ \frac{i}{\hbar} S(E) \right]
                f' (E-\mu) \nonumber \\
        & = &
        \frac{2 i \pi}{\beta} \frac{i \tau(E_1)}{\hbar}
        A(E_1) \exp \left[ \frac{i}{\hbar} S(E_1)
        \right] \; .  \label{contour1}
        \end{eqnarray}

At low temperatures, the function $f'(E-\mu)$ is essentially zero everywhere
in the complex plane, except for
a narrow band of width $\beta^{-1}$ near the line ${\rm Re}(E) = \mu$, 
therefore the vertical portions of the contour ($E=0$ and $E\gg\mu$) 
give negligible
contributions. Noticing that $f'(E-\mu)$ has a periodicity of $2i\pi/\beta$
we can ignore the complex part of $E$ in the factor $f'$ of the second
integral. Finally, since ${\rm Im} (E_1)=\pi/\beta$ and ${\rm Im} (E)=2\pi/\beta$ 
along the upper portion, we can evaluate the prefactors at $\mu$ and 
expand the actions (which are multiplied by $1/\hbar$) as $S(E) = S(\mu) + 
\tau(\mu)(E - \mu)$ in leading order in $\beta^{-1}$ and $\hbar$,
obtaining
        \[ I(T)
        \left(1 -  \exp\left[-\frac{2\pi\tau(\mu)}{\beta\hbar}\right] \right)
        =
        - \frac{2 \pi \tau(\mu)}{\beta \hbar}
        A(\mu) \exp \left[ \frac{i}{\hbar} S(\mu)
                - \frac{\pi\tau(\mu)}{\beta \hbar} \right] \; .\]
That is,
        \[ I(T)
        \left(1 -  \exp\left[-\frac{2\tau(\mu)}{\tau_c}\right] \right)
        =
        I_0  \frac{2 \tau(\mu)}{\tau_c}
        \exp \left[ - \frac{\tau(\mu)}{\tau_c} \right] \; ,\]
from which one readily obtains the result of Eq.~(\ref{I_beta}).

\subsubsection*{Further comments}

We would like to use the above calculation to motivate 
some choices made in section~\ref{sec:TherFor} which might have appeared
rather arbitrary. Concerning for instance the grand potential 
$\Omega(\mu)$ two equivalent expressions have been introduced: The
usual Eq.~(\ref{eq:therpot}) and  Eq.~(\ref{smoothc}) which is 
obtained from integration by parts. On the other hand, we have used only 
   \be \label{right_Oosc}
   \Oosc (\mu)  =  - \int_0^{\infty} \dif E \  \oosc(E) \ f'(E-\mu) 
   \ee
as the ``operational'' definition of the oscillating part $\Oosc$ 
of $\Omega$, and one might wonder whether an integral analogue to 
the one of Eq.~(\ref{eq:therpot}) like
	\begin{equation} \label{wrong_Oosc}
	- \int \dosc(E) f^{(-1)} (E-\mu) \dif E \; ,
	\end{equation}
(where  $f^{(-1)}(E-\mu) = \ln{(1+\exp{[\beta(\mu\!-\!E)]})}/\beta$ is the
primitive of the Fermi function) could not be used as well.
This is not the case for the two following reasons:
\begin{itemize}
\item[i)] First, the oscillating functions $\dosc$ or $\oosc$ are usually
obtained in a semiclassical approach and are therefore valid only
for large energies.  If the chemical potential $\mu$ is in the semiclassical
regime and $\beta^{-1} \ll \mu$, which is always the case for the problems
we consider, only the neighborhood of $\mu$ in which $\oosc$ can be used
safely, contributes significantly to the integrals of
Eq.~(\ref{right_Oosc}).  On the contrary, the
integral~(\ref{wrong_Oosc}) involves energies close to zero. Therefore
there is no reason that $\dosc$ is accurate, being quite often a diverging
function.
\item[ii)] In addition, even if one has at hand an equation
as (\ref{DOSLL}) which is a non--semiclassically exact expression,
the integrals of Eq.~(\ref{right_Oosc})  and Eq.~(\ref{wrong_Oosc})
are, strictly speaking, not equivalent.  The latter form contains some
boundary terms not present in the former, which obviously have 
to be removed from $\Oosc$ since they do not average to zero under
 a local smoothing. 
\end{itemize}

In a semiclassical treatment
the derivative $f'$ of the Fermi function is superior 
to any of its integrated versions since it is significant only at energies
where semiclassical approximations can be used safely.

 \section{Semiclassical expansion of the mean density of states}
 \label{app:wigner}

In this appendix we calculate the first two terms in an $\hbar$
expansion of the smooth part of the density of states $\bar d(E)$.
We follow the standard approach introduced by Wigner
\cite{wig32} using the notion of the Wigner transform of an operator.
The Wigner transform of a quantum operator $\hat {\cal O}$ 
is defined by
	\be \label{wigner}
	[{\cal O}]_{\cal W} (\bq,\bp)  	 =
	\int \dif {\bf x} \, e^{-i \bp \cdot {\bf x} / \hbar}
            \langle{ \bq + {{\bf x}\over2} \vert \hat{\cal O} \vert
                        \bq - {{\bf x}\over 2} }\rangle \; .
	\ee
Among different properties of the Wigner transform we will essentially make
use of the following two: First, the trace of an operator
is related to the integral over phase space of its transform by means of
	\be \label{eq:trace}
	{\rm Tr} (\hat{\cal O}) = \frac{1}{(2\pi \hbar )^d}
	\int \dif \bq \dif \bp\ [{\cal O}]_{\cal W} (\bq , \bp) \; .
	\ee
(We stress that this is an exact, not semiclassical, relation.)
Secondly, for any operator function of the position and momentum quantum
operators ${\cal F}(\hat \bq, \hat \bp)$ 
(with some specified ordering), the semiclassical leading order
approximation to its Wigner transform is just the related classical function,
that is
	\be \label{leading_wigner}
	[{\cal F}(\hat \bq, \hat \bp)]_{\cal W} = {\cal F}(\bq, \bp)  + O(\hbar) \; .
	\ee
When $\cal F$ depends only on $\hat \bq$ or $\hat \bp$, the relation between
the Wigner transform and the classical function is exact (no corrective terms in
$\hbar$), as can be directly
checked from Eq.~(\ref{wigner}).

We will follow closely the presentation of Ref.~\cite{boh93} to which
the reader is referred to for further details.  The first step in the calculation
of $\bar d(E)$ is to consider the Laplace transform of the level density,
Eq.~(\ref{partition}), which due to the property (\ref{eq:trace}) can be written as
	\be \label{part}
	Z(\lambda )  =
	\frac{{\gs}}{(2\pi\hbar )^d}
	\int \dif \bq \dif \bp \, [ e^{-\lambda \Hch} ]_{\cal W} (\bq,\bp)
	\ee
(${\sf g_s}=2$ is the spin degeneracy factor).
Using Eq.~(\ref{leading_wigner}), the leading order [Weyl] term in $\hbar$,
$ Z_{\rm W}$, is obtained by replacing
$[ e^{-\lambda \Hch} ]_{\cal W} (\bq,\bp)$ by $e^{-\lambda {\cal H} (\bq,\bp)}$,
where
	\be \label{classical_H}
	{\cal H} = \frac{1}{2m} \ \left(\bp - \frac{e}{c} \bA\right)^2 \ + \ V(\bq) 
	\ee
 is the classical Hamiltonian.  At this level of approximation
$Z(\lambda)$ is given by
	\be \label{ZW}
	Z_{\rm W}(\lambda) =
	\frac{{\sf g_s}}{(2\pi \hbar )^d} \int \dif \bq \dif \bp \,
	\exp{\left( -\lambda \left[\frac{(\bp - \frac{e}{c} \bA)^2}{2m} + V(\bq)
	\right]\right)} \ . \ee
Since this term is field independent (see the change of variable
(\ref{change}) in the text) we need to go to the next order in $\lambda$ in
order to obtain non-vanishing contributions to the magnetic susceptibility.
Therefore we consider the asymptotic
semiclassical expansion
	\be \label{expansion}
	[ e^{-\lambda \Hch} ]_{\cal W} (\bq,\bp) =
	  e^{-\lambda [\Hch]_{\cal W} (\bq,\bp)}
	\sum_{n=0}^\infty \left ( {{-\hbar^2}\over 4} \right) ^n
	{{C_n}(\bq,\bp, \lambda) \over{(2n)!}} \ .
	\ee
We have already seen that $C_0 = 1$. The following coefficients $C_n$ can be
obtained recursively by grouping terms according to their power in $\hbar$.
In particular, the first coefficient is given by \cite{wig32,boh93}
	\be \label{A1}
	\frac{\partial C_1}{\partial \lambda} = 
	- e^{\lambda [\Hch]_{\cal W}}
	\left( [\Hch]_{\cal W} \ {\stackrel{\leftrightarrow}{\Lambda}}^2 \
	e^{-\lambda [\Hch]_{\cal W}}
	\right) \ , \ee
where $\displaystyle \stackrel{\leftrightarrow}{\Lambda} =
\sum_{i=1}^{d} \stackrel{\leftarrow}{\partial q_i}
\stackrel{\rightarrow}{\partial p_i} - \stackrel{\leftarrow}{\partial p_i}
\stackrel{\rightarrow}{\partial q_i}$ is the Moyal bracket.

Inserting the classical Hamiltonian (\ref{classical_H}) at the place of
$[\Hch]_{\cal W}$ we obtain $C_1$ as the sum of three terms
	\[C_1 = (C_1^0) + (C_1^{\rm a}) + (C_1^{\rm s}) \; , \]
where
	\be
	(C_1^0) = {1 \over m}
	\left[ \lambda^2 \nabla^2 V(\bq) - {\lambda^3 \over 3m}
	\left( m \nabla V(\bq) \cdot \nabla V(\bq) +
	\left( (\bp - {\scriptstyle \frac{e}{c}} \bA) \cdot {\nabla} \right )^2
	V(\bq) \right ) \right]
	\ee
is, up to the change of variable Eq.~(\ref{change}) in the integration over the
phase space, the first order correction without magnetic field
$C_1({\scriptstyle H\!=\!0})$ given in \cite{wig32}.
$(C_1^{\rm a})$ is given by terms being antisymmetric in $\bp$ that vanish
when taking the trace over phase space. Finally,
	\begin{eqnarray}
	(C_1^{\rm s}) & = & \frac{e^2}{c^2} \sum_{i,k=1}^{d} \ \left\{
	\frac{\lambda^2}{m^2} \
	\left[ (\partial q_i A_k)(\partial q_i A_k) -
	(\partial q_i A_k)(\partial q_k A_i) \right]
		+ \right. \nonumber \\
		& +  & \left.
	\frac{2 \lambda^3}{3 m^3} \ \left[
	(\partial q_i A_k)(\partial q_k A_i)
	(p_i- {\scriptstyle \frac{e}{c}}A_i)^2 -
	(\partial q_i A_k)(\partial q_i A_k)
	\frac{(p_k- {\scriptstyle \frac{e}{c}} A_k)^2+
	(p_i- {\scriptstyle \frac{e}{c}} A_i)^2}{2} \right]
	\right\}  \; .
	\label{C_1inter}
	\end{eqnarray}
The first correction to the Laplace transform of the density of states
is given by
	\be \label{Z1}
	Z_1(\lambda) =  \frac{{- \sf g_s}}{(2\pi \hbar )^d} 
	        \int \dif \bq \dif \bp \,
		\left[ \lambda^2 \frac{\mu_B^2 H^2}{6}
		+ \frac{\hbar^2}{8}C_1^0 \right]
		e^{-\lambda {\cal H}} 
	\ee
($\mu_B = (e\hbar) / (2mc)$ is the Bohr magneton).
Eqs.~(\ref{Z1}) and (\ref{Z_1}) are obtained from
Eq.~(\ref{C_1inter}) by using the identity
	\be
	H^2 = \sum_{jk} \left[ 
	\left(  \frac{\partial A_j}{\partial q_k} \right)^2 -
	 \frac{\partial A_j}{\partial q_k}
	\frac{\partial A_k}{\partial q_j} \right]
	\ee
and a few transformations that leave the integral
over $\bp$ unchanged, namely :
(i) the change of variables Eq.~(\ref{change}) (allowing the substitution
of ${\cal H}$ by ${\cal H}^0  = \bp^2/2m+V(\bq)$)
(ii) the
elimination of all terms antisymmetric in $\bp$, (iii) the replacement
of all terms of the form $p_i^2 e^{-\lambda \bp^2 / 2 m}$ by
$(m/\lambda)e^{-\lambda \bp^2 / 2 m}$.
Note finally that the field appears only in the term
$-\lambda^2 \mu_B^2 H^2 / 6$, which is independent of the confining
potential $V(\bq)$.  This is at the root of the very general applicability
of the Landau result.

\section{Calculation of ${\em \lowercase{g}_E}$ for a ring billiard}
 \label{app:ring_g}

In this appendix we derive the explicit form $I_2=g_E(I_1)$ of the 
energy surface $E$ in action space for a ring geometry. The calculation
reduces to the evaluation of the integral of Eq.~(\ref{eq:action}) along
two independent paths on the invariant torus. The only subtlety arises
from the difficulty of visualizing the integration paths in our 
four-dimensional phase-space where the tori are discontinuous due to the
presence of hard walls. We closely follow the procedure used by Keller
and Rubinow \cite{keller} for the circular billiard, and we refer to this
work for further details. 

In a circular ring (with outer and inner radii $a$ and $b$) we can
distinguish two types of periodic trajectories: those which 
do not touch the
inner disk (type-I, Fig.~\ref{fig:circle}a) and those which do hit it 
(type-II, Fig.~\ref{fig:circle}b).
Type-I trajectories have their caustics outside the inner
disk and therefore they are unaffected by their presence.
They have an angular momentum $pc$, with $b<c<a$ \ ($p=\sqrt{2mE}$). 
Taking as the integration path ${\cal C}_1$ the
concentric circle of radius
$R$, we have $\bp\, \dif \bq = (pc/R) \dif q$ and then
	\be \label{eq:action1}
        I_1 = \frac{1}{2\pi} \oint_{{\cal C}_1} \frac{c}{R} \ p \, \dif q
        = p \ c \; \ .
        \ee
The action variable $I_1$ is just the angular momentum. The straight
part of the path ${\cal C}_2$ of Fig.~\ref{fig:circle}a is chosen along 
a classical 
trajectory, where $\bp$ is constant and collinear with $\dif \bq$. For the
part along the outer circle $\bp \dif \bq = - (pc/a) \dif q$.
Combining both contributions we have
 	\be \label{eq:action3}
        I^{(I)}_2 = \frac{p}{\pi} \left\{ \left[a^2-c^2\right]^{1/2} - \
        c \ \arccos\left(\frac{c}{a}\right) \right\} \ .
        \ee
Elimination $c$ between (\ref{eq:action1}) and (\ref{eq:action3}) leads
to Eq.~(\ref{circle:gE}) of the text, valid for the description
in action space of the energy surface of the circular billiard 
\cite{keller} and the energy surface associated with type-I trajectories 
in the ring billiard ($pb<I_1<pa$). We have chosen the integration paths
for type-I trajectories different from those of Ref.~\cite{keller}
because slight modifications of them are applicable for 
type-II trajectories.
 
Type-II trajectories have their caustics in the interior of the inner 
disk, that is, they have an angular momentum $pc$, with $c<b$.
Integration along the path ${\cal C}_1$ of Fig.~\ref{fig:circle}b
leads to the identification of $I_1$ with the angular momentum 
$pc$ (similarly to Eq.~(\ref{eq:action1})). By choosing 
the path ${\cal C}_2$ as shown on Fig.~\ref{fig:circle}b,
the action integral along this path is simply the 
difference $I^{(I)}_2(a) - I^{(I)}_2(b)$ (where both terms
are given by Eq.~(\ref{eq:action3}), except that for the second
$a$ should be replaced by $b$). This yields
        \be \label{eq:action2}
        I^{(II)}_2 = \frac{p}{\pi} \left\{\left[a^2-c^2\right]^{1/2}  \ -
 	\ \left[b^2-c^2\right]^{1/2} \ - \
       c \left[ \arccos\left(\frac{c}{a}\right)  \ - \
       \arccos\left(\frac{c}{b}\right) \right] \right\}   \ .
        \ee 
Eliminating $c$ between (\ref{eq:action1}) and (\ref{eq:action2}) 
leads to Eq.~(\ref{ring:gE}) of the text.
 
\section{Calculation of the determinant $D_\bM$ at zero field for
a generic integrable system}
\label{app:D_M}

In the semiclassical approximation of the Green function
(Eq.~(\ref{eq:green})) the amplitude $D_t$ associated with a classical
trajectory $t$ is given by \cite{gutz_book}
	
	\be \label{def:D}
	D_t = \left| \ba[tb]{cc}
	\displaystyle \frac{\partial^2 S_t}{\partial \bq \partial \bq'} &
	\displaystyle \frac{\partial^2 S_t}{\partial \bq \partial E} \\
	& \\
	\displaystyle \frac{\partial^2 S_t}{\partial E \partial \bq'} &
	\displaystyle \frac{\partial^2 S_t}{\partial E \partial E}
		\ea \right|^{1/2}
        =
	\frac{1}{\left| \dot{q}_2 \dot{q}'_2 \right|^{1/2}}
	\left|- \frac{\partial^2 S}{\partial q_1 \partial q'_1} 
		\right|^{1/2}
     	\; .
	\ee
Note the second equality holds not only  when $q_2$ is taken
along the orbit and
$q_1$ in the perpendicular  direction, as supposed by Gutzwiller
in its original derivation \cite{gut71}, but also, as shown by
Littlejohn, in any coordinate system
(see Sec.~III.C in Ref.~\cite{little90} and Sec.~III in \cite{cre90}).
Although a priori $q_1$ and $q_2$ play a similar role,
their non--symmetrical appearance on the right hand side of Eq.~(\ref{def:D})
($q_1$ and $q_2$ can be exchanged without affecting
the value of $D_\bM$) is due to the fact that one coordinate
(here $q_1$) is chosen as a Poincar\'e surface of section,
and the dependence of  the other coordinate (here $q_2$)
just expresses the conservation of energy.

Turning now to the particular problem we are concerned with, i.e.\ an integrable
system at zero field and the diagonal part of the Green function, the above
Eq.~(\ref{def:D}) applies to $D_\bM$ (except for a change 
$t \rightarrow \bM$ in the label of the orbits).  Moreover,
the measure $D_\bM \dif q_1 \dif q_2$ in Eq.~(\ref{contribution}) is 
invariant under the transformation $(q_1,q_2) \rightarrow (\theta_1,\theta_2)$ 
at zero magnetic field (see Ref.~\cite{gri95} for
a more detailed discussion of this point).   In other words, noting 
in Eq.~(\ref{def:D}) $D_\bM(\bq)$ the determinant in the original
$\bq$ coordinate and  $D_\bM(\theta)$ the determinant defined in the same
way but in the system of coordinates given by Eq.~(\ref{canonical_transform}),
one has $D_\bM(\bq) dq_1 dq_2 = D_\bM(\theta) d\theta_1 d\theta_2$.
Therefore
	\be \label{Dm2}
	D_\bsM (\bq) \ \left| \left(\frac{\partial \bq}{\partial \theta}
		\right) \right|
	= \frac{1}{\left| \dot{\theta_2} \dot{\theta_2'}
						\right|^{1/2}}
	\left| - \frac{\partial^2 S}{\partial \theta_1 \partial \theta_1'}
      	\right|^{1/2}
	=
	\frac{1}{\dot{\theta}_2}
	\left| \left( \frac{\partial \theta_1'}{\partial J_1}
              \right)_{\theta_1} \right|^{-1/2} \; ,
	\ee
where the derivatives have to be taken at $E$, $\theta_2$,
and $\theta_2' = \theta_2 + 2\pi M_2$ constant. To compute the r.h.s.\
of Eq.~(\ref{Dm2}) one just needs the expression of the Poincar\'e mapping
$(\theta_1,J_1) \rightarrow (\theta_1',J_1') $
between the two ($\theta_2 = {\rm const.}$) Poincar\'e surfaces of section.
Since the motion is integrable, $J_1'=J_1$, and from 
Eq.~(\ref{canonical_transform}) we obtain
 	\be \label{mapping}
	\theta_1' (J_1,\theta_1) =
	\theta_1 + 2\pi r u_2^2 (\alpha(J_1) - u_1/u_2) \; ,
	\ee
where $\alpha(J_1)$ is the winding number of the torus labeled by $J_1$.
Thus
	\be
	\left( \frac{\partial \theta_1'}{\partial J_1} \right)_{\theta_1}
	= 2 \pi r u_2^2
	\frac{d \alpha}{d J_1} \; .
	\ee
We recall that the function $g_E$ introduced in
section~\ref{sec:integrable} is defined by the implicit relation
$H(I_1,\, I_2 \!=\! g_E(I_1)) = E$,
which yields after differentiation  $\dif g_E/ \dif I_1 = -\alpha$.
Therefore
	\be
	\frac{d \alpha}{d J_1} =
	u_2 \frac{d \alpha}{d I_1} =
	-u_2 \frac{\dif^2 g_E}{\dif I_1^2} \; ,
	\ee
from which one finally obtains
	\be
	D_\bsM \
	\left| \left( \frac{\partial \bq}{\partial \theta} \right) \right|
	= \frac{1}{\dot{\theta}_2}
	   	\frac{1}{ \left| 2 \pi r u_2^3 g_E^{''} \right|^{1/2}} \; .
	\ee
$D_\bsM$ is inversely proportional to the square root
of the curvature of the line
$H(I_1,I_2) = E = {\rm const.}$ and independent of $\theta$.

\section{Diagonal part of the Green function for a free electron
in a constant magnetic field}
\label{app:highB}

In this section, we calculate semiclassically the diagonal part of the Green
function $G(\br,\br)$ for a free electron moving in a plane in a perpendicular
magnetic field.  The resulting classical cyclotron motion is extremely simple,
but yields slight complications in the semiclassical evaluation
of the diagonal part of the Green function because all trajectories
starting at some point $\br$ are refocused precisely at $\br$.
The calculation of the prefactors deserves special attention but
can be done using a slight variation of the standard techniques and
yields for unconstrained systems the usual result Eq.~(\ref{intro:susdHvA}).
In addition to provide an alternative (semiclassical) derivation
of the de Haas--van Alphen effect, our procedure allows to
compute correctly the contribution of the cyclotron orbits for billiard
systems, i.e.~it takes into account the corrections due to the boundaries
which appear to be necessary if the cyclotron radius $r_c$ is not small compared to 
the typical dimension $a$ of the system.

\subsubsection{Computation of the prefactor of a Green function near a focal
point}

It is an old problem to obtain a correct semiclassical solution of wave 
equations valid also near turning points, focal points, caustics, etc., where the usual
expressions are diverging. A general solution for this problem can
for instance be found in the book of Maslov \cite{maslov}.
In this subsection, we will give the explicit form of this general theory
when applied to the calculation of a two--dimensional
Green function and consider in the next subsection the particular
problem of a free electron in a constant magnetic field.
To avoid confusion we will slightly modify our usual notations, writing
$G(\br|\br')$ instead of $G(\br,\br')$. In addition, we will make more
explicit what are the initial and final points by using
$\br^i = (x^i,y^i)$ for the initial (source) point and,
and $\br^f = (x^f,y^f)$  for the final (observation) point.
As already stated, the semiclassical evaluation of the
Green function $G(\br^i|\br^f)$ yields a sum over
all classical trajectories $t$ joining $\br^i$ to $\br^f$ at energy
$E$
	\be \label{sum_t}
	G(\br^i|\br^f) = \sum_t G_t(\br^i|\br^f) \; .
	\ee
For a trajectory $t$ starting at $\br^i$  such that $\br^f$ {\em is not a focal point},
one can use (cf.~Appendix~\ref{app:D_M},
and in particular the discussion concerning the non--symmetric role of $x$ and
$y$)
	\begin{eqnarray}
	G_t(\br^i|\br^f) & =  &
            \frac{1}{i \hbar} \frac{1}{\sqrt{ 2 i \pi \hbar}}
	       D_t  \exp{\left[\frac{i}{\hbar} S_t -
                            \eta_t\frac{\pi}{2}\right]} \ , \label{Gt}\\
	D_t & = & \frac{1}{\left| \dot{y}^i \dot{y}^f \right|^{1/2}}
	\left| \frac{\partial^2 S}{\partial x^i \partial x^f} \right|^{1/2}
	\; . \label{Dt}
	\end{eqnarray}
However, the above expression is not valid near focal points where [locally,
and at fixed $y^f$] $x^f$ becomes independent of $p_x^i$.
The use of the action integral $S(\br^i,\br^f)$ supposes that $\br^i$ and $\br^f$ can be
taken as independent variables, and ${\partial^2 S}/{\partial x^i \partial x^f}$ 
is a priori not meaningful since $x^f$ is entirely determined by $x^i$.  
Writing
${\partial^2 S}/{\partial x^i \partial x^f} = -(\partial x^f / \partial
p_x^i)^{-1} = -\infty$ one sees moreover  that $D_t$ is, as 
mentioned above, in fact diverging.

To overcome this difficulty Maslov proposed a procedure 
to compute $G_t(\br^i|\br^f)$ using a momentum (or mixed
position/momentum) representation, by defining  (omitting
for a moment the source point $\br^i$)
	\begin{equation} \label{alt_Gt}
	G_t(x^f,y^f) = {\cal F}^{-1}_{p_x^f \rightarrow x^f}
		[ \tilde G_t(p_x^f,y^f) ] \; ,
	\end{equation}
where ${\cal F}^{-1}_{p_x^f \rightarrow x^f}$ is the inverse Fourier
transform
	\begin{equation} \label{fourier}
	{\cal F}^{-1}_{p_x^f \rightarrow x^f} [\cdot]
 	= \frac{1}{\sqrt{-2i\pi \hbar}} \int d p_x^f [\cdot]
	\exp \left(\frac{i}{\hbar} x^f p_x^f \right) \cdot
	\end{equation}
performing quantum mechanically the change
from the mixed representation $(p_x,y)$ to the position
representation $(x,y)$.

Eq.~(\ref{alt_Gt}) is just the definition
of $\tilde G_t$ which remains to be evaluated semiclassically.
The general theory presented in Ref.~\cite{maslov} (Section 5.1)
can be however applied to our problem, giving

	\be \label{tGt}
  	\tilde G_t(p_x^f,y^f)  =
	\frac{1}{i \hbar} \frac{1}{\sqrt{ 2 i \pi \hbar}}
	\tilde D_t  \exp{\left[\frac{i}{\hbar} \tilde S_t -
                            \tilde \eta_t \frac{\pi}{2}\right]}
	\ee
where
 	\begin{eqnarray}
	\tilde S & = & S - p_x^f x^f \label{tSt}  \; ,\\
	\nonumber \\
	\tilde D_t & = & D_t
	\left|\frac{\partial x^f}{\partial p_x^f}
			\right|^{1/2}_{\br^i = {\rm const.}}
	\label{tDt} \;, \\
	\nonumber \\
	\tilde \eta_t  & = &  \left\{ \ba{r}
	 \eta_t   \quad {\rm if} \quad \partial p_x^f / \partial x^f > 0 \\
	 \eta_t + 1 \quad {\rm if} \quad \partial p_x^f / \partial x^f < 0
			\ea \right. \; .\label{t_eta}
	\end{eqnarray}
Without entering into a derivation of this semiclassical formula for
$\tilde G_t$, it can be checked that starting from
Eq.~(\ref{tGt}) the evaluation of  the inverse Fourier
transform Eq.~(\ref{alt_Gt}) using stationary phase approximation
readily yields Eq.~(\ref{Gt}).  Far from any  focal point
both expressions are equivalent at the semiclassical
level.  Near a focal point however, Eq.~(\ref{tGt}) still provides an accurate
approximation for $\tilde G_t$ because the Lagrangian
manifold, on which the Green function is constructed semiclassically, has
a non singular projection onto the plane $(p_x,y)$.  Therefore,
contrarily to Eq.~(\ref{Gt}) which is diverging, Eq.~(\ref{alt_Gt})
is still a valid semiclassical approximation for $G_t$, provided
the inverse Fourier transformation {\em is evaluated exactly}
(or using uniform techniques going beyond stationary point approximation
\cite{ozor87}).

>From Eqs.~(\ref{Dt}) and (\ref{tDt}) one has
       	\begin{equation} \label{tDt2}
	\tilde D_t = 	\frac{1}{\left| \dot{y}^i \dot{y}^f \right|^{1/2}}
	\ \left| \frac{\partial^2 \tilde S}{\partial x^i \partial p_x^f}
	\right|^{1/2} 	\; .
	\end{equation}
This explains why the Legendre transform $\tilde S_t$ of $S_t$ has to be
understood as a function of $x^i$ and $p_x^f$. In practice, this means that,
to compute $\tilde S_t$ from Eq.~(\ref{tSt}), the action integral $S_t$ has
to be calculated for a trajectory starting at position $x^i$ and ending with a
momentum $p_x^f$, and that in the additional term $x^f p_x^f$, $x^f$ has to
be interpreted as  $x^f \, (x^i,p_x^f)$.  Finally, note that for a
Hamiltonian, which can be decomposed into a kinetic energy plus potential part
(including the case where a magnetic field is present),
 $\partial p_x^f / \partial x^f $ is always negative just in front of
a focal point and always positive directly after the focal point.
Therefore
	\[
	\tilde \eta_t  = \left\{ \ba{r}
		\eta_t  \quad  {\rm right~after~~a~focal~point} \\
		\eta_t+1 \quad {\rm just~before~a~focal~point}
			\ea \right. \; . \]
Since precisely at focal points $\eta_t$ is incremented by
one unit (for a kinetic energy plus potential Hamiltonian), this
implies that, when crossing a focal point, $\tilde \eta_t$ remains
constant, keeping the value which $\eta_t$ acquires after the focal point.
This latter has to be taken into account for the computation of Maslov indices of 
a trajectory at a focal point.

\subsubsection{Application to the cyclotron motion}

Turning now to the specific problem we are concerned with, i.e.~cyclotron 
motion and diagonal elements of the Green function, we need to calculate
$\tilde S (x^i,y^i,p_x^f,y^f)$, where we can however restrict 
ourselves to $y^f=y^i$ since the partial derivatives are taken only in the 
$x$ direction.  Eq.~(\ref{tSt}) states that, omitting the $y$'s,
$\tilde S_n(x^i,p_x^f) = S_n(x^i,p_x^f) - x^f(x^i,p_x^f) p_x^f$, where
$S_n(x^i,p_x^f)$ is the action integral along a trajectory starting at
the abscissa $x^i$ and arriving with a momentum $p_x^f$.  But here
the Poincar\'e map $(x^i,p_x^i) \rightarrow (x^f,p_x^f)$ is just the
identity, and therefore
	\begin{eqnarray*}
	\tilde S_n(x^i,p_x^f) & = & n S_0 - x^i p_x^f \; , \\
	\left|\frac{\partial^2 \tilde S_n}{\partial x^i \partial p_x^f}
	    \right| & = & 1 \; ;
	\end{eqnarray*}
where $S_0$ is given by Eq.~(\ref{eq:LASdHvA}).
Noting moreover that $\dot{y}^i = \dot{y}^f$ for all trajectories,
and that they pass through two focal points at each turn, one
has from Eq.~(\ref{alt_Gt}) (omitting the Weyl part of G)
	\begin{equation} \label{G3}
	G(\br^i | \br^f \! = \! \br^i) = \sum_n
  		\frac{1}{i \hbar} \frac{1}{\sqrt{ 2 i \pi \hbar}}
		\frac{(-1)^n}{\sqrt{ - 2 i \pi \hbar}}
	       \exp{\left[\frac{i}{\hbar} n S_0 \right]}
		\int \frac{d p_x^f}{\dot{y}^f} \; .
	\end{equation}
At fixed position, $d p_x^f = m \, d \dot{x}^f$ (the vector potential
eliminates). Therefore the remaining integral in Eq.~(\ref{G3}) is just,
up to a multiplication by the mass $m$ of the electron,
an integral over the angle $\theta$ specifying the
direction of the trajectory at $\br$.  For unbounded motion, it simply gives
a factor $2 \pi m$, yielding the expected result
	\begin{equation} \label{app:G_dHvA}
	G(\br^i | \br^f \! = \! \br^i) = \sum_n (-1)^n \frac{m}{\hbar^2}
	\exp(i n S_0/\hbar - i \pi/2) \; .
	\end{equation}
In billiard systems the contribution to $G$ of the
cyclotron orbits is the same as for the unbounded motion,
except that for points $\br$ close to the boundary, Eq.~(\ref{app:G_dHvA})
has to be reduced by a multiplicative factor $\theta_{\rm eff} / (2 \pi)$,
where $\theta_{\rm eff} $ is the angular measure of the trajectories not
affected by the boundary.  The contribution to the density of states of
the cyclotron orbits is thus
	\begin{equation}
	\dosc(E,H) = s(H) \, \frac{\gs A m}{\pi \hbar^2}
	\sum_n (-1)^n  	\cos(n \pi k r_c  ) \; .
	\end{equation}
The multiplicative factor $s(H)$ is given by
	\be
	s(H) =\frac{1}{2\pi A} \int d \br d\theta \zeta(\br,\theta) \; .
	\ee
The function $\zeta(\br,\theta)$ is defined such that
$\zeta = 1$ if the trajectory started at $\br$ with initial velocity
along $\theta$ does not hit the boundary, and $\zeta = 0$ otherwise.
Substituting in the integral above the variables $(\br,\theta)$ by
$(\tilde \br, \tilde \theta)$, where $\tilde \br$ specifies the center
of the cyclotron orbit and $\tilde \theta$ the position on this orbit
(the Jacobian of the transformation is equal to one) and performing
the integral over $\tilde \theta$ since then $\zeta$ depends only on $\tilde
\br$, one obtains
	\be  \label{app:s(H)}
	s(H) =\frac{1}{A} \int d \br \zeta(\tilde \br) \; ,
	\ee
which yields Eq.~(\ref{eq:s(H)}) for the square geometry.

As a final comment, we note that the approach described here for a
two--dimensional electron gas can be generalized in a straightforward manner
to three dimensional systems, including cases with non-spherical Fermi
surfaces.

\newpage

\begin{figure}
\caption{
(a) Evolution of the first 200 energy levels (of one symmetry
class (see section \protect\ref{sec:numeric})) of a square billiard in a
uniform perpendicular magnetic field $H$ as a function of the normalized
flux $\varphi = H a^2/\Phi_0$  $(\Phi_0 = h c/e)$. The energies
are scaled such that the zero field limit gives $\protect E=n_x^2+n_y^2$. 
At high fields the levels converge successively to the Landau levels
while the non--integrable intermediate field regime exhibits a 
complex spectral structure.
(b) Full line:  Numerically calculated susceptibility of the square 
at finite temperature at an energy corresponding to $\sim$1100
occupied independent particle states. The susceptibility, being 
strongly enhanced with respect to the bulk value $\protect\chi_L$, exhibits 
pronounced oscillations which are accurately reproduced by analytical
semiclassical expressions (dashed line) based on families of flux--enclosing 
electron orbits  (shown in the upper insets for
the different magnetic field regimes).}
\label{f1}
\end{figure}


\begin{figure}
\caption{Schematic illustration of the separation of the 
density of states $D(\mu)$ (solid line) into a smooth part
$\bar D$ (dashed line) and an oscillating component.
The total number of electrons $N$ is indicated by the shaded
area, and equal to the product of $\bar D$ and $\bar \mu$.
}
\label{f2}
\end{figure}


\begin{figure}
\caption{
Magnetic susceptibility at zero field for a circular billiard of radius $a$
as a function of $\kf a$ (solid line) from Eq.~(\protect\ref{chicir1}) and
as obtained by keeping only the first two terms of the sum (dashed line).
The typical susceptibility from Eq.~(\protect\ref{chicir1t}) is represented
by the dotted horizontal line.}
\label{fig:chi_circle}
\end{figure}

\begin{figure}
\caption{
First harmonic of the persistent current in a ring with $r=b/a=0.9 \ $
 as a function of $\kf a$ (solid line) for a cut-off length $L_c=6a$ according to 
Eqs.~(\protect\ref{I1})-(\protect\ref{allI1Rs}) together with 
the contribution coming from type-I trajectories (dashed line). 
The typical persistent current
from Eq.~(\protect\ref{i1t}) is represented by the dotted horizontal line.}
\label{fig:chi_ring}
\end{figure}

\begin{figure}
\caption{
First harmonic of the typical persistent current in rings of 
different thickness ($r=b/a$) for
various cut-off lengths $L_c = 30 a$ (circles), $6a$ (diamonds) and $3a$
(triangles) according to Eq.~(\protect\ref{i1t}). Filled symbols correspond
to the total persistent current and lay approximately on a horizontal line
for each $L_c$, consistent with the asymptotic behavior of 
Eq.~(\protect\ref{I1TRn5}) indicated by arrows on the extreme right of 
the plot. Unfilled symbols represent the contributions from each type
of trajectory and are joined by dotted lines (type-I) and dashed lines
(type-II). These guide-to-the-eye exhibits the approximate behavior of 
Eqs.~(\protect\ref{allJtlos}) and shows how the
$r$ characteristic of the switching from one type of trajectories 
to the other increases with temperature.}
\label{fig:chi_ty}
\end{figure}


\begin{figure}
\caption{ 
a) Trajectory from the family~(1,1) of the square
billiard.  The abscissa $x_0$ of the intersection of the trajectory
with the lower side of the square, together with the label $\epsilon=\pm1$
precising the sense of the motion, label the trajectories inside the
family.  
b)  Trajectory from the family~(2,1) of the square billiard, illustrating
the flux cancelation occurring for other periodic trajectories than
those in the (1,1) family (or their repetitions).}
\label{fig:fam11}
\end{figure}

\begin{figure}
\caption{The method of images: The Green function $G(q,q')$ is constructed
from the free Green function $G^0$ by placing a source point at each
mirror image $q_i$ of the actual source $q$.  To each of the resulting
contribution $G^0(q_i,q')$ is associated a classical trajectory
(solid line).  This latter is obtained from the straight line joining
 $q_i$ to $q'$ (dash line) by mapping all its intersected images
back onto the original billiard.}
\label{fig:image}
\end{figure}

\begin{figure}
\caption{
(a) Magnetic susceptibility of a square as a function of $\protect
k_{\scriptscriptstyle F} a$ from numerical calculations (dotted line)
at zero field and at a temperature equal to 10 level-spacings. The number of
electrons is $N=\gs (\kf a)^2/(4 \pi)$. The
full line shows our semiclassical approximation
 (Eq.~(\protect\ref{square:chi1}))
taking into account only the family (11) of shortest orbits
with the temperature correction factor $R_T(L_{11})$. 
The period $\pi/ \protect\sqrt{2}$ of the quantum result indicates the
dominance of the shortest periodic orbits enclosing non-zero area
with length $L_{11}=2\protect\sqrt{2} a$. (b) Susceptibility $\chi$ as a
function of the normalized flux through the sample (at a 
Fermi energy corresponding to $\sim 400$ enclosed electrons) 
from Eq.~(\protect\ref{square:chi1}) (solid)
and numerics (dotted). The susceptibility arising from
the stationary-phase integration $\protect {\cal C}^{\rm S}$ 
(Eq.~(\protect\ref{eq:statphas})) shown as the dashed line diverges at $\protect \varphi
\rightarrow 0$.
}
\label{fig:chi1}
\end{figure}

\begin{figure}
\caption{
(a) Average magnetic susceptibility of an ensemble of squares differing in
size as a function of $\protect k_{\scriptscriptstyle F} a$ for various
temperatures (8, 6 and 4 level spacings for the three triplets of curves
from below) and a flux $\protect\varphi=0.15$. 
Solid line: average of the semiclassical approximation to $\protect\chi^{(2)}$
according to the analytical result of Eq.~(\protect\ref{square:chi}). Dotted line: average of 
$\protect\chi^{(2)}$ obtained by using Eq.~(\protect\ref{DF2}) and exact diagonalization.
Dashed curve: average of the canonical susceptibility calculated directly
from Eq.~(\protect\ref{chiqm}) after the exact diagonalization. The considerable agreement 
between the solid and dotted curves illustrates the precision of the
semiclassical apprroximation, while the agreement between the dotted and
dashed lines shows the applicability of the thermodynamical expansion 
Eq.~(\protect \ref{allDF}).
(b) Flux dependence of the averaged susceptibility normalized to 
$\protect\chi_N = \chi_L k_{\scriptscriptstyle F} a R^2_T(L_{11})$ at 
$\protect k_{\scriptscriptstyle F} a \simeq 70$ from the semiclassical
expression Eq.~(\protect\ref{square:chi}) (solid) and numerical 
calculations
(dashed). The thick solid (dashed) curve denotes an average of
the semiclassical (numerical) result over an ensemble with a large 
dispersion of sizes which is denoted by $\protect \langle \chi \rangle$
(see text). The shift
of the numerical with respect to the semiclassical results
reflects the Landau susceptibility (due to $\protect F^0$ 
in Eq.~(\protect\ref{eq:fd}))
and effects from bouncing--ball orbits (see section \protect
\ref{sec:interfield}) not included in the semiclassical trace. 
}
\label{fig:chi2}
\end{figure}

\begin{figure}
\caption{Solid: low temperature limit of the average susceptibility 
$\overline{ \chi^{(2)}} $ of an ensemble of squares, as given by
Eq.~(\protect \ref{square:chi_tot}) and normalized by $\chi_N$ as defined
in the previous figure caption.
Dashed: contribution of the family (1,1) to this result.  
Even in the very low temperature regime the magnetic response is 
dominated by the 
(1,1) family except for the singularity which develops at zero 
magnetic field.}
\label{fig:repetition}
\end{figure}


\begin{figure}
\caption{Solid: exact (Fresnel) function $\C(\varphi)$ as given
by Eq.~(\protect \ref{Csimple}).  Dashed: approximation of $\C(\varphi)$
by the Bessel function $J_0(32\varphi / \pi^2)$ (see text).}
\label{FresnelvsBessel}
\end{figure}

\begin{figure}
\caption{Solid: function ${\sf F}(\xi)$ (see 
Eq.~(\protect{\ref{eq:Fzeta}}))
describing the magnetic field dependence of the average susceptibility 
for an ensemble of chaotic microstructures.
Dashed: quadratic approximation ($\pi^2/6 - (3\zeta(3) + 2\pi^4/15)\xi^2$)
of ${\sf F}(\xi)$.}
\label{fig:Fzeta}
\end{figure}


\begin{figure}
\caption{
Grand-canonical susceptibility  of a square potential well
as a function of magnetic flux $\protect \varphi =Ha^2/\Phi_0$.
The full lines always denote the quantum mechanical results. Panel (a):
$\protect \chi/\chi_L$ calculated at a Fermi energy of
2140 enclosed electrons at a temperature $\protect
kT/\Delta=8$. Dashed (dotted) line:
Semiclassical result due to bouncing--ball orbits
from Eq.~(\protect\ref{eq:chibb}) with action $\protect S_{10}$
according to the exact expression of Eq.~(\protect\ref{eq:LAS10}), (quadratic 
approximation Eq.~(\protect\ref{eq:S10exp})).
Dashed--dotted line: Susceptibility contribution from family (11) from
Eq.~(\protect\ref{square:chi1}) with offset of $-80$ for reasons of 
representation.
(b)  Dashed line: Semiclassical contribution (Eq.~(\protect\ref{eq:chibbr})) 
from bouncing--ball orbits for 1440 electrons and $\protect kT/\Delta=7$. 
The lower value of $\kf$ makes it necessary to describe the actions by
Eq.~(\protect \ref{eq:LAS10}). 
(c) same as in (b) but for a low temperature $\protect kT/\Delta = 2$ for
which repetitions are important and the use of Eq.~(\protect\ref{eq:chibbr}) is
necessary to approach the quantum results. }
\label{fig:chibb}
\end{figure}

\begin{figure}
\caption{
Schematic representation of a typical flux--enclosing bouncing--ball orbit with
cyclotron radius $\protect r_c$. The dashed lines denote the limits of
the $H$--dependent range of bouncing--ball orbits.  }
\label{fig:bbschema}
\end{figure}

\begin{figure}
\caption{De Haas--van Alphen like oscillations of the susceptibility 
of a square at magnetic fluxes corresponding to $\protect r_c < a/2$ 
for 2140 electrons at $\protect kT/\Delta = 8$. Full line: 
quantum calculations; dashed line: analytical semiclassical
result from cyclotron orbits according to
Eq.~(\protect \ref{sec7:susdHvA}).  }
\label{fig:dhva}
\end{figure}


\begin{figure}
\caption{ Contour of integration in the complex energy plane used
to evaluate the integral Eq.~(\protect \ref{convolution}).  The derivative
$f'(E-\mu)$ of the Fermi function has a periodicity of $2i\pi/\beta$
and double poles located at $E_n = \mu +i(2n+1)\pi/\beta$  ($n$ being
a positive or negative integer).  Moreover, at low temperature,
$f'(E-\mu)$ is essentially zero except for a narrow band of width
$\beta^{-1}$ near the line $\protect {\rm  Re}(E) = \mu$.  
With this contour
of integration, the integrand of Eq.~(\protect \ref{convolution}) has to be
evaluated only in the small domain $[\mu - \beta^{-1},\mu + \beta^{-1}]
\times [0,2i\pi\beta^{-1}]$ where a linearized approximation
of the action is accurate.}
\label{contour}
\end{figure}

\begin{figure}
\caption{
   Integration paths on the invariant tori used to 
   compute the action integrals $I_1$ and $I_2$ for the circular and 
   the ring geometries. a) Path ${\cal C}_1$ (thick dashed) and 
   ${\cal C}_2$ (thick solid) for the circle and type-I
   trajectories of the ring. The straight part of  ${\cal C}_2$ is
   along a classical trajectory (thin solid), whose caustic (dotted)
   is outside the inner disk. b) Path  ${\cal C}_2$ (thick solid) for
   type-II trajectories of the ring. Path ${\cal C}_1$ is similar as
   in a) and therefore not shown. The straight parts of ${\cal C}_2$
   are along classical trajectories. We indicated one of them (solid
   thin) and its caustic (dotted) laying inside the inner disk.
   The thick-dashed line joining the straight  parts of ${\cal C}_2$
   is a guide-to-the-eye putting in evidence the simple form of 
   Eq.~(\protect \ref{eq:action2}).
}
\label{fig:circle}
\end{figure}

\newpage

\begin{table}
\begin{tabular}{c|ccc}
         & $\Dosc(E)/\bar D$ &  $\chi/\cl$ & $\bar{\chi}/\cl$ \\
\hline 
chaotic &  $(\kf a)^{-1~~}$ &  $(\kf a)^{~~~}$  & $(\kf a)^0$         \\
regular &  $(\kf a)^{-1/2}$   &  $(\kf a)^{3/2}$  & $(\kf a)^{~}$      
\end{tabular}
\caption{$(\kf a)$ dependence of the oscillating part of
the density of states and of the magnetic response
depending on the absence (chaotic case) or the presence (regular
case) of continuous families of periodic orbits for two-dimensional
billiard-like microstructures. ($\bar D = (\gs m A)/(2\pi\hbar^2)$ is 
independent of the nature of the dynamics.)}
\label{tab:I} 
\end{table}


\begin{references}



\bibitem{Land} L.D.~Landau, Z.~Phys. {\bf 64}, 629 (1930).

\bibitem{LanLip} L.D.~Landau and E.M.~Lifshitz, {\it Statistical Physics}
                 (Pergamon Press, 1985).

\bibitem{vLeeu21} J.~H.~van Leeuwen, J.~Phys.~(Paris) {\bf 2}, 361 (1921).

\bibitem{Peierls} R.E.~Peierls, {\it Quantum Theory of Solids} (Oxford
University Press, 1964); and {\it Surprises in Theoretical Physics}
(Princeton University Press, Princeton NJ, 1979).

\bibitem{SondWil} E.~H.~Sondheimer and A.~H.~Wilson, 
Proc.~Roy.~Soc.~{\bf 210 A}, 173 (1952).

\bibitem{Pei33}  R.E.~Peierls, Z.~Phys.~{\bf 80}, 763 (1933).

\bibitem{Shoe} D.~Shoenberg, Proc.~Roy.~Soc.~{\bf 170 A}, 341 (1939).

\bibitem{vanLthesis} D.A.~van Leeuwen, Ph.D.~thesis
                  (University of Leiden, unpublished, 1993);
	J.M.~ van Ruitenbeek and D.A.~ van Leeuwen, 
        Mod.~Phys.~Lett. B {\bf 7}, 1053 (1993).

\bibitem{plate} A.~Papapetrou, Z.~Phys.\ {\bf 107}, 387 (1937),
                L.~Friedman, Phys.~Rev.\ {\bf 134 A}, 336 (1964),
                S.~S.~Nedorezov, Zh.~Eksp.~Teor.~Fiz.\ {\bf 64}, 
                624 (1973), [Sov.~Phys.~JETP {\bf 37}, 317 (1973)].

\bibitem{Dingle52} R.B.~Dingle, Proc.~Roy.~Soc.~{\bf 212 A}, 47 (1952).

\bibitem{Denton73} R.V.~Denton, Z.~Phys.~{\bf 265}, 119 (1973).

\bibitem{parabol} R.~N\'emeth, Z.~Phys. B {\bf 81}, 89 (1990).

\bibitem{Bog} E.N.~Bogacheck and G.A.~Gogadze, Pis'ma Zh. Eksp. Teor. Fiz.
{\bf 63}, 1839 (1972), [Sov. Phys. JETP {\bf 36}, 973 (1973)].

\bibitem{Bivin77} D.B.~Bivin and J.~W.~McClure, Phys.~Rev.~{\bf B 16},
                  762 (1977).
                
\bibitem{LOS95} W.~Lehle, Yu.~N.~Ovchinnikov, and A.~Schmid, unpublished.

\bibitem{Sha} B.~Shapiro, Physica A, {\bf 200}, 498 (1993).

\bibitem{rob86} M.~Robnik, J.~Phys. {\bf A 19} 3619 (1986).

\bibitem{antoine}  M.~Antoine, thesis (Universit\'e Paris~VI, unpublished,
                  1991).

\bibitem{vRuitenb} J.M.~van Ruitenbeek, Z.~Phys.~D {\bf 19}, 247 (1991);
	J.M.~ van Ruitenbeek and D.A.~ van Leeuwen, Phys. Rev.
        Lett.~{\bf 67}, 641 (1991).

\bibitem{Imrymeso} Y.~Imry, in {\it Directions in Condensed Matter
Physics}, ed. by G.~Grinstein and G.~Mazenko (World Scientific
Press, Singapore, 1986).

\bibitem{gutz_book} M.C.~Gutzwiller, {\it Chaos in Classical and Quantum
Mechanics} (Springer-Verlag, Berlin, 1990).

\bibitem{RevBoh}O.~Bohigas, in {\it Chaos and
Quantum Physics}, Proceedings of Les Houches summer school,
Session LII, 1989, edited  by M.-J.~Giannoni, A.~Voros, and J.~Zinn-Justin
(North-Holland, Amsterdam, 1991).

\bibitem{Chaos} C.M.~Marcus, R.M~Westervelt, P.F.~Hopkings, 
and A.C.~Gossard, Chaos, {\bf 3}, 643 (1993).

\bibitem{Chaost} H.U.~Baranger, R.A.~Jalabert, and A.D.~Stone, 
Chaos, {\bf 3}, 665 (1993).

\bibitem{levy93} L.P.~L\'evy, D.H.~Reich, L.~Pfeiffer, and K.~West, 
Physica B {\bf 189}, 204 (1993).

\bibitem{BenMailly} D.~Mailly, C.~Chapelier, and A.~Benoit,
                    Phys. Rev. Lett. {\bf 70}, 2020 (1993).

\bibitem{NakTho88} K.~Nakamura and H.~Thomas, Phys. Rev. Lett.
                   {\bf 61}, 247 (1988).

\bibitem{Rezak_etal91} K.~Rezakhanlou, H.~Kunz, and A.~Crisanti,
                  Europhys. Lett. {\bf 16}, 629 (1991).

\bibitem{BIL} M.~B\"{u}ttiker, Y.~Imry, and R.~Landauer, Phys. Lett. A
{\bf 96}, 365 (1983).

\bibitem{von_thesis} F.~von~Oppen, Ph.D. thesis (University
of Washington, unpublished, 1993).

\bibitem{LanMori} R.~Landauer, in {\it Coulomb and Interference Effects in
Small Electronic Structures}, ed. by D.C.~Glattli, M.~Sanquer, and 
J.~Tr\^{a}n Thanh V\^{a}n (Frontiers, Gif-sur-Yvette, 1994).

\bibitem{RiGe} H.F.~Cheung, Y.Gefen, E.K.~Riedel, IBM J.~Res.
Develop. {\bf 32}, 359 (1988).

\bibitem{percon} L.~P.~L\'evy, G.~Dolan, J.~Dunsmuir, and H.~Bouchiat,
                 Phys. Rev. Lett. {\bf 64}, 2074 (1990).

\bibitem{BM} H.~Bouchiat and G.~Montambaux, J.~Phys. (Paris) {\bf 50},
2695 (1989).

\bibitem{Imry} Y.~Imry, in {\it Coherence Effects in Condensed Matter
Systems}, edited by B.~Kramer (Plenum, 1991).

\bibitem{ensemble}  A.~Schmid, Phys. Rev. Lett. {\bf 66}, 80 (1991);
F.~von Oppen and E.~K.~Riedel, {\it ibid} 84; B.~L.~Altshuler, Y.~Gefen,
and Y.~Imry, {\it ibid} 88.

\bibitem{Chandra91} V.~Chandrasekhar, R.~A.~Webb, M.~J.~Brady, 
M.~B.~Ketchen, W.~J.~Gallagher, and A.~Kleinsasser, 
Phys.~Rev.~Lett.~{\bf 67}, 3578 (1991).

\bibitem{Illinois}R.L.~Schult, {\em et al}, Superlattices and
Microstructures, {\bf 11}, 73 (1991).

\bibitem{RivO} F.~von~Oppen, and E.K.~Riedel, Phys.~Rev.~B {\bf 48}, 
9170 (1993).

\bibitem{BerryKeat} M.V.~Berry and J.P.~Keating, J.~Phys.~A {\bf 27}, 
6167 (1994).

\bibitem{Che} O.D.~Cheishvili, Pis'ma Zh. Eksp. Teor. Fiz. {\bf 48}, 206
(1988), [JETP Lett. {\bf 48}, 225 (1988)].

\bibitem{disor} S.~Oh, A.~Yu. Zyuzin, and A.~Serota Phys. Rev. B {\bf 44},
8858 (1991); A.~Raveh and B.~Shapiro, Europhys. Lett. {\bf 19}, 109 (1992);
B.L.~Altshuler, Y.~Gefen, Y.~Imry, and G.~Montambaux, 
Phys.~Rev.~B {\bf 47}, 10340 (1993).

\bibitem{rod2000} K.~Richter, D.~Ullmo, and R.A.~Jalabert unpublished.

\bibitem{AsMe} N.W.~Ashcroft and N.D.~Mermin, {\em Solid State Physics}
(Saunders College, Philadelphia, 1976).

\bibitem{Eisen} J.P.~Eisenstein {\em et al}, Phys. Rev. Lett. {\bf 55}, 
875 (1985).

\bibitem{SivanImry} U.~Sivan and Y.~Imry, Phys.~Rev.~Lett. {\bf 61},
1001, (1988).

\bibitem{jap} D.~Yoshioka, J. Phys. Soc. Japan {\bf 62}, 3198 (1993).

\bibitem{HajShap} J.~Hajdu and B.~Shapiro, Europhys. Lett. {\bf 28},
                  61 (1994).

\bibitem{Harsh} H.~Mathur, M.~G\"ok\c cedag and A.D.~Stone
Phys.~Rev.~Lett {\bf 74}  1855 (1995).

\bibitem{aga94} 0.~Agam, J.~Phys. I (France) {\bf 4} 694 (1994).

\bibitem{gut71} M.C. Gutzwiller, J.~Math.~Phys.~{\bf 11}, 1791 (1970)
J.~Math.~Phys.~{\bf 12}, 343 (1971).

\bibitem{ber76} M.V. Berry, M. Tabor, Proc. R. Soc. Lond.
                A.~{\bf 349}, 101 (1976).

\bibitem{ber77} M.V. Berry, M. Tabor, J.~Phys.~A {\bf 10} 371 (1977).

\bibitem{bal69} R.~Balian and C.~Bloch, Ann.~Phys. {\bf 69}, 76 (1972),
reprinted in: {\em Claude Bloch, Oeuvre Scientifique}, ed. by 
R.~Balian, C.~de Dominicis, V.~Gillet, and A.~Messiah  (North Holland /
American Elsevier, 1975).

\bibitem{ozor86} A.M. Ozorio de Almeida, in {\em Quantum Chaos and
Statistical Nuclear Physics}, Lecture Notes in Physics {\bf 263}, 
ed.~T.~Seligman (New~York, Springer, 1986).

\bibitem{ozor:book} A.M. Ozorio de Almeida, {\em Hamiltonian Systems:
Chaos and Quantization} (Cambridge University Press, 1988).

\bibitem{URJ95}  D.~Ullmo, K.~Richter, and R.A.~Jalabert,
Phys. Rev.~Lett.~{\bf 74}, 383 (1995).

\bibitem{JRU94}  R.A.~Jalabert, K.~Richter, and  D.~Ullmo, in
Ref.~\cite{LanMori}.

\bibitem{vO95} F.~von~Oppen, Phys.~Rev.~B {\bf 50}, 17151 (1994).

\bibitem{gef94} Y.~Gefen, D.~Braun, and G.~Montambaux,
Phys. Rev.~Lett.~{\bf 73}, 154 (1994).

\bibitem{altgef95} A.~Altland and  Y.~Gefen, Phys. Rev.~B {\bf 51}, 10671 (1995).




\bibitem{Prado} S.~D.~Prado, M.~A.~M.~de~Aguiar,
J.P.~Keating and R.~Egydio de Carvalho, J.~Phys.~A {\bf 27}, 6091 (1994).

\bibitem{Kubo64} R.~Kubo, J.~Phys.~Soc.~Japan {\bf 19}, 2127 (1964).

\bibitem{seeley} M.~Seeley, Am.~J.~Math. {\bf 91}, 889 (1969).

\bibitem{wig32} E.P.~Wigner, Phys.~Rev. {\bf 40}, 749 (1932).




\bibitem{Kulik} I.O.~Kulik, Zh. Eksp. Teor. Fiz. {\bf 58} 2171 (1970)
[Sov. Phys. JETP {\bf 31}, 1172 (1970)]; and ZhETF Pis. Red.
{\bf 11} 407 (1970) [JETP Lett. {\bf 11}, 275 (1970)].

\bibitem{arnold:book} V.I.~Arnold, {\em Mathematical Methods of
Classical Mechanics}, (Springer-Verlag, New York, 1984).

\bibitem{keller} J.B.~Keller and S.I.~Rubinow, Annals of Phys. {\bf 9},
24 (1960).

\bibitem{boh95} O.~Bohigas, M.-J. Giannoni, A.M.~Ozorio de Almeida,
and C.~Schmit, Nonlinearity {\bf 8}, 203 (1995).


\bibitem{Gradshtein} I.S.~Gradshteyn and I.M.~Ryzhik, {\em Table of
Integrals, Series and Products}, (Academic Press, New York, 1980).

\bibitem{bra91} M.~Brack, O.~Genzken, and K.~Hansen, Z.~Phys. D {\bf 21},
655 (1991).


\bibitem{gri95} M.~Grinberg, S.~Tomsovic, and D.~Ullmo, unpublished.

\bibitem{han84} J.H.~Hannay and A.M.~Ozorio de Almeida J.~Phys.\
{\bf A17} 3429 (1984).

\bibitem{dor91} E.~Doron, U.~Smilansky and A.~Frenkel, Physica~{\bf D50}
367 (1991), R.~Jensen, Chaos {\bf 1} 101 (1991).


\bibitem{rob86b} M.~Robnik in {\em Nonlinear Phenomena and Chaos},
ed.\ S.~Sarkar (Adam Hilger, Bristol, 1986).

\bibitem{ERWS88} U.~Eichmann, K.~Richter, D.~Wintgen, and W.~Sandner,
Phys.\ Rev.\ Lett. {\bf 61}, 2438 (1988).


\bibitem{inter} V.~Ambegaokar and U.~Eckern, Phys.~Rev.~Lett.~{\bf 65}, 381
 (1991).

\bibitem{JDS}R.~Jalabert and S.~Das~Sarma Phys. Rev. B {\bf 40}, 
9723 (1989).



\bibitem{boh93} O. Bohigas, S. Tomsovic, D. Ullmo,
                Phys.~Rep.~{\bf 223}, 43 (1993).

\bibitem{little90} R. G. Littlejohn,
                J.~Math.~Phys. {\bf 31}, 2952 (1990).

\bibitem{cre90} S.C.~Creagh, J.M.~Robbins, R.G.~Littlejohn,
                Phys.~Rev.~A~{\bf 42}, 1907 (1990).


\bibitem{maslov} V.P.~Maslov and M.V.~Fedoriuk, {\em Semiclassical
Approximation in Quantum Mechanics} (D.~Reidel Publishing Company,
Dordrecht, 1981).

\bibitem{ozor87} A.M. Ozorio de Almeida and J.H.~Hannay,
J.~Phys.~{\bf A20} 5873 (1987).

\end{references}
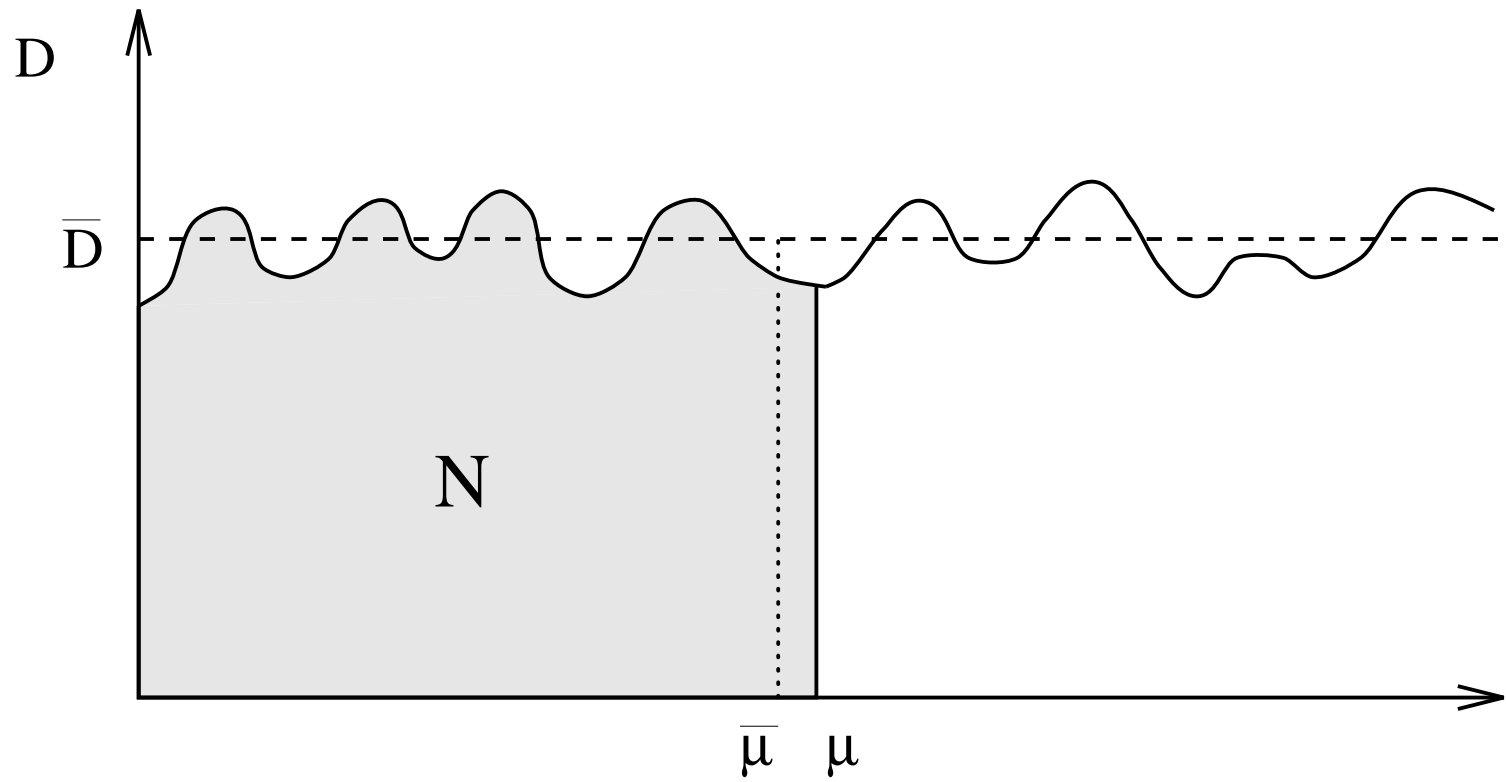
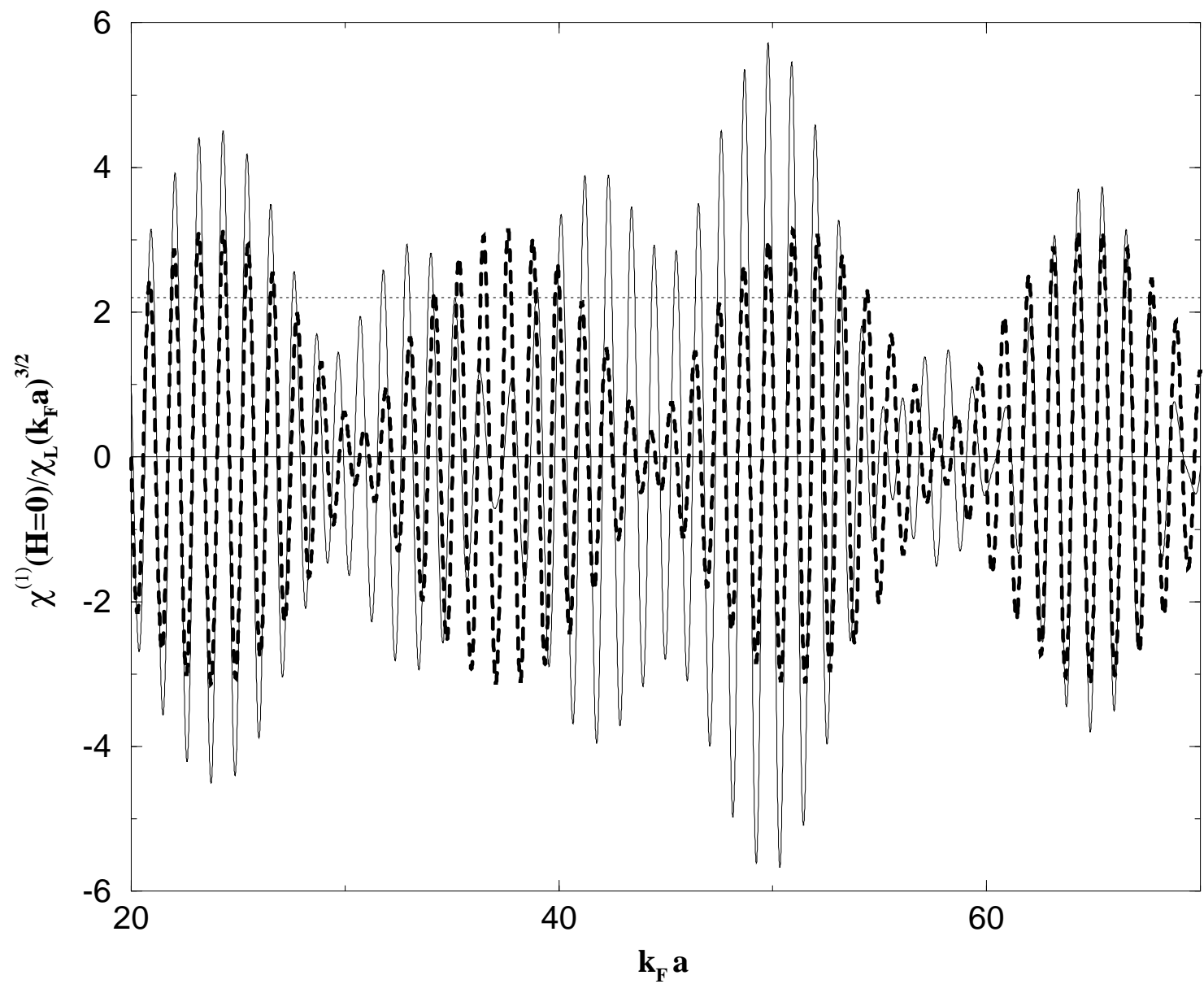
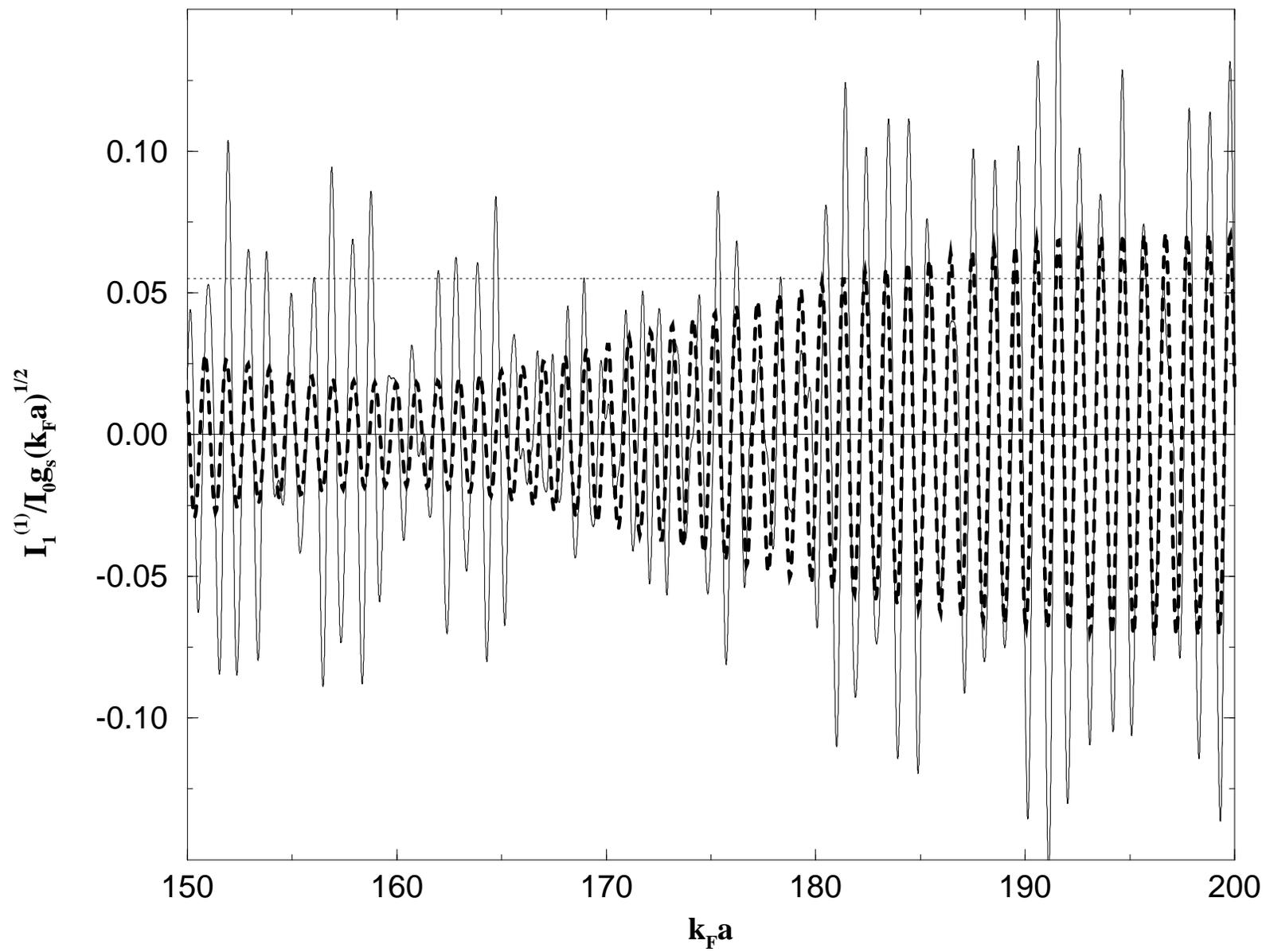
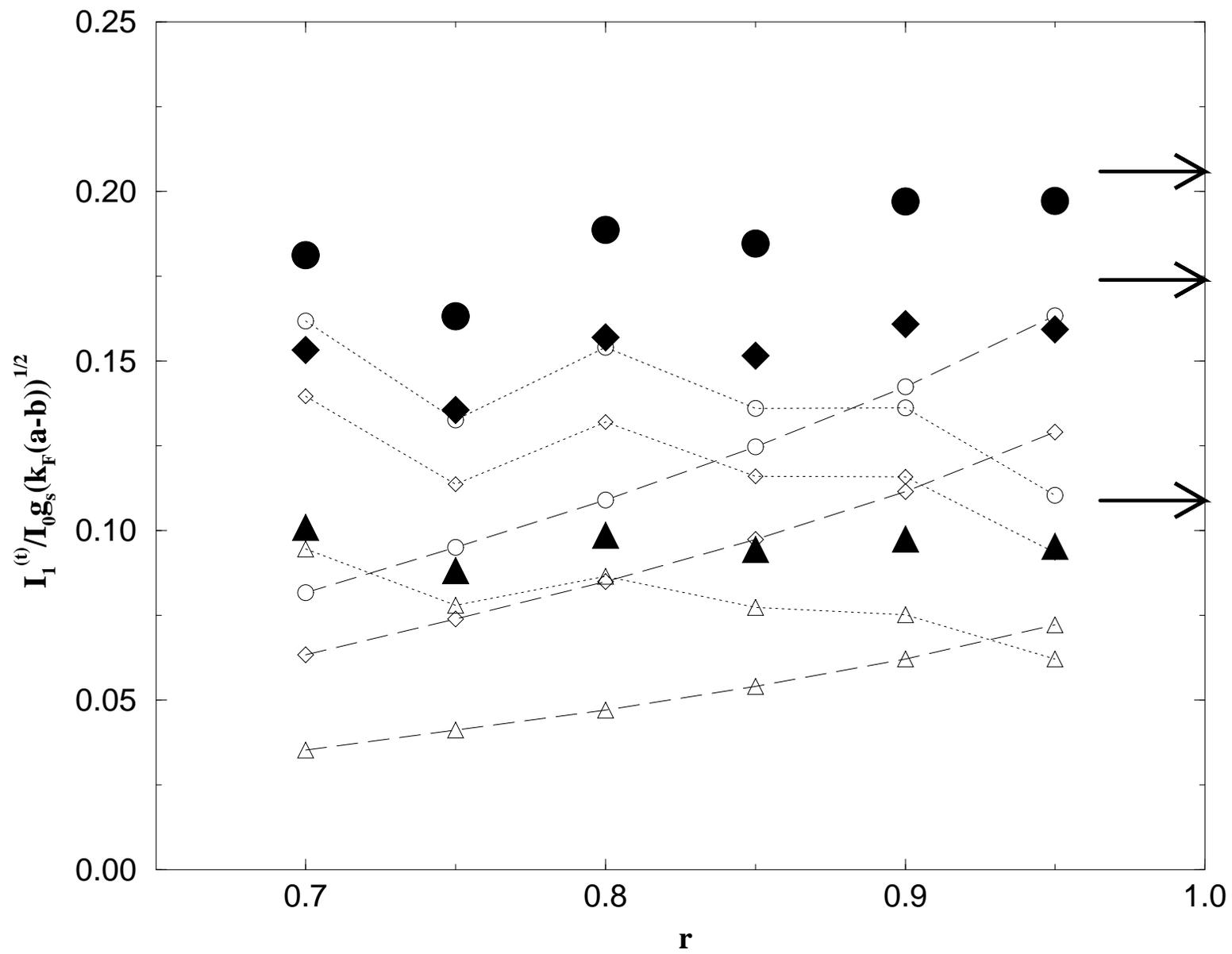
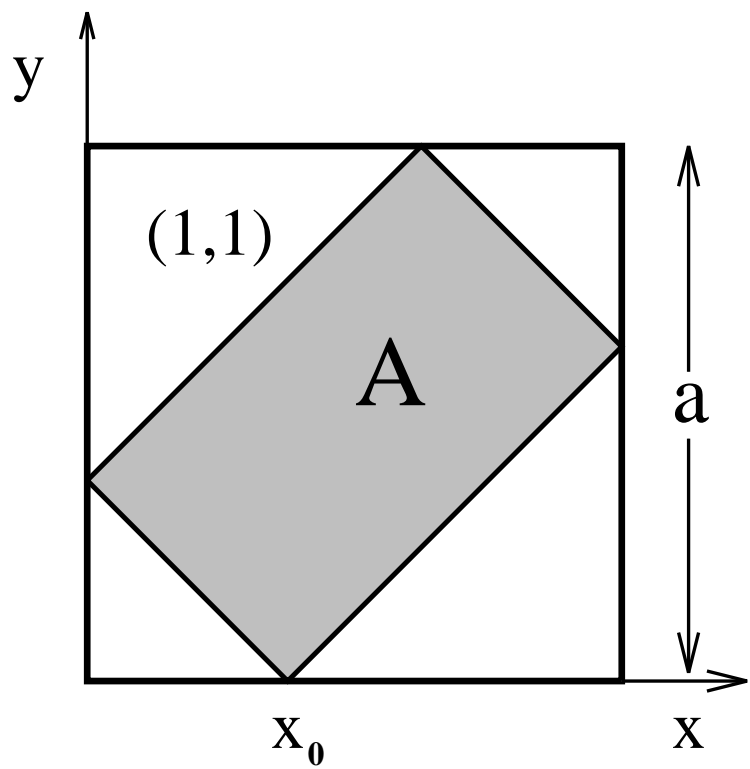
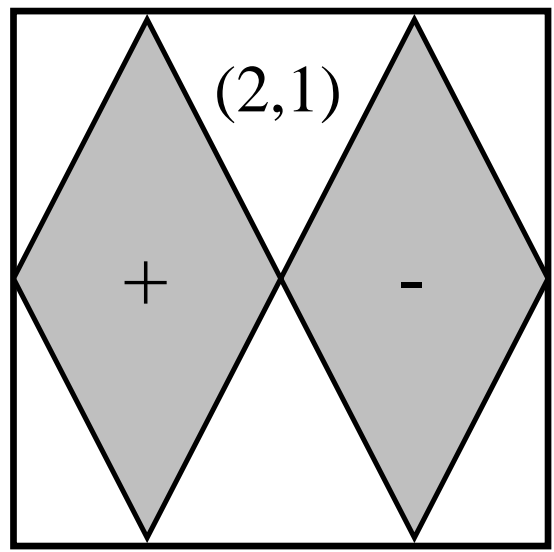
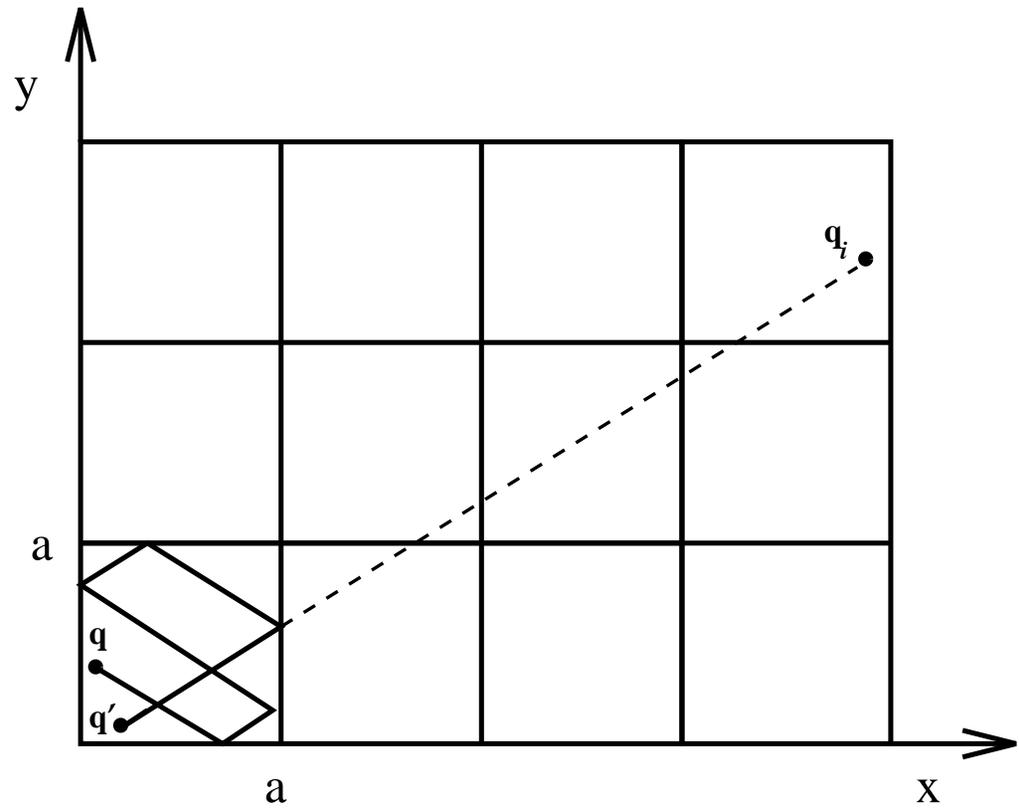
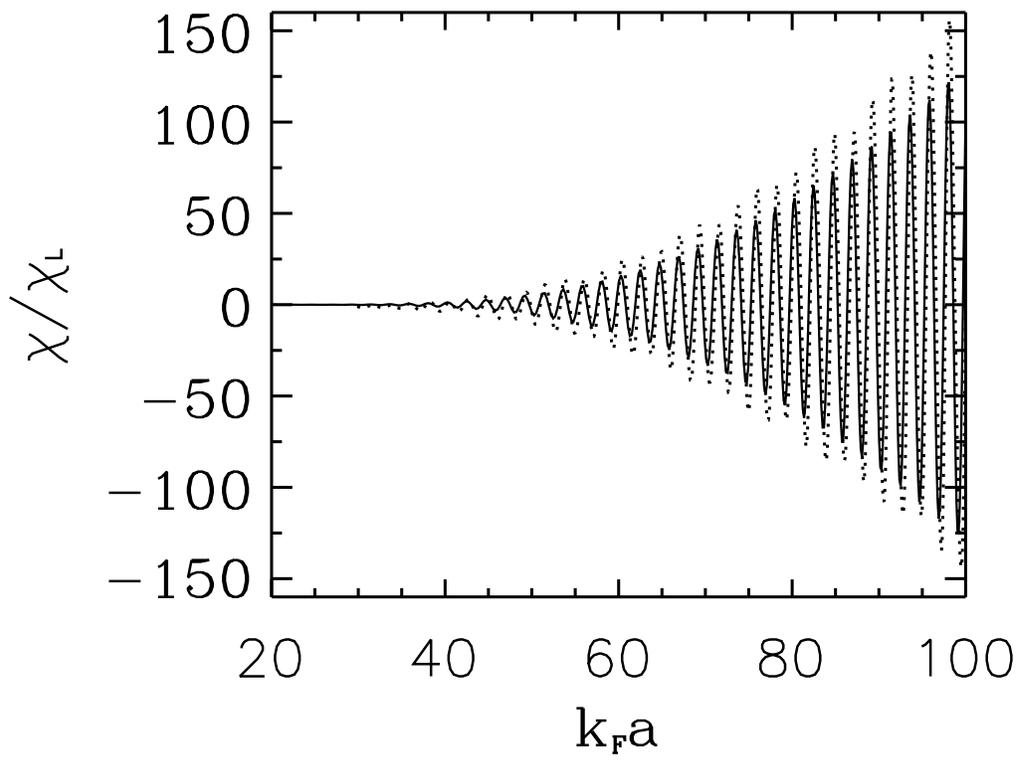
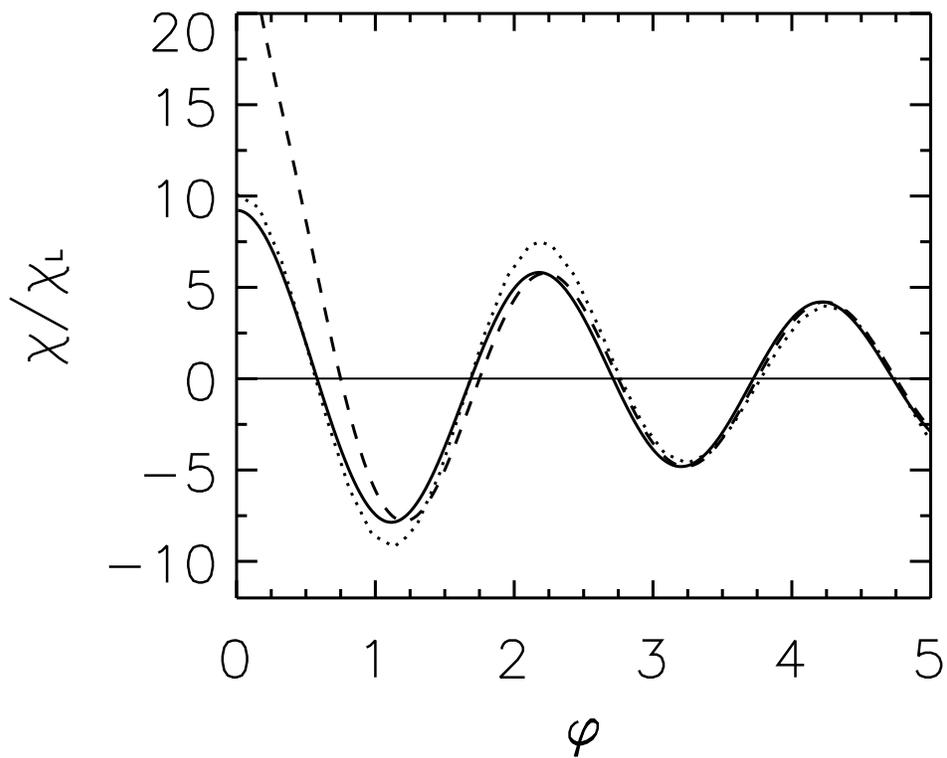
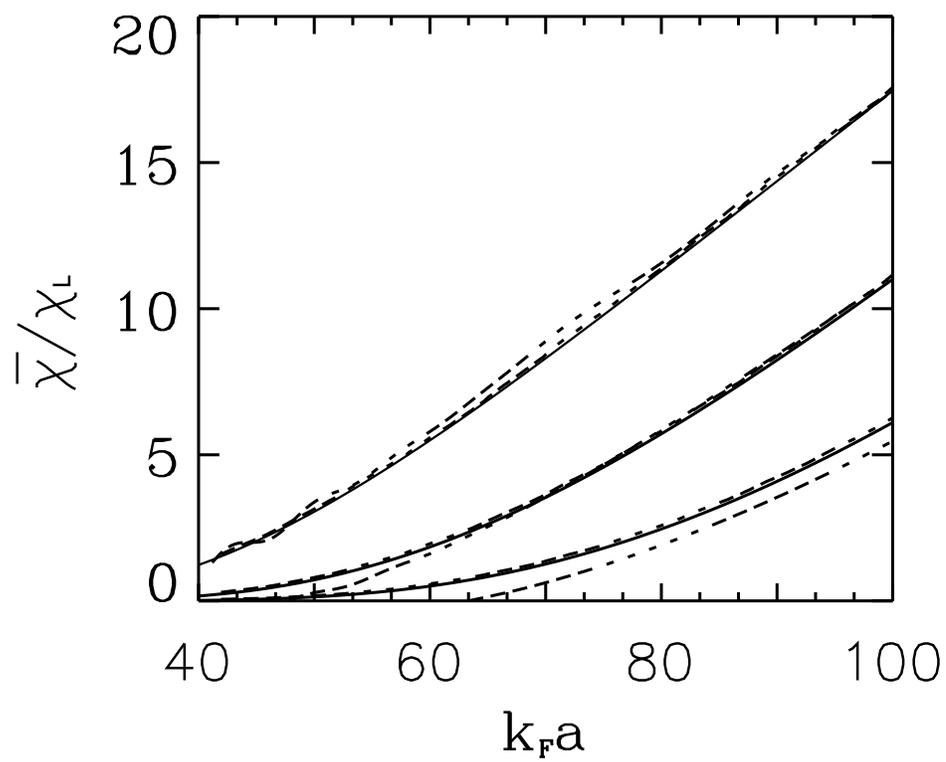
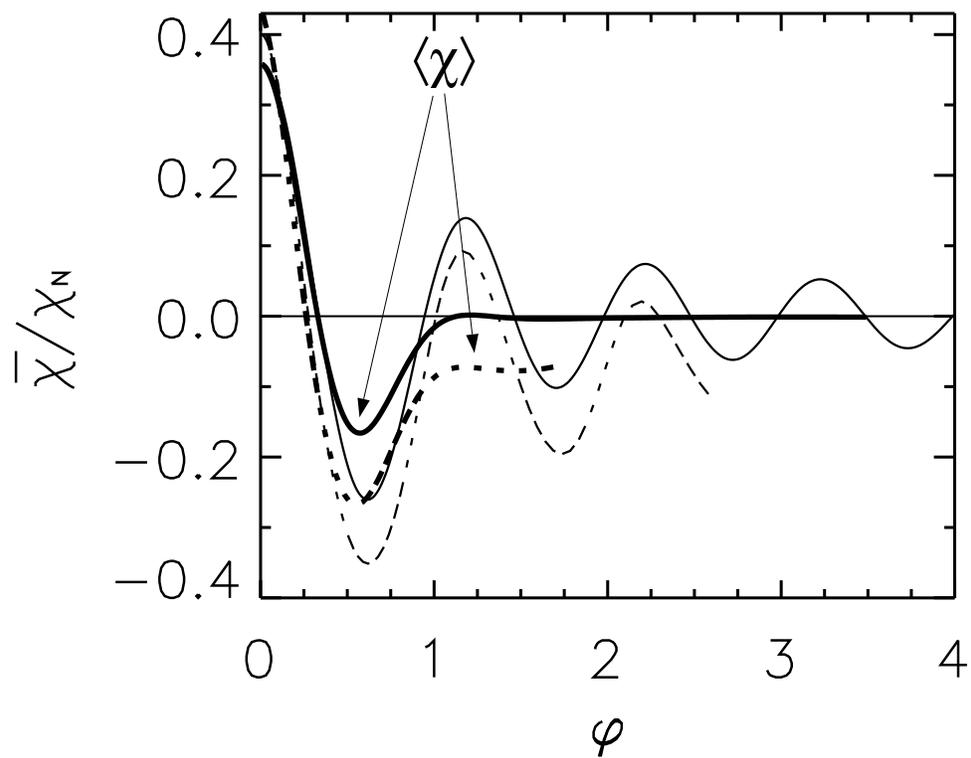
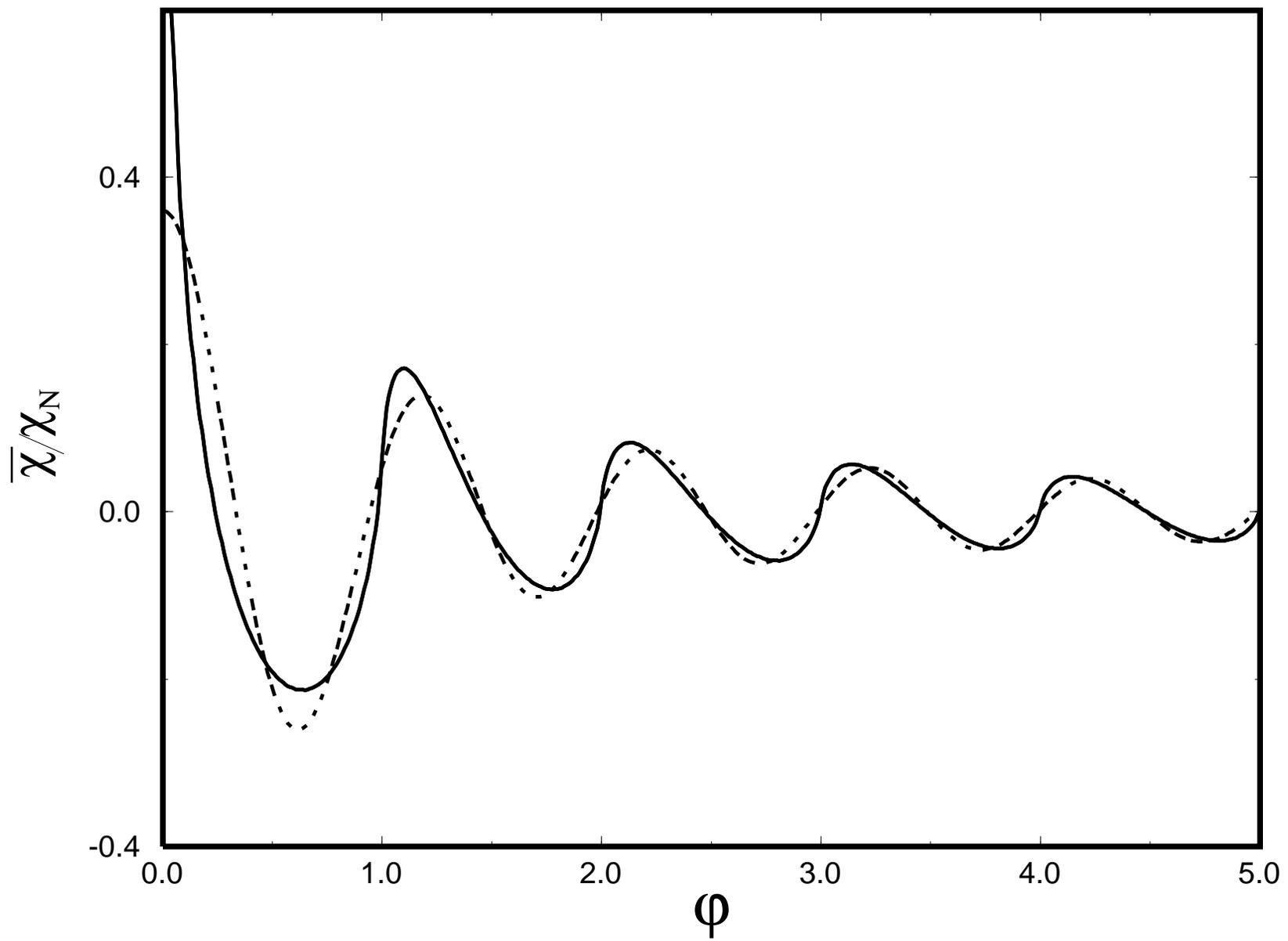
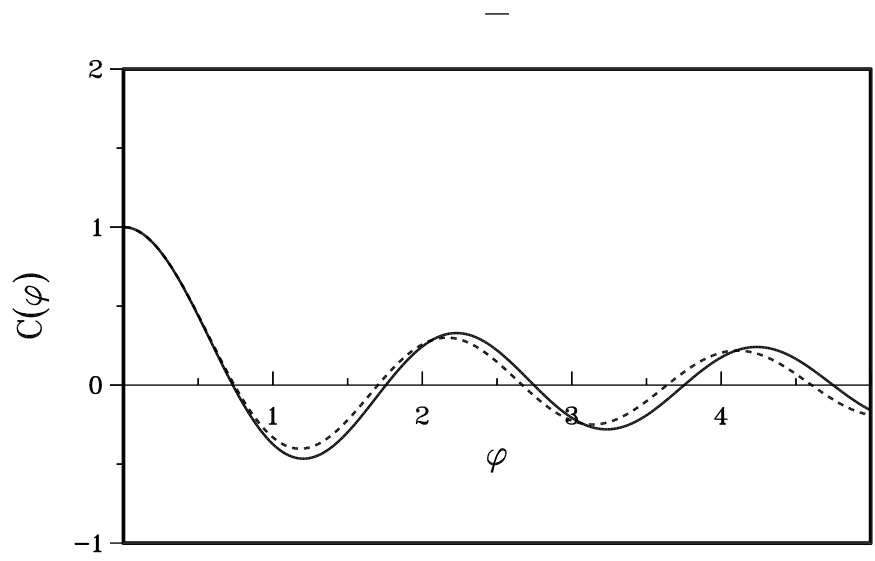
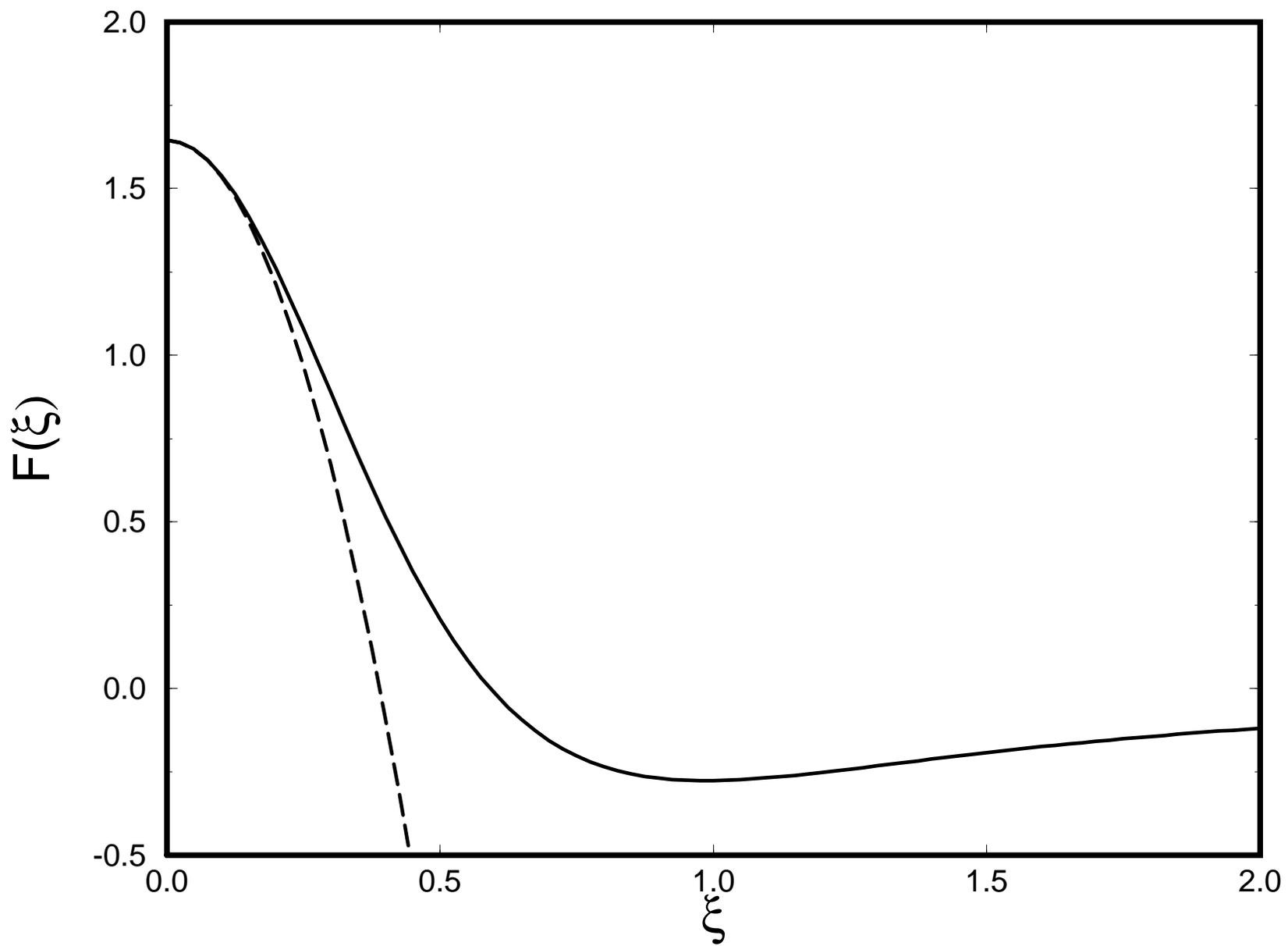
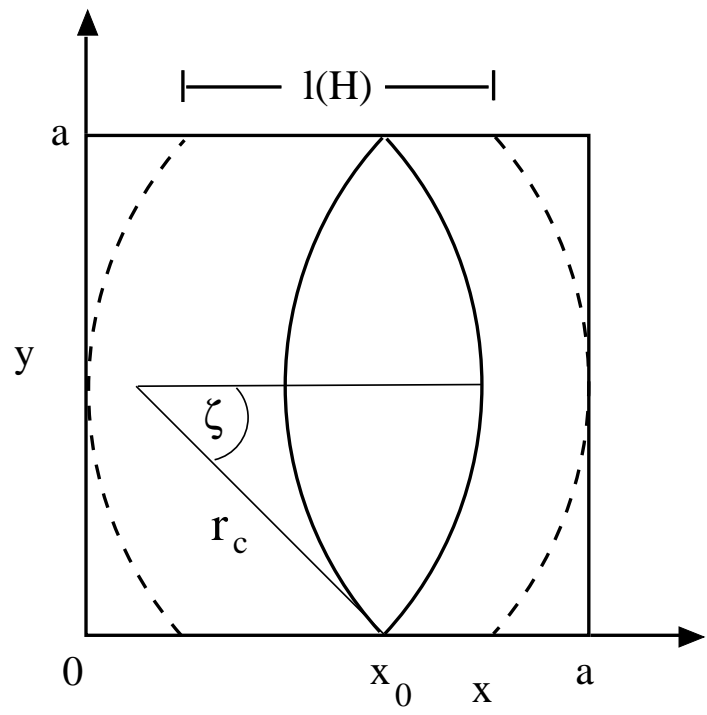
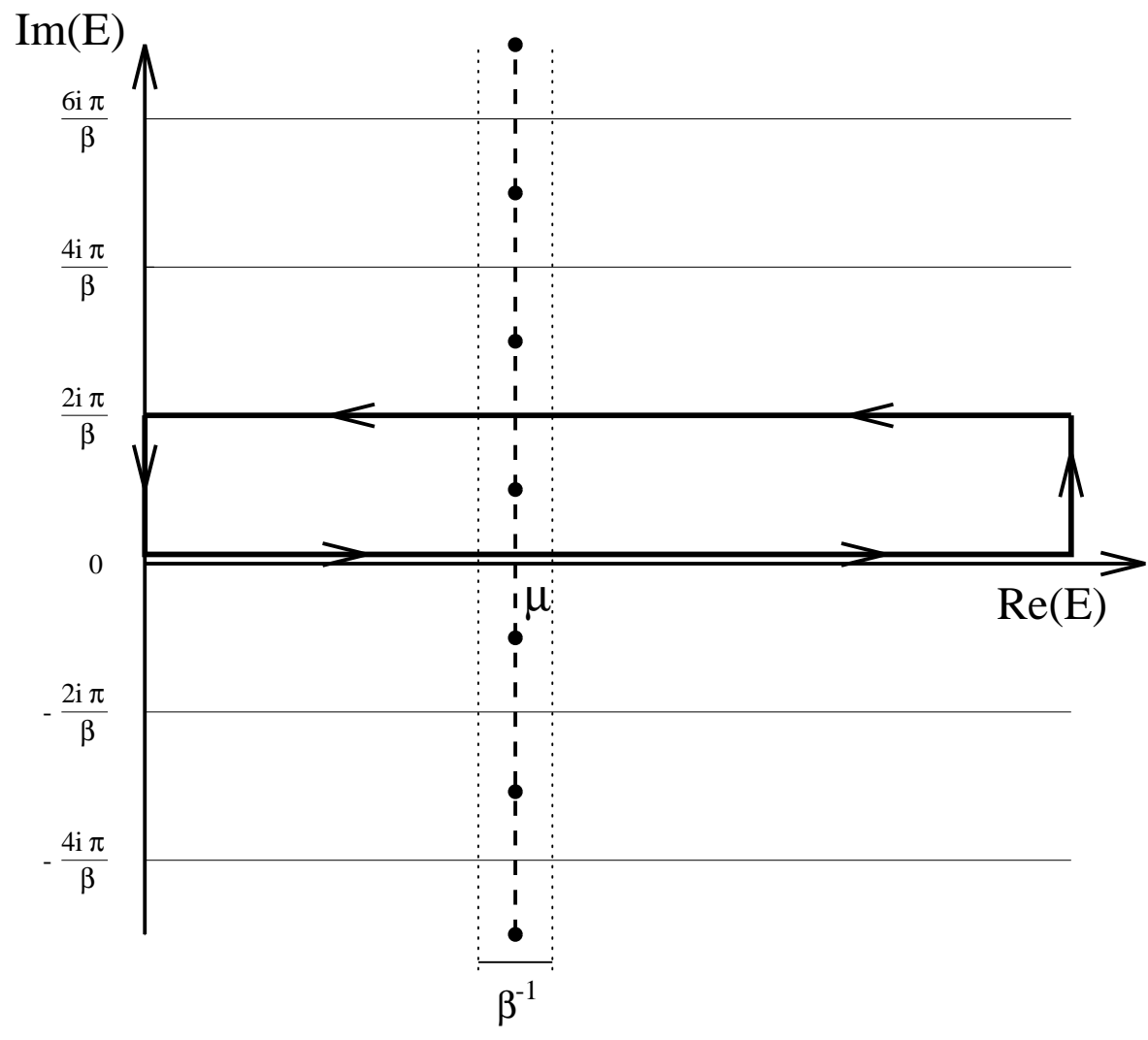
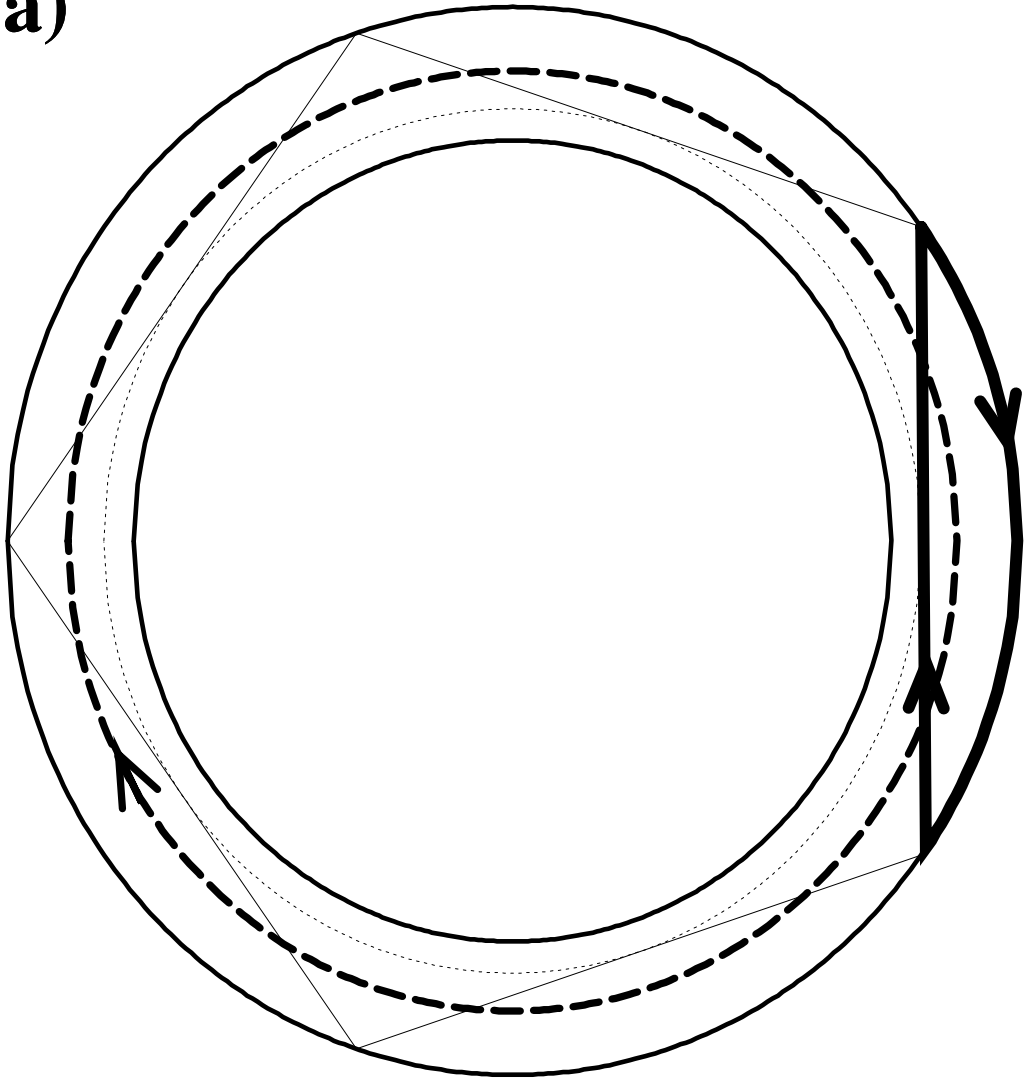
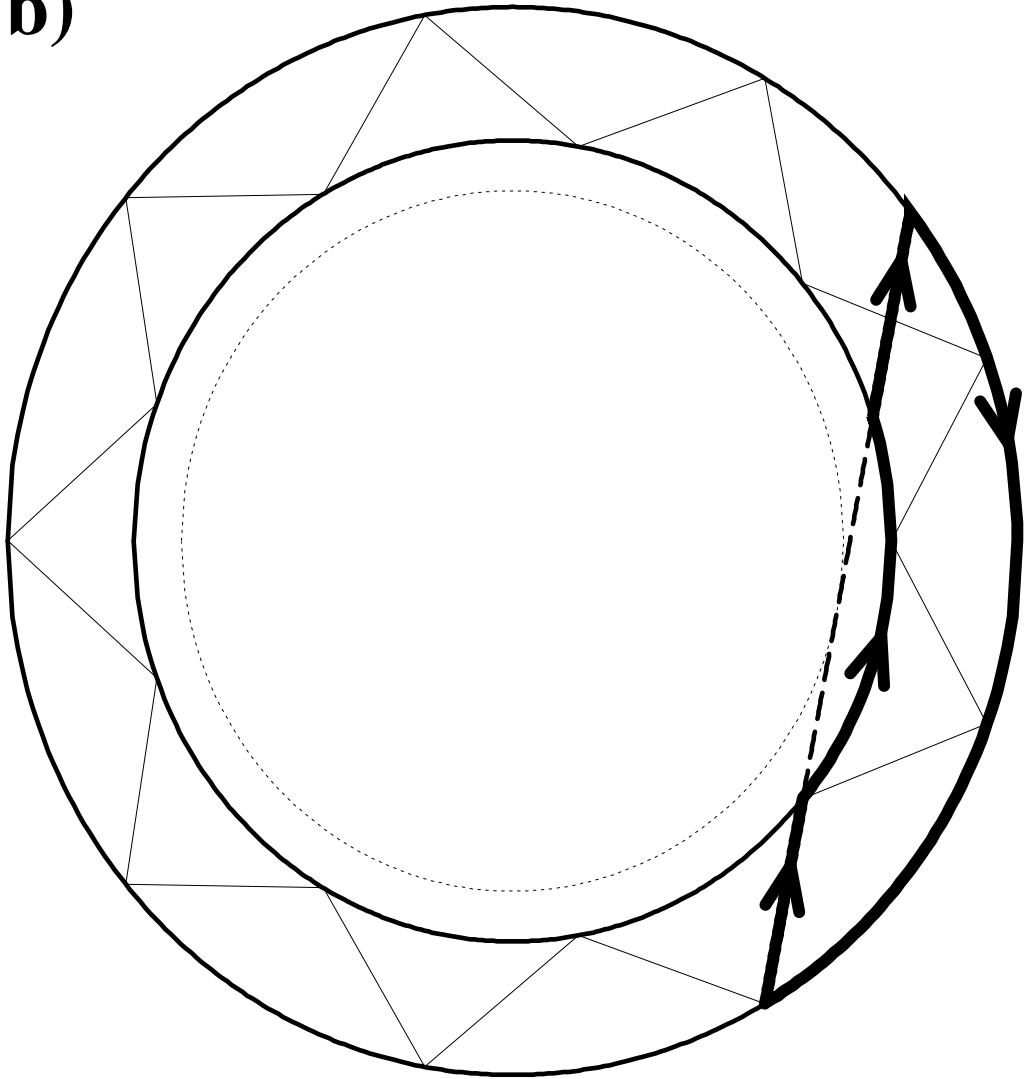
\end{document}